\documentclass[preprint]{ptephy_om}

\preprintnumber{KYUSHU-HET-355} 

\usepackage{graphics}

\usepackage{amsmath} 
\usepackage{amsthm} 
\usepackage{url} 
\usepackage{stmaryrd}
\usepackage{tikz}
\usepackage{tikz-feynhand}
\usepackage{mathtools}
\usepackage{bm}
\usepackage{booktabs}
\usepackage{subcaption}
\usepackage{color}

\DeclareMathOperator{\tr}{tr}

\newcommand{\Slash}[1]{{\ooalign{\hfil/\hfil\crcr$#1$}}}
\numberwithin{equation}{section}

\allowdisplaybreaks

\begin{document}

\title{Gauge-invariant nonperturbative Wilson action in quantum electrodynamics}


\author[1]{Sorato Nagao}
\affil[1]{Department of Physics, Kyushu University, 744 Motooka, Nishi-ku,
Fukuoka 819-0395, Japan}

\author[1,2]{Hiroshi Suzuki}
\affil[2]{Quantum and Spacetime Research Institute (QuaSR), Kyushu University,
744 Motooka, Nishi-ku, Fukuoka 819-0395, Japan}





\begin{abstract}%
By employing the gradient flow exact renormalization group (GFERG), we study
the renormalization group (RG) flow of a manifestly gauge- or
Becchi--Rouet--Stora--Tyutin (BRST)-invariant nonperturbative ansatz of the
one-particle irreducible (1PI) Wilson action in quantum electrodynamics.
The gauge invariance of the Wilson action is \emph{exactly\/} preserved under
the RG flow. We explicitly solve the GFERG equation in the leading and
partially next-to-leading orders of the large-$N_f$ approximation, where $N_f$
is the number of flavors. We obtain gauge-invariant critical exponents and the
gauge-invariant 1PI Wilson action at an infrared (IR) fixed point for~$D<4$,
where $D$ is the spacetime dimension.
\end{abstract}

\subjectindex{B05,B31,B32}

\maketitle

\section{Introduction}
\label{sec:1}
The Wilson exact renormalization group (ERG) or functional renormalization
group dictates the behavior of a physical system under the scale transformation
and also provides a general idea of how quantum field theory in continuum
space-time should be defined~\cite{Wilson:1973jj,Morris:1993qb,Becchi:1996an,Pawlowski:2005xe,Igarashi:2009tj,Rosten:2010vm,Dupuis:2020fhh}. In the context of
elementary particle physics, on the other hand, the gauge symmetry is a
fundamental principle for identifying the correct physical degrees of freedom
and how the gauge symmetry is preserved under the ERG flow is a fundamental
issue. Although ERG is conventionally formulated by employing a momentum cutoff,
which is naively incompatible with the gauge symmetry, the gauge or
Becchi--Rouet--Stora--Tyutin (BRST) symmetry still survives in a modified
form~\cite{Becchi:1996an} as a gauge or BRST Ward--Takahashi (WT) identity (the
quantum master equation). Therefore, we would say that there is no fundamental
issue with the gauge symmetry in ERG.

However, if one wants to find any solution to the ERG flow equation, the
situation poses a serious problem. The ERG flow equation for the Wilson action
is a nonlinear functional differential equation and an exact solution in an
infinite-dimensional theory-space cannot be expected. A possible approach is to
set a plausible ansatz for the Wilson action, which corresponds to a certain
truncation of the theory-space, and determine the flow of unknown parameters
and functions in the ansatz by the ERG flow equation. The ERG equation is then
solved approximately. The trouble with this approach in the conventional ERG is
that the gauge WT identity is also a nonlinear functional differential
equation for the Wilson action. We do not expect an exact solution and there is
no general guarantee that the postulated ansatz fulfills the gauge WT identity.
Therefore, in the conventional ERG framework, \emph{both\/} the ERG flow
equation and the gauge WT identity are only approximately fulfilled.\footnote{%
In perturbation theory, one can solve both equations simultaneously to a given
order of perturbation theory. See~Ref.~\cite{Igarashi:2009tj}.}
This breaking of the gauge symmetry calls into question the physical relevance
of fixed points and associated critical exponents thus obtained.

A manifest gauge symmetry in ERG is thus highly desirable.

One possible implementation of ERG in gauge theory with a manifest gauge
symmetry is the block spin transformation in lattice gauge theory. However,
bearing in mind gravity, which is a more general gauge theory, we think that
ERG in continuous space-time should also be explored.

The gradient flow exact renormalization group (GFERG) considered
in~Refs.~\cite{Sonoda:2020vut,Miyakawa:2021hcx,Miyakawa:2021wus,Sonoda:2022fmk,Miyakawa:2023yob,Sonoda:2025tyu,Nagao:2025sri} is a variant of ERG in gauge
theory that aims to preserve such a manifest gauge symmetry.\footnote{%
For the works of Morris et~al.\ with the same motivation,
see Refs.~\cite{Morris:1999px,Morris:2000fs,Arnone:2005fb,Morris:2005tv}}.
See~Refs.~\cite{Abe:2022smm,Miyakawa:2022qbz,Haruna:2023spq} for related works.
The idea of GFERG is based on an observation that the block-spin process in ERG
for scalar field theory can be understood in terms of heat
diffusion~\cite{Sonoda:2019ibh}. This observation naturally leads to a
formulation of ERG by a gauge covariant diffusion equation, the gradient flow
equation~\cite{Narayanan:2006rf,Luscher:2010iy,Luscher:2011bx,Luscher:2013cpa}.
It turns out that the resulting ERG flow equation preserves a \emph{manifest\/}
gauge- or BRST-invariance. It can also be seen that correlation functions
computed by the Wilson action in GFERG coincide with correlation functions in
the gradient flow formulation~\cite{Luscher:2010iy,Luscher:2011bx,Luscher:2013cpa} up to contact terms (see~Ref.~\cite{Sonoda:2025tyu} for a readable
exposition).

In this paper, we study the renormalization group (RG) flow of a manifestly
gauge- or BRST-invariant nonperturbative ansatz of the one-particle
irreducible (1PI) Wilson action in quantum electrodynamics (QED)\footnote{%
In this paper, we consider space-time dimensions other than four and use the
term QED loosely.} by employing GFERG.\footnote{%
The QED RG functions in the gradient flow renormalization
scheme~\cite{Luscher:2010iy} have recently been studied to very high orders
in~Ref.~\cite{Georg:2026ozz}.} For studies of QED (with four-Fermi
interactions) in the conventional ERG framework, see~Refs.~\cite{Aoki:1996fh,Gies:2004hy,Igarashi:2016gcf,Gies:2020xuh,Igarashi:2021zml,Echigo:2025dia}. The
flow of GFERG \emph{exactly\/} preserves the invariance of the 1PI action under
the \emph{conventional $U(1)$ gauge transformation\/} in~Eq.~\eqref{eq:(2.47)}.
No gauge-noninvariant term such as the photon mass term, except the gauge
fixing term, is induced by the RG flow. In this paper, we postulate a possibly
simplest gauge-invariant nonperturbative form of the 1PI Wilson action in QED;
no four-Fermi interaction is included. We can claim that our simple ansatz
\emph{exactly\/} fulfills the gauge WT identity thanks to the manifest gauge
invariance of GFERG. That is, the gauge WT identity is exactly solved in our
approach.

This paper is organized as follows: In~Section~\ref{sec:2}, we recapitulate the
formulation of GFERG, in particular, in the case of QED. Emphasis is placed on
gauge symmetry and its consequences. We work things out in the 1PI action
language~\cite{Nicoll:1977hi,Wetterich:1992yh,Wetterich:1993ne,Morris:1993qb}.
After some preparation in~Section~\ref{sec:3} for the computation of functional
derivatives, in~Section~\ref{sec:4}, we write down the GFERG flow equation for
the ansatz. The right-hand side of the GFERG flow equation is given
by~Eqs.~\eqref{eq:(4.3)} and~\eqref{eq:(4.4)}; these are the most important
results in this paper. However, since Eqs.~\eqref{eq:(4.3)}
and~\eqref{eq:(4.4)} are quite complicated, we could not solve the GFERG flow
of the ansatz in full generality; this is left as a subject for future work.
Instead, in this paper, we explicitly solve the GFERG flow equation in the
leading and partially next-to-leading orders of the large-$N_f$ approximation,
where $N_f$ is the number of fermion flavors~\cite{Appelquist:1988sr}. We
obtain gauge-invariant critical exponents and the gauge-invariant 1PI Wilson
action at an infrared (IR) fixed point for~$D<4$, where $D$ is the space-time
dimension; this problem has recently been studied by the conformal
bootstrap~\cite{Nakayama:2025mrm}. Section~\ref{sec:5} is devoted to the
conclusion. In~Appendix~\ref{sec:A}, we study the explicit form of integration
vertices in the 1PI action that resulted from our ansatz. Appendix~\ref{sec:B}
gives a certain trick which is helpful in the computation
of~Eq.~\eqref{eq:(4.4)}. Appendix~\ref{sec:C} gives a neat way to manipulate
the gauge WT identities of the vertex functions; this is useful to show the
transversality of~$\mathcal{R}_{\mu\nu}(k)$ in~Eq.~\eqref{eq:(4.3)}.

\section{GFERG in QED}
\label{sec:2}
\subsection{GFERG for the Wilson action}
\label{sec:2.1}
In this paper, we basically follow the formulation of GFERG summarized
in~Ref.~\cite{Sonoda:2025tyu}. For $U(1)$ gauge theory, setting the gauge
group generator as~$T^a\to-i$, Eq.~(5.1) of~Ref.~\cite{Sonoda:2025tyu} yields
the integral representation of the Wilson action~$S_\Lambda$ in QED, i.e.,
a $U(1)$ gauge theory coupled to a Dirac fermion,\footnote{%
Throughout this paper, $D$ denotes the space-time dimension and we set
$\epsilon:=(4-D)/2$.}
\begin{align}
   &e^{S_\Lambda[A,\Bar{c},c,\Bar{\psi},\psi]}
\notag\\
   &:=\mathcal{N}(\Lambda)
   \int[dA'][dc'][d\Bar{c}'][d\psi'][d\Bar{\psi}']\,
\notag\\
   &\qquad\qquad{}
   \times\exp\left\{
   -\frac{\Lambda^2}{2}\int d^Dx\,
   \left[
   A_\mu(x)-\Lambda^{-\epsilon}Z_A(\Lambda)A_\mu'(t,x)
   \right]^2\right\}
\notag\\
   &\qquad\qquad{}
   \times\exp\left\{
   -\Lambda^2\int d^Dx\,
   \left[
   \Bar{c}(x)-Z_{\Bar{c}}(\Lambda)\Lambda^{-\epsilon}Z_A(\Lambda)\Bar{c}'(x)\right]
   \left[c(x)-\Lambda^{-\epsilon}Z_A(\Lambda)c'(t,x)\right]
   \right\}
\notag\\
   &\qquad\qquad{}
   \times\exp\left\{
   i\Lambda\int d^Dx\,
   \left[\Bar{\psi}(x)-Z_\psi(\Lambda)\Bar{\psi}'(t,x)\right]
   \left[\psi(x)-Z_\psi(\Lambda)\psi'(t,x)\right]
   \right\}
\notag\\
   &\qquad\qquad{}
   \times e^{S[A',\Bar{c}',c',\Bar{\psi}',\psi']},
\label{eq:(2.1)}
\end{align}
where $\Lambda$ is the ultraviolet (UV) cutoff of the Wilson action and the
``flow time'' $t$ is related to~$\Lambda$ by
\begin{equation}
   t:=\frac{1}{\Lambda^2}-\frac{1}{\Lambda_0^2},
\label{eq:(2.2)}
\end{equation}
$\Lambda_0$ being the UV cutoff for the bare
theory~$S[A',\Bar{c}',c',\Bar{\psi}',\psi']$. $\mathcal{N}(\Lambda)$ is a
normalization constant defined by
\begin{align}
   \mathcal{N}(\Lambda)
   &:=\biggl\{
   \int[dA][dc][d\Bar{c}][d\psi][d\Bar{\psi}]\,
\notag\\
   &\qquad\qquad{}
   \times
   \exp\left[
   -\frac{\Lambda^2}{2}\int d^Dx\,A_\mu(x)^2\right]
   \exp\left[
   -\Lambda^2\int d^Dx\,\Bar{c}(x)c(x)\right]
\notag\\
   &\qquad\qquad{}
   \times
   \exp\left[
   i\Lambda\int d^Dx\,\Bar{\psi}(x)\psi(x)\right]
   \biggr\}^{-1}.
\label{eq:(2.3)}
\end{align}
Here, we have changed the normalization of fields from that
of~Ref.~\cite{Sonoda:2025tyu} as
$A_\mu'(t,x)\to\Lambda^{-\epsilon}Z_A(\Lambda)A_\mu'(t,x)$,
$c'(t,x)\to\Lambda^{-\epsilon}Z_A(\Lambda)c'(t,x)$,
and~$\Bar{c}'(t,x)\to\Lambda^{-\epsilon}Z_A(\Lambda)\Bar{c}'(t,x)$. GFERG in QED
is also studied in detail in~Ref.~\cite{Miyakawa:2021wus}.

Equation~\eqref{eq:(2.1)}, the ``Reuter formula,'' is a quite general
expression for the Wilson action (see, Eqs. (2.4) and (2.5)
of~Ref.~\cite{Reuter:2019byg}). In this formula, the relation between the
``smeared variables'' such as~$A_\mu'(t,x)$ and the integration
variables such as~$A_\mu'(x)$ specifies the block-spin or coarse graining
procedure. In GFERG, we set it by gauge-covariant diffusion equations, the
gradient flow equations in~Refs.~\cite{Luscher:2010iy,Luscher:2013cpa}:
\begin{align}
   \partial_tA_\mu'(t,x)&=\partial_\nu'F_{\nu\mu}'(t,x)
   +\alpha_0\partial_\mu'\partial_\nu A_\nu'(t,x),&
   A_\mu'(t=0,x)&=A_\mu'(x),
\notag\\
   \partial_t\psi'(t,x)
   &=\left[D_\mu'D_\mu'-\alpha_0\partial_\mu A_\mu'(t,x)\right]\psi'(t,x),&
   \psi'(t=0,x)&=\psi'(x),
\notag\\
   \partial_t\Bar{\psi}'(t,x)
   &=\Bar{\psi}'(t,x)
   \left[\overleftarrow{D}_\mu'\overleftarrow{D}_\mu'
   +\alpha_0\partial_\mu A_\mu'(t,x)\right],&
   \Bar{\psi}'(t=0,x)&=\Bar{\psi}'(x),
\label{eq:(2.4)}
\end{align}
where $\alpha_0$ is a constant and
\begin{equation}
   F_{\mu\nu}'(t,x):=\partial_\mu A_\nu'(t,x)-\partial_\nu A_\mu'(t,x),
\label{eq:(2.5)}
\end{equation}
and
\begin{equation}
   D_\mu':=\partial_\mu-iA_\mu'(t,x),\qquad
   \overleftarrow{D}_\mu':=\overleftarrow{\partial}_\mu+iA_\mu'(t,x).
\label{eq:(2.6)}
\end{equation}
Because of the gauge covariance of the gradient flow equations, the resulting
Wilson action is gauge- or
BRST-invariant~\cite{Sonoda:2020vut,Miyakawa:2021hcx}. The conventional ERG, on
the other hand, corresponds to a use of simple gauge-noncovariant diffusion
equations. For ERG from the perspective of diffusion equations,
see~Refs.~\cite{Abe:2018zdc,Carosso:2018bmz,Carosso:2018rep,Carosso:2019qpb,Matsumoto:2020lha,Tanaka:2022pwt}. For the Faddeev--Popov (FP) ghost field, we set
\begin{equation}
   \partial_tc'(t,x)=\alpha_0\partial^2 c'(t,x),\qquad
   c'(t=0,x)=c'(x).
\label{eq:(2.7)}
\end{equation}
The FP antighost field is not diffused; see Ref.~\cite{Sonoda:2025tyu} for our
reasoning for this, hinted by the prescription in~Ref.~\cite{Luscher:2013cpa}.

The flow equation for the Wilson action~\eqref{eq:(2.1)} (the GFERG equation)
is then simply obtained by taking the derivative with respect
to~$\Lambda$ as\footnote{%
The parameter~$\alpha_0$  does not affect the physics as long
as~$\alpha_0\neq0$ (see~Ref.~\cite{Sonoda:2025tyu}) and we have set~$\alpha_0=1$
in deriving this expression.}
\begin{align}
   &-\Lambda\frac{\partial}{\partial\Lambda}
   e^{S_\Lambda[A,\Bar{c},c,\Bar{\psi},\psi]}
\notag\\
   &=\int d^Dx\,
   \biggl(
   \frac{\delta}{\delta A_\mu(x)}
   \left\{
   \left[-\frac{2}{\Lambda^2}
   \partial^2-\epsilon-\gamma_A(\Lambda)\right]\Hat{A}_\mu(x)
   +\frac{1}{\Lambda^2}\frac{\delta}{\delta A_\mu(x)}\right\}
\notag\\
   &\qquad{}
   +\frac{\delta}{\delta c(x)}
   \left\{
   \left[\frac{2}{\Lambda^2}
   \partial^2+\epsilon+\gamma_A(\Lambda)\right]\Hat{c}(x)
   -\frac{2}{\Lambda^2}
   \frac{\delta}{\delta\Bar{c}(x)}\right\}
\notag\\
   &\qquad{}
   +\frac{\delta}{\delta\Bar{c}(x)}
   \left[\gamma_{\Bar{c}}(\Lambda)+\epsilon+\gamma_A(\Lambda)\right]
   \Hat{\Bar{c}}(x)
\notag\\
   &\qquad{}
   +\frac{\delta}{\delta\Bar{\psi}(x)}
   \left\{
   \frac{2}{\Lambda^2}
   \left[\partial^2
   +2i\Lambda^\epsilon Z_A(\Lambda)^{-1}\Hat{A}_\mu(x)\partial_\mu
   -\Lambda^{2\epsilon}Z_A(\Lambda)^{-2}\Hat{A}_\mu(x)\Hat{A}_\mu(x)\right]
   +\gamma_\psi(\Lambda)\right\}
   \Hat{\Bar{\psi}}(x)
\notag\\
   &\qquad{}
   +\frac{\delta}{\delta\psi(x)}
   \left\{
   \frac{2}{\Lambda^2}
   \left[\partial^2
   -2i\Lambda^\epsilon Z_A(\Lambda)^{-1}\Hat{A}_\mu(x)\partial_\mu
   -\Lambda^{2\epsilon}Z_A(\Lambda)^{-2}\Hat{A}_\mu(x)\Hat{A}_\mu(x)\right]
   +\gamma_\psi(\Lambda)\right\}
   \Hat{\psi}(x)
\notag\\
   &\qquad{}
   -\frac{i}{\Lambda}\frac{\delta}{\delta\psi(x)}
   \frac{\delta}{\delta\Bar{\psi}(x)}
   \biggr)
   e^{S_\Lambda[A,\Bar{c},c,\Bar{\psi},\psi]},
\label{eq:(2.8)}
\end{align}
where we have noted Eq.~\eqref{eq:(2.2)} and introduced
\begin{align}
   \Hat{A}_\mu(x)
   &:=A_\mu(x)+\frac{1}{\Lambda^2}\frac{\delta}{\delta A_\mu(x)},&&
\notag\\
   \Hat{c}(x)
   &:=c(x)+\frac{1}{\Lambda^2}\frac{\delta}{\delta\Bar{c}(x)},&
   \Hat{\Bar{c}}(x)
   &:=\Bar{c}(x)-\frac{1}{\Lambda^2}\frac{\delta}{\delta c(x)},
\notag\\
   \Hat{\psi}(x)
   &:=\psi(x)+\frac{i}{\Lambda}\frac{\delta}{\delta\Bar{\psi}(x)},&
   \Hat{\Bar{\psi}}(x)
   &:=\Bar{\psi}(x)-\frac{i}{\Lambda}\frac{\delta}{\delta\psi(x)}.
\label{eq:(2.9)}
\end{align}
Anomalous dimensions in~Eq.~\eqref{eq:(2.8)} are defined by
\begin{align}
   \gamma_A(\Lambda)&:=-\Lambda\frac{d}{d\Lambda}\ln Z_A(\Lambda),&
   \gamma_{\Bar{c}}(\Lambda)
   &:=-\Lambda\frac{d}{d\Lambda}\ln Z_{\Bar{c}}(\Lambda),
\notag\\
   \gamma_{\psi}(\Lambda)
   &:=-\Lambda\frac{d}{d\Lambda}\ln Z_{\psi}(\Lambda).&&
\label{eq:(2.10)}
\end{align}

\subsection{Ghost sector and the reduced WT identity}
\label{sec:2.2}
In the GFERG equation in the present Abelian theory, Eq.~\eqref{eq:(2.8)}, the
FP ghost sector is completely decoupled and, assuming that the Wilson action is
bilinear in the ghost and antighost fields, its dependence can be exactly
determined. That is, setting\footnote{%
Throughout this paper, we use the abbreviation,
$\int_p:=\int\,d^Dp/(2\pi)^D$.}
\begin{equation}
   S_\Lambda=-\int_p\,
   \Bar{c}(-p)K_c(p)c(p)+(\text{terms independent of $c$ and~$\Bar{c}$}),
\label{eq:(2.11)}
\end{equation}
Eq.~\eqref{eq:(2.8)} implies
\begin{align}
   -\Lambda\frac{\partial}{\partial\Lambda}K_c(p)
   &=\left[
   \frac{2p^2}{\Lambda^2}-\gamma_{\Bar{c}}(\Lambda)-2\epsilon-2\gamma_A(\Lambda)
   \right]K_c(p)
\notag\\
   &\qquad{}
   -\left[\frac{2}{\Lambda^2}p^2+2-\gamma_{\Bar{c}}(\Lambda)
   -2\epsilon-2\gamma_A(\Lambda)\right]
   \frac{1}{\Lambda^2}K_c(p)^2.
\label{eq:(2.12)}
\end{align}
The general solution to this is given by (here, we have
used~Eq.~\eqref{eq:(2.10)})
\begin{equation}
   K_c(p)
   =\frac{p^2}{f(p^2/\mu^2)Z_{\Bar{c}}(\Lambda)
   (\Lambda/\mu)^{-2\epsilon}Z_A(\Lambda)^2e^{-p^2/\Lambda^2}
   +p^2/\Lambda^2},
\label{eq:(2.13)}
\end{equation}
where $f(x)$ is an arbitrary function that is analytic in~$x$; $\mu$ is an
arbitrary mass scale. We then choose the boundary condition for GFERG and the
normalization of the ghost-antighost fields such that\footnote{%
In the dimensionless formulation in which all quantities are measured by
using~$\Lambda$ as the unit, we set $p:=\Bar{p}\Lambda$ and $\mu:=e^\tau\Lambda$
and, in the IR limit $\tau\to\infty$,
$f(p^2/\mu^2)=f(\Bar{p}^2e^{-2\tau})\to\text{const.}$ Therefore, the first
condition in~Eq.~\eqref{eq:(2.14)} corresponds to keeping only this marginal
operator among infinitely many derivative operators produced by the expansion
of~$f(p^2/\mu^2)$.}
\begin{equation}
   f(p^2/\mu^2)=\text{const.},\qquad
   f(p^2/\mu^2)Z_{\Bar{c}}(\Lambda)=(\Lambda/\mu)^{2\epsilon}Z_A(\Lambda)^{-2}.
\label{eq:(2.14)}
\end{equation}
Under these choices,\footnote{%
This is somewhat different from the ghost action
in~Ref.~\cite{Miyakawa:2021wus} (see Eq.~(61) there) because in the present
formulation, imitating the prescription in~Ref.~\cite{Luscher:2013cpa}, the
antighost field~$\Bar{c}$ is not diffused.}
\begin{equation}
   S_\Lambda=-\int_p\,
   \Bar{c}(-p)\frac{p^2}{e^{-p^2/\Lambda^2}+p^2/\Lambda^2}
   c(p)+(\text{terms indep. of $c$ and~$\Bar{c}$}).
\label{eq:(2.15)}
\end{equation}
Equation~\eqref{eq:(2.14)} implies an equality among the anomalous dimensions
in~Eq.~\eqref{eq:(2.10)}:
\begin{equation}
   \gamma_{\Bar{c}}(\Lambda)=-2\epsilon-2\gamma_A(\Lambda).
\label{eq:(2.16)}
\end{equation}

Now, the GFERG equation, Eq.~\eqref{eq:(2.8)}, preserves the WT identity
associated with the BRST symmetry which we assume for the bare action. By using
the ghost action~\eqref{eq:(2.15)}, the WT identity, Eq.~(5.7)
of~Ref.~\cite{Sonoda:2025tyu}, reduces to
\begin{align}
   &k_\mu\frac{\delta S_\Lambda}{\delta A_\mu(k)}
   +\frac{1}{\xi(\Lambda)}e^{2k^2/\Lambda^2}k^2k_\mu
   \left[A_\mu(-k)+\frac{1}{\Lambda^2}\frac{\delta S_\Lambda}{\delta A_\mu(k)}
   \right]
\notag\\
   &\qquad\qquad{}
   -\Lambda^\epsilon Z_A(\Lambda)^{-1}\int_p\,\biggl[
   \Bar{\psi}(-p-k)\frac{\delta}{\delta\Bar{\psi}(-p)}S_\Lambda
   -S_\Lambda\frac{\overleftarrow{\delta}}{\delta\psi(p+k)}\psi(p)
   \biggr]
   =0,
\label{eq:(2.17)}
\end{align}
where $S_\Lambda=S_\Lambda[A,\Bar{\psi},\psi]$. In deriving this, we have noted
that the inverse mapping in~Eq.~(5.8) of~Ref.~\cite{Sonoda:2025tyu}, is
simply given by~$\mathcal{I}_\mu[A;(t,k)]=e^{tk^2}A_\mu(k)$ in the present
Abelian theory with~$\alpha_0=1$. We have also introduced the renormalized
gauge-fixing parameter~$\xi(\Lambda)$ from the bare gauge fixing
parameter~$\xi_0$ by
\begin{equation}
   \frac{1}{\xi(\Lambda)}e^{k^2/\Lambda^2}
   :=\frac{Z_{\Bar{c}}(\Lambda)}{\xi_0}e^{tk^2}.
\label{eq:(2.18)}
\end{equation}
Here, $t$ is identified with~$1/\Lambda^2$ in the limit~$\Lambda_0\to\infty$;
recall~Eq.~\eqref{eq:(2.2)}. From this and Eqs.~\eqref{eq:(2.2)},
\eqref{eq:(2.10)} and~\eqref{eq:(2.16)}, we have
\begin{equation}
   -\Lambda\frac{d}{d\Lambda}\xi(\Lambda)
   =\left[2\epsilon+2\gamma_A(\Lambda)\right]\xi(\Lambda).
\label{eq:(2.19)}
\end{equation}

We emphasize that the WT identity~\eqref{eq:(2.17)}, which is identical
to~Eq.~(65) of~Ref.~\cite{Miyakawa:2021wus}, is linear in the Wilson
action~$S_\Lambda$. It shows that, under the infinitesimal transformations,
\begin{align}
   \delta A_\mu(x)
   &=\left[1
   -\frac{1}{\xi(\Lambda)}e^{-2\partial^2/\Lambda^2}\frac{\partial^2}{\Lambda^2}
   \right]
   \partial_\mu\chi(x),&&
\notag\\
   \delta\psi(x)&=i\Lambda^\epsilon Z_A(\Lambda)^{-1}\chi(x)\psi(x),&
   \delta\Bar{\psi}(x)&=-i\Lambda^\epsilon Z_A(\Lambda)^{-1}\chi(x)\Bar{\psi}(x),
\label{eq:(2.20)}
\end{align}
the Wilson action exhibits a breaking of the particular form,
\begin{equation}
   \delta
   S_\Lambda[A,\Bar{\psi},\psi]
   =\int d^Dx\,A_\mu(x)\frac{1}{\xi(\Lambda)}e^{-2\partial^2/\Lambda^2}
   \partial^2\partial_\mu\chi(x).
\label{eq:(2.21)}
\end{equation}
In the 1PI formulation in~Section~\ref{sec:2.4}, the
transformation~\eqref{eq:(2.20)} takes the form of the conventional $U(1)$
gauge transformation; see~Eq.~\eqref{eq:(2.27)}. This exact symmetry property
of the Wilson action, which is preserved under the flow of GFERG, is crucial in
our study.

\subsection{Chiral symmetry}
\label{sec:2.3}
In~Eq.~\eqref{eq:(2.1)}, we consider the change of variables in the form
of the infinitesimal chiral transformation,
$\psi'(x)\to(1+i\alpha\gamma_5)\psi'(x)$
and~$\Bar{\psi}'(x)\to\Bar{\psi}'(x)(1+i\alpha\gamma_5)$. We assume that the
bare action~$S$ is invariant under these. Then, since flowed fields
$\psi'(t,x)$ and~$\Bar{\psi}'(t,x)$ are transformed in the same way under
the chiral transformation, we find that the Wilson action obeys the chiral WT
identity,
\begin{equation}
   \int d^Dx\,
   \left\{
   \frac{\delta}{\delta\psi(x)}i\gamma_5\Hat{\psi}(x)
   +\tr\left[\frac{\delta}{\delta\Bar{\psi}(x)}\Hat{\Bar{\psi}}(x)i\gamma_5
   \right]
   \right\}e^{S_\Lambda}=0.
\label{eq:(2.22)}
\end{equation}
When the Wilson action is quadratic in fermion fields, this chiral WT identity
is nothing but the Ginsparg--Wilson relation~\cite{Ginsparg:1981bj}.

\subsection{1PI formulation}
\label{sec:2.4}
In what follows, we consider the Wilson action~$S_\Lambda$ in which the
dependence on the ghost fields in~Eq.~\eqref{eq:(2.11)} is subtracted;
$S_\Lambda$ denotes the Wilson action which depends only on the gauge and
fermion fields. Then, the 1PI Wilson action~$\mathit{\Gamma}_\Lambda$ is defined
by the Legendre transformation,
\begin{align}
   \mathit{\Gamma}_\Lambda[\mathcal{A},\Bar{\Psi},\Psi]
   &:=\frac{\Lambda^2}{2}\int d^Dx\,
   \left[\mathcal{A}_\mu(x)-A_\mu(x)\right]^2
   -i\Lambda\int d^Dx\,
   \left[\Bar{\Psi}(x)-\Bar{\psi}(x)\right]\left[\Psi(x)-\psi(x)\right]
\notag\\
   &\qquad{}
   +S_\Lambda[A,\Bar{\psi},\psi],
\label{eq:(2.23)}
\end{align}
where variables of the 1PI action are given by
\begin{align}
   \mathcal{A}_\mu(x)
   &:=A_\mu(x)
   +\frac{1}{\Lambda^2}\frac{\delta S_\Lambda}{\delta A_\mu(x)},&&
\notag\\
   \Psi(x)&:=\psi(x)
   +\frac{i}{\Lambda}\frac{\delta}{\delta\Bar{\psi}(x)}S_\Lambda,&
   \Bar{\Psi}(x)&:=\Bar{\psi}(x)
   -\frac{i}{\Lambda}\frac{\delta}{\delta\psi(x)}S_\Lambda.&
\label{eq:(2.24)}
\end{align}
The relations dual to these are
\begin{align}
   A_\mu(x)
   &=\mathcal{A}_\mu(x)
   -\frac{1}{\Lambda^2}
   \frac{\delta\mathit{\Gamma}_\Lambda}{\delta\mathcal{A}_\mu(x)},&&
\notag\\
   \psi(x)&=\Psi(x)
   -\frac{i}{\Lambda}\frac{\delta}{\delta\Bar{\Psi}(x)}\mathit{\Gamma}_\Lambda,&
   \Bar{\psi}(x)&=\Bar{\Psi}(x)
   +\frac{i}{\Lambda}\frac{\delta}{\delta\Psi(x)}\mathit{\Gamma}_\Lambda.&
\label{eq:(2.25)}
\end{align}
Hence, we obtain
\begin{align}
   \frac{\delta S_\Lambda}{\delta A_\mu(x)}
   &=\frac{\delta\mathit{\Gamma}_\Lambda}{\delta\mathcal{A}_\mu(x)},&&
\notag\\
   \frac{\delta}{\delta\psi(x)}S_\Lambda
   &=\frac{\delta}{\delta\Psi(x)}\mathit{\Gamma}_\Lambda,&
   \frac{\delta}{\delta\Bar{\psi}(x)}S_\Lambda
   &=\frac{\delta}{\delta\Bar{\Psi}(x)}\mathit{\Gamma}_\Lambda.
\label{eq:(2.26)}
\end{align}

It is now interesting to see what transformation is induced on variables in the
1PI action by the transformation~\eqref{eq:(2.20)}~\cite{Sonoda:2022fmk}.
From~Eq.~\eqref{eq:(2.24)}, we have
\begin{align}
   \delta\mathcal{A}_\mu(x)&=\delta A_\mu(x)+\frac{1}{\Lambda^2}
   \frac{\delta\delta S_\Lambda}{\delta A_\mu(x)}
   =\partial_\mu\chi(x),
\notag\\
   \delta\Psi(x)&=
   i\Lambda^\epsilon Z_A(\Lambda)^{-1}\chi(x)\psi(x)
   +i\Lambda^\epsilon Z_A(\Lambda)^{-1}\chi(x)\frac{\delta}{\delta\Bar{\psi}(x)}
   S_\Lambda
\notag\\
   &=ie(\Lambda)\chi(x)\Psi(x),
\notag\\
   \delta\Bar{\Psi}(x)
   &=-ie(\Lambda)\chi(x)\Bar{\Psi}(x),
\label{eq:(2.27)}
\end{align}
where we have used Eq.~\eqref{eq:(2.21)} and introduced the dimensionful gauge
coupling by~\cite{Sonoda:2022fmk}
\begin{equation}
   e(\Lambda):=\Lambda^\epsilon Z_A(\Lambda)^{-1}.
\label{eq:(2.28)}
\end{equation}
Remarkably, these take the form \emph{identical to the conventional $U(1)$
gauge transformation}. Moreover, from~Eq.~\eqref{eq:(2.23)}, we find
\begin{equation}
   \delta\mathit{\Gamma}_\Lambda[\mathcal{A},\Bar{\Psi},\Psi]
   =-\frac{1}{\xi(\Lambda)}\int d^Dx\,
   e^{-2\partial^2/\Lambda^2}\partial^2\chi(x)
   \cdot\partial_\mu\mathcal{A}_\mu(x).
\label{eq:(2.29)}
\end{equation}
This is a property of the 1PI action that is exactly preserved under the GFERG
evolution. This shows that $\mathit{\Gamma}_\Lambda$ has the structure,
\begin{equation}
   \mathit{\Gamma}_\Lambda[\mathcal{A},\Bar{\Psi},\Psi]
   =\mathit{\Gamma}_\Lambda^{\text{inv}}[\mathcal{A},\Bar{\Psi},\Psi]
   -\frac{1}{2\xi(\Lambda)}\int d^Dx\,
   \left[e^{-\partial^2/\Lambda^2}\partial_\mu\mathcal{A}_\mu(x)\right]^2,
\label{eq:(2.30)}
\end{equation}
where $\mathit{\Gamma}_\Lambda^{\text{inv}}$ is invariant under the $U(1)$ gauge
transformation~\eqref{eq:(2.27)}. Since Eq.~\eqref{eq:(2.27)} takes the same
form as the conventional $U(1)$ gauge transformation, it is easy to write down
possible candidates of~$\mathit{\Gamma}_\Lambda^{\text{inv}}$; for the locality of
the theory, $\mathit{\Gamma}_\Lambda^{\text{inv}}$ must be a local functional.

Finally in this subsection, we note that the chiral WT
identity~\eqref{eq:(2.22)} is written in terms of the 1PI variables as
\begin{align}
   &\int d^Dx\,
   \left\{
   \frac{\delta}{\delta\Psi(x)}\mathit{\Gamma}_\Lambda\cdot
   i\gamma_5\Psi(x)
   +\tr\left[
   \frac{\delta}{\delta\Bar{\Psi}(x)}\mathit{\Gamma}_\Lambda\cdot
   \Bar{\Psi}(x)i\gamma_5
   \right]
   \right\}
\notag\\
   &\qquad{}
   +\int d^Dx\,
   \left\{
   \frac{\delta}{\delta\psi(x)}i\gamma_5\Psi(x)
   +\tr\left[
   \frac{\delta}{\delta\Bar{\psi}(x)}\Bar{\Psi}(x)i\gamma_5
   \right]
   \right\}=0.
\label{eq:(2.31)}
\end{align}
In this expression, the first line corresponds to the change
of~$\mathit{\Gamma}_\Lambda$ under the naive chiral transformations,
$\delta\Psi(x)=i\gamma_5\Psi(x)$
and~$\delta\Bar{\Psi}(x)=\Bar{\Psi}(x)i\gamma_5$. The second line is more
subtle and contains a quantum correction to the chiral symmetry; the chiral
anomaly will arise from the second line.

\subsection{Dimensionless formulation}
\label{sec:2.5}
The GFERG equation, Eq.~\eqref{eq:(2.8)}, describes how the Wilson action
changes under the change of the UV cutoff~$\Lambda$. Here, instead, we want to
know how the Wilson action changes under the scale transformation on field
variables.

This can be achieved by measuring all quantities using the UV cutoff~$\Lambda$
as the unit; with these dimensionless variables, the UV cutoff itself becomes
unity and the change in~$\Lambda$ is converted into the scale change in
momenta and coordinates.

Thus, we introduce dimensional variables as
\begin{align}
   \Bar{x}&:=\Lambda x,&&
\notag\\
   \Bar{A}_\mu(\Bar{x})
   &:=\Lambda^{-(D-2)/2}A_\mu(x),&&
\notag\\
   \Bar{\psi}(\Bar{x})
   &:=\Lambda^{-(D-1)/2}\psi(x),&
   \Bar{\Bar{\psi}}(\Bar{x})
   &:=\Lambda^{-(D-1)/2}\Bar{\psi}(x),
\label{eq:(2.32)}
\end{align}
and set
\begin{equation}
   \Bar{S}_\tau[\Bar{\phi}]:=S_\Lambda[\phi],\qquad
   e^{-\tau}:=\frac{\Lambda}{\mu},
\label{eq:(2.33)}
\end{equation}
where $\mu$ is an arbitrary mass scale, and
\begin{equation}
   \gamma_A(\tau):=\frac{d}{d\tau}\ln Z_A(\Lambda),\qquad
   \gamma_{\psi}(\tau)
   :=\frac{d}{d\tau}\ln Z_{\psi}(\Lambda).
\label{eq:(2.34)}
\end{equation}
In this way, the change in~$\Lambda$ or~$\tau$, by keeping $\Bar{x}$ held
fixed, tells us the change of the Wilson action that corresponds to the scale
transformation. For instance, the correlation length measured in~$\Bar{x}$ is
shortened by~$e^{-\delta\tau}$ under~$\tau\to\tau+\delta\tau$. We also define
dimensionless couplings by (recall~Eq.~\eqref{eq:(2.28)})
\begin{equation}
   e_\tau:=\Lambda^{-\epsilon}e(\Lambda)=Z_A(\Lambda)^{-1},\qquad
   \xi_\tau:=\xi(\Lambda).
\label{eq:(2.35)}
\end{equation}
Then, it can be seen that the GFERG equation in the dimensionless variables
yields, removing all bars from variables~\cite{Sonoda:2025tyu}
\begin{align}
   &\frac{\partial}{\partial\tau}
   e^{S_\tau[A,\Bar{\psi},\psi]}
\notag\\
   &=\int d^Dx\,
   \biggl(
   \frac{\delta}{\delta A_\mu(x)}
   \left\{
   \left[-2
   \partial^2-\epsilon-\gamma_A(\tau)\right]\Hat{A}_\mu(x)
   +\frac{\delta}{\delta A_\mu(x)}\right\}
\notag\\
   &\qquad\qquad{}
   +\left(-\frac{D-2}{2}-x\cdot\frac{\partial}{\partial x}\right)
   A_\mu(x)\cdot\frac{\delta}{\delta A_\mu(x)}
\notag\\
   &\qquad\qquad\qquad{}
   +\frac{\delta}{\delta\Bar{\psi}(x)}
   \left\{
   2\left[\partial^2
   +2ie_\tau\Hat{A}_\mu(x)\partial_\mu
   -e_\tau^2\Hat{A}_\mu(x)\Hat{A}_\mu(x)\right]
   +\gamma_\psi(\tau)\right\}
   \Hat{\Bar{\psi}}(x)
\notag\\
   &\qquad\qquad\qquad{}
   +\frac{\delta}{\delta\psi(x)}
   \left\{
   2\left[\partial^2
   -2ie_\tau\Hat{A}_\mu(x)\partial_\mu
   -e_\tau^2\Hat{A}_\mu(x)\Hat{A}_\mu(x)\right]
   +\gamma_\psi(\tau)\right\}
   \Hat{\psi}(x)
\notag\\
   &\qquad\qquad\qquad{}
   -i\frac{\delta}{\delta\psi(x)}
   \frac{\delta}{\delta\Bar{\psi}(x)}
\notag\\
   &\qquad\qquad\qquad{}
   +\left(-\frac{D-1}{2}-x\cdot\frac{\partial}{\partial x}\right)
   \psi(x)\cdot\frac{\delta}{\delta\psi(x)}
\notag\\
   &\qquad\qquad\qquad{}
   +\left(-\frac{D-1}{2}-x\cdot\frac{\partial}{\partial x}\right)
   \Bar{\psi}(x)\cdot\frac{\delta}{\delta\Bar{\psi}(x)}
   \biggr)
   e^{S_\tau[A,\Bar{\psi},\psi]},
\label{eq:(2.36)}
\end{align}
where
\begin{align}
   \Hat{A}_\mu(x)
   &:=A_\mu(x)+\frac{\delta}{\delta A_\mu(x)},&&
\notag\\
   \Hat{\psi}(x)
   &:=\psi(x)+i\frac{\delta}{\delta\Bar{\psi}(x)},&
   \Hat{\Bar{\psi}}(x)
   &:=\Bar{\psi}(x)-i\frac{\delta}{\delta\psi(x)}.
\label{eq:(2.37)}
\end{align}

The 1PI action in the dimensionless formulation is given by
\begin{align}
   \mathit{\Gamma}_\tau[\mathcal{A},\Bar{\Psi},\Psi]
   &:=\frac{1}{2}\int d^Dx\,
   \left[\mathcal{A}_\mu(x)-A_\mu(x)\right]^2
   -i\int d^Dx\,
   \left[\Bar{\Psi}(x)-\Bar{\psi}(x)\right]\left[\Psi(x)-\psi(x)\right]
\notag\\
   &\qquad{}
   +S_\tau[A,\Bar{\psi},\psi],
\label{eq:(2.38)}
\end{align}
where
\begin{align}
   \mathcal{A}_\mu(x)
   &:=A_\mu(x)
   +\frac{\delta S_\tau}{\delta A_\mu(x)}
   =e^{-S_\tau}\Hat{A}_\mu(x)e^{S_\tau},&&
\notag\\
   \Psi(x)&:=\psi(x)
   +i\frac{\delta}{\delta\Bar{\psi}(x)}S_\tau
   =e^{-S_\tau}\Hat{\psi}(x)e^{S_\tau},&
   \Bar{\Psi}(x)&:=\Bar{\psi}(x)
   -i\frac{\delta}{\delta\psi(x)}S_\tau
   =e^{-S_\tau}\Hat{\Bar{\psi}}(x)e^{S_\tau},&
\label{eq:(2.39)}
\end{align}
and
\begin{align}
   A_\mu(x)
   &=\mathcal{A}_\mu(x)
   -\frac{\delta\mathit{\Gamma}_\tau}{\delta\mathcal{A}_\mu(x)},&&
\notag\\
   \psi(x)&=\Psi(x)
   -i\frac{\delta}{\delta\Bar{\Psi}(x)}\mathit{\Gamma}_\tau,&
   \Bar{\psi}(x)&=\Bar{\Psi}(x)
   +i\frac{\delta}{\delta\Psi(x)}\mathit{\Gamma}_\tau,&
\label{eq:(2.40)}
\end{align}
and, therefore,
\begin{align}
   \frac{\delta S_\tau}{\delta A_\mu(x)}
   &=\frac{\delta\mathit{\Gamma}_\tau}{\delta\mathcal{A}_\mu(x)},&&
\notag\\
   \frac{\delta}{\delta\psi(x)}S_\tau
   &=\frac{\delta}{\delta\Psi(x)}\mathit{\Gamma}_\tau,&
   \frac{\delta}{\delta\Bar{\psi}(x)}S_\tau
   &=\frac{\delta}{\delta\Bar{\Psi}(x)}\mathit{\Gamma}_\tau.
\label{eq:(2.41)}
\end{align}

We want to have the GFERG equation for the 1PI action~$\mathit{\Gamma}_\tau$.
First, since the Legendre transformation~\eqref{eq:(2.38)} does not
contain~$\tau$, we have
\begin{equation}
   \frac{\partial}{\partial\tau}
   \mathit{\Gamma}_\tau[\mathcal{A},\Bar{\Psi},\Psi]
   =\frac{\partial}{\partial\tau}
   S_\tau[A,\Bar{\psi},\psi]
\label{eq:(2.42)}
\end{equation}
and the right-hand side of this equation is given by~Eq.~\eqref{eq:(2.36)}.
Employing Eqs.~\eqref{eq:(2.39)}--\eqref{eq:(2.41)},
from~Eq.~\eqref{eq:(2.36)}, we have
\begin{align}
   &\frac{\partial}{\partial\tau}
   \mathit{\Gamma}_\tau[\mathcal{A},\Bar{\Psi},\Psi]
\notag\\
   &=\int d^Dx\,
   \biggl(
   \left[\frac{\delta}{\delta A_\mu(x)}
   +\frac{\delta\mathit{\Gamma}_\tau}{\delta\mathcal{A}_\mu(x)}\right]
   \left\{
   \left[-2
   \partial^2-\epsilon-\gamma_A\right]\mathcal{A}_\mu(x)
   +\frac{\delta\mathit{\Gamma}_\tau}{\delta\mathcal{A}_\mu(x)}\right\}
\notag\\
   &\qquad\qquad\qquad{}
   +\left(-\frac{D-2}{2}-x\cdot\frac{\partial}{\partial x}\right)
   \left[\mathcal{A}_\mu(x)
   -\frac{\delta\mathit{\Gamma}_\tau}{\delta\mathcal{A}_\mu(x)}\right]
   \cdot\frac{\delta\mathit{\Gamma}_\tau}{\delta\mathcal{A}_\mu(x)}
\notag\\
   &\qquad\qquad\qquad{}
   +\left[
   \frac{\delta}{\delta\Bar{\psi}(x)}
   +\frac{\delta}{\delta\Bar{\Psi}(x)}\mathit{\Gamma}_\tau
   \right]
\notag\\
   &\qquad\qquad\qquad{}
   \times
   \left\{
   2\left[\partial^2
   +2ie_\tau\left[
   \mathcal{A}_\mu(x)
   +\frac{\delta}{\delta A_\mu(x)}\right]\partial_\mu
   -e_\tau^2\left[
   \mathcal{A}_\mu(x)
   +\frac{\delta}{\delta A_\mu(x)}
   \right]^2
   \right]
   +\gamma_\psi\right\}
   \Bar{\Psi}(x)
\notag\\
   &\qquad\qquad\qquad{}
   +\left[
   \frac{\delta}{\delta\psi(x)}
   +\frac{\delta}{\delta\Psi(x)}\mathit{\Gamma}_\tau
   \right]
\notag\\
   &\qquad\qquad\qquad{}
   \times
   \left\{
   2\left[\partial^2
   -2ie_\tau\left[
   \mathcal{A}_\mu(x)
   +\frac{\delta}{\delta A_\mu(x)}
   \right]\partial_\mu
   -e_\tau^2\left[
   \mathcal{A}_\mu(x)
   +\frac{\delta}{\delta A_\mu(x)}
   \right]^2
   \right]
   +\gamma_\psi\right\}
   \Psi(x)
\notag\\
   &\qquad\qquad\qquad{}
   -i\left[
   \frac{\delta}{\delta\psi(x)}
   +\frac{\delta}{\delta\Psi(x)}\mathit{\Gamma}_\tau
   \right]
   \frac{\delta}{\delta\Bar{\Psi}(x)}\mathit{\Gamma}_\tau
\notag\\
   &\qquad\qquad\qquad{}
   +\left(-\frac{D-1}{2}-x\cdot\frac{\partial}{\partial x}\right)
   \left[
   \Psi(x)-i\frac{\delta}{\delta\Bar{\Psi}(x)}\mathit{\Gamma}_\tau
   \right]
   \cdot\frac{\delta}{\delta\Psi(x)}\mathit{\Gamma}_\tau
\notag\\
   &\qquad\qquad\qquad{}
   +\left(-\frac{D-1}{2}-x\cdot\frac{\partial}{\partial x}\right)
   \left[
   \Bar{\Psi}(x)+i\frac{\delta}{\delta\Psi(x)}\mathit{\Gamma}_\tau
   \right]
   \cdot\frac{\delta}{\delta\Bar{\Psi}(x)}
   \mathit{\Gamma}_\tau
   \biggr).
\label{eq:(2.43)}
\end{align}
This expression contains the functional derivatives of the Legendre transformed
variables $(\mathcal{A}_\mu,\Psi,\Bar{\Psi})$ with respect to the original field
variables $(A_\mu,\psi,\Bar{\psi})$. By the chain rule of differentiation, the
first derivatives satisfy
\begin{align}
   &\begin{pmatrix}
   \delta_{\mu\rho}&0&0\\
   0&\delta_A^C&0\\
   0&0&\delta_C^A\\
   \end{pmatrix}\delta(x-z)
   =\begin{pmatrix}
   \frac{\delta A_\rho(z)}{\delta A_\mu(x)}&
   \frac{\delta\Bar{\psi}^C(z)}{\delta A_\mu(x)}&
   \frac{\delta\psi_C(z)}{\delta A_\mu(x)}\\
   \frac{\delta}{\delta\Bar{\psi}^A(x)}A_\rho(z)&
   \frac{\delta}{\delta\Bar{\psi}^A(x)}\Bar{\psi}^C(z)&
   \frac{\delta}{\delta\Bar{\psi}^A(x)}\psi_C(z)\\
   \frac{\delta}{\delta\psi_A(x)}A_\rho(z)&
   \frac{\delta}{\delta\psi_A(x)}\Bar{\psi}^C(z)&
   \frac{\delta}{\delta\psi_A(x)}\psi_C(z)\\
   \end{pmatrix}
\notag\\
   &=\int d^Dy\,
   \begin{pmatrix}
   \frac{\delta\mathcal{A}_\nu(y)}{\delta A_\mu(x)}&
   \frac{\delta\Bar{\Psi}^B(y)}{\delta A_\mu(x)}&
   \frac{\delta\Psi_B(y)}{\delta A_\mu(x)}\\
   \frac{\delta}{\delta\Bar{\psi}^A(x)}\mathcal{A}_\nu(y)&
   \frac{\delta}{\delta\Bar{\psi}^A(x)}\Bar{\Psi}^B(y)&
   \frac{\delta}{\delta\Bar{\psi}^A(x)}\Psi_B(y)\\
   \frac{\delta}{\delta\psi_A(x)}\mathcal{A}_\nu(y)&
   \frac{\delta}{\delta\psi_A(x)}\Bar{\Psi}^B(y)&
   \frac{\delta}{\delta\psi_A(x)}\Psi_B(y)\\
   \end{pmatrix}
\notag\\
   &\qquad\qquad\qquad{}
   \times
   \begin{pmatrix}
   \frac{\delta A_\rho(z)}{\delta\mathcal{A}_\nu(y)}&
   \frac{\delta\Bar{\psi}^C(z)}{\delta\mathcal{A}_\nu(y)}&
   \frac{\delta\psi_C(z)}{\delta\mathcal{A}_\nu(y)}\\
   \frac{\delta}{\delta\Bar{\Psi}^B(y)}A_\rho(z)&
   \frac{\delta}{\delta\Bar{\Psi}^B(y)}\Bar{\psi}^C(z)&
   \frac{\delta}{\delta\Bar{\Psi}^B(y)}\psi_C(z)\\
   \frac{\delta}{\delta\Psi_B(y)}A_\rho(z)&
   \frac{\delta}{\delta\Psi_B(y)}\Bar{\psi}^C(z)&
   \frac{\delta}{\delta\Psi_B(y)}\psi_C(z)\\
   \end{pmatrix}.
\label{eq:(2.44)}
\end{align}
Therefore, the first derivatives are given by the inverse of
\begin{align}
   &\begin{pmatrix}
   \frac{\delta A_\rho(z)}{\delta\mathcal{A}_\nu(y)}&
   \frac{\delta\Bar{\psi}^C(z)}{\delta\mathcal{A}_\nu(y)}&
   \frac{\delta\psi_C(z)}{\delta\mathcal{A}_\nu(y)}\\
   \frac{\delta}{\delta\Bar{\Psi}^B(y)}A_\rho(z)&
   \frac{\delta}{\delta\Bar{\Psi}^B(y)}\Bar{\psi}^C(z)&
   \frac{\delta}{\delta\Bar{\Psi}^B(y)}\psi_C(z)\\
   \frac{\delta}{\delta\Psi_B(y)}A_\rho(z)&
   \frac{\delta}{\delta\Psi_B(y)}\Bar{\psi}^C(z)&
   \frac{\delta}{\delta\Psi_B(y)}\psi_C(z)\\
   \end{pmatrix}
\notag\\
   &=\begin{pmatrix}
   \delta_{\nu\rho}\delta(y-z)
   -\dfrac{\delta^2\mathit{\Gamma}_\tau}
   {\delta\mathcal{A}_\nu(y)\delta\mathcal{A}_\rho(z)}&
   -i\dfrac{\delta}{\delta\mathcal{A}_\nu(y)}\mathit{\Gamma}_\tau
   \dfrac{\overleftarrow{\delta}}{\delta\Psi_C(z)}&
   -i\dfrac{\delta}{\delta\mathcal{A}_\nu(y)}
   \dfrac{\delta}{\delta\Bar{\Psi}^C(z)}\mathit{\Gamma}_\tau\\
   -\dfrac{\delta}{\delta\Bar{\Psi}^B(y)}
   \dfrac{\delta\mathit{\Gamma}_\tau}{\delta\mathcal{A}_\rho(z)}&
   \delta_B^C\delta(y-z)-i\dfrac{\delta}{\delta\Bar{\Psi}^B(y)}
   \mathit{\Gamma}_\tau\dfrac{\overleftarrow{\delta}}{\delta\Psi_C(z)}&
   -i\dfrac{\delta}{\delta\Bar{\Psi}^B(y)}
   \dfrac{\delta}{\delta\Bar{\Psi}^C(z)}\mathit{\Gamma}_\tau\\
   \dfrac{\delta\mathit{\Gamma}_\tau}{\delta\mathcal{A}_\rho(z)}
   \dfrac{\overleftarrow{\delta}}{\delta\Psi_B(y)}&
   i\mathit{\Gamma}_\tau\dfrac{\overleftarrow{\delta}}{\delta\Psi_B(y)}
   \dfrac{\overleftarrow{\delta}}{\delta\Psi_C(z)}&
   \delta_C^B\delta(y-z)-i
   \dfrac{\delta}{\delta\Bar{\Psi}^C(z)}\mathit{\Gamma}_\tau
   \dfrac{\overleftarrow{\delta}}{\delta\Psi_B(y)}\\
   \end{pmatrix},
\label{eq:(2.45)}
\end{align}
where we have used~Eq.~\eqref{eq:(2.40)}.

The GFERG equation, Eq.~\eqref{eq:(2.43)}, contains functional derivatives up
to third order, such
as~$\delta^3/(\delta\Bar{\psi}\delta A_\mu\delta A_\mu)\Bar{\Psi}$, whereas the
conventional ERG equation would contain only the first-order functional
derivative. This difference arises from the nonlinear terms in the underlying
diffusion equations, Eqs.~\eqref{eq:(2.4)}, which are required for the gauge
invariance. Those higher derivative terms make computations in GFERG quite
complicated (see the next section) as the price for preserving manifest gauge
invariance.

We note that in the present dimensionless formulation, Eq.~\eqref{eq:(2.30)}
reads
\begin{equation}
   \mathit{\Gamma}_\tau[\mathcal{A},\Bar{\Psi},\Psi]
   =\mathit{\Gamma}_\tau^{\text{inv}}[\mathcal{A},\Bar{\Psi},\Psi]
   -\frac{1}{2\xi_\tau}\int d^Dx\,
   \left[e^{-\partial^2}\partial_\mu\mathcal{A}_\mu(x)\right]^2
\label{eq:(2.46)}
\end{equation}
and $\mathit{\Gamma}_\tau^{\text{inv}}$ is invariant under
\begin{align}
   \delta\mathcal{A}_\mu(x)&=\partial_\mu\chi(x),&&
\notag\\
   \delta\Psi(x)&=ie_\tau\chi(x)\Psi(x),&
   \delta\Bar{\Psi}(x)
   &=-ie_\tau\chi(x)\Bar{\Psi}(x).
\label{eq:(2.47)}
\end{align}
This structure of the gauge invariance of the 1PI action is preserved under
the flow of GFERG~\eqref{eq:(2.43)}. In particular, besides the last
gauge-fixing term of~Eq.~\eqref{eq:(2.46)} with a renormalized gauge
parameter~$\xi_\tau$, no gauge-noninvariant term, such as the photon mass term,
is induced by our ERG flow. From Eqs.~\eqref{eq:(2.10)} and~\eqref{eq:(2.19)},
we also have
\begin{equation}
   \frac{d}{d\tau}e_\tau=-\gamma_A(\tau)e_\tau,\qquad
   \frac{d}{d\tau}\xi_\tau
   =\left[2\epsilon+2\gamma_A(\tau)\right]\xi_\tau.
\label{eq:(2.48)}
\end{equation}

The chiral WT identity in the dimensionless formulation takes the same form
as~Eq.~\eqref{eq:(2.31)}:
\begin{align}
   &\int d^Dx\,
   \left\{
   \frac{\delta}{\delta\Psi(x)}\mathit{\Gamma}_\tau\cdot
   i\gamma_5\Psi(x)
   +\tr\left[
   \frac{\delta}{\delta\Bar{\Psi}(x)}\mathit{\Gamma}_\tau\cdot
   \Bar{\Psi}(x)i\gamma_5
   \right]
   \right\}
\notag\\
   &\qquad{}
   +\int d^Dx\,
   \left\{
   \frac{\delta}{\delta\psi(x)}i\gamma_5\Psi(x)
   +\tr\left[
   \frac{\delta}{\delta\Bar{\psi}(x)}\Bar{\Psi}(x)i\gamma_5
   \right]
   \right\}=0.
\label{eq:(2.49)}
\end{align}

This completes a recapitulation of GFERG in QED.

\section{nonperturbative ansatz for the 1PI action}
\label{sec:3}
\subsection{Ansatz and the $-1$ variables}
\label{sec:3.1}
In this paper, we study the 1PI Wilson action~\eqref{eq:(2.46)} with the
following particular gauge-invariant part:
\begin{align}
   &\mathit{\Gamma}_\tau^{\text{inv}}[\mathcal{A},\Bar{\Psi},\Psi]
\notag\\
   &=-\int d^Dx\,
   \frac{1}{4}
   \left[\partial_\mu\mathcal{A}_{-1\nu}(x)
   -\partial_\nu\mathcal{A}_{-1\mu}(x)\right]
   \mathcal{K}_\tau(-\partial^2)
   \left[\partial_\mu\mathcal{A}_{-1\nu}(x)
   -\partial_\nu\mathcal{A}_{-1\mu}(x)\right]
\notag\\
   &\qquad{}
   +i\int d^Dx\,\Bar{\Psi}_{-1}(x)
   \left[\Slash{\partial}-ie_\tau\Slash{\mathcal{A}}_{-1}(x)
   -m_\tau\right]\Psi_{-1}(x).
\label{eq:(3.1)}
\end{align}
This would be the simplest but still nontrivial gauge-invariant
nonperturbative ansatz or truncation for the 1PI Wilson action. Note that this
expression is written in field variables with the subscript~$-1$, the ``$-1$
variables'' introduced in~Ref.~\cite{Sonoda:2022fmk}. These are given from the
original variables as the solution to flow equations,
\begin{align}
   \partial_t\mathcal{A}_{t\mu}(x)
   &=\partial_\nu
   \left[
   \partial_\nu\mathcal{A}_{t\mu}(x)-\partial_\mu\mathcal{A}_{t\nu}(x)
   \right]
   +\alpha_0\partial_\mu\partial_\nu\mathcal{A}_{t\nu}(x),&
   \mathcal{A}_{0\mu}(x)&=\mathcal{A}_\mu(x),
\notag\\
   \partial_t\Psi_t(x)
   &=\left\{
   \left[\partial_\mu-ie\mathcal{A}_{t\mu}(x)\right]^2
   +ie\alpha_0\left[\partial_\mu\mathcal{A}_{t\mu}(x)\right]
   \right\}\Psi_t(x),&
   \Psi_0(x)&=\Psi(x),
\notag\\
   \partial_t\Bar{\Psi}_t(x)
   &=\left\{
   \left[\partial_\mu+ie\mathcal{A}_{t\mu}(x)\right]^2
   -ie\alpha_0\left[\partial_\mu\mathcal{A}_{t\mu}(x)\right]
   \right\}\Bar{\Psi}_t(x),&
   \Bar{\Psi}_0(x)&=\Bar{\Psi}(x),
\label{eq:(3.2)}
\end{align}
by solving \emph{backward\/} from~$t=0$ to~$t=-1$ (in the present paper, we
set~$\alpha_0=1$).\footnote{%
The time duration $\Delta t$ of the dimensionful flow time~\eqref{eq:(2.2)}
between~$\Lambda=\Lambda_0$ (the true UV cutoff) and $\Lambda$
is~$\Delta t=1/\Lambda^2-1/\Lambda_0^2$ which
becomes, for~$\Lambda_0\to\infty$, $\Delta\Bar{t}=1$ in the dimensionless flow
time~$\Bar{t}:=\Lambda^2t$ assumed in~Eq.~\eqref{eq:(3.2)}. Therefore, the $-1$
variables can be naturally interpreted as initial conditions of the flow
equations at~$\Lambda_0\to\infty$, which reproduces the original variables at
the cutoff~$\Lambda$.} In these differential equations, the parameter~$e$ is
regarded as a constant and, after solving the equations from~$t=0$ to~$t=-1$,
the parameter~$e$ is replaced by~$e_\tau$. We note that for~$\alpha_0=1$,
$\mathcal{A}_{-1\mu}(x)$ is simply given by
\begin{equation}
   \mathcal{A}_{-1\mu}(x)=e^{-\partial^2}\mathcal{A}_\mu(x),
\label{eq:(3.3)}
\end{equation}
whereas $\Psi_{-1}(x)$ and~$\Bar{\Psi}_{-1}(x)$ are infinite order
in~$\mathcal{A}_\mu(x)$ through flow equations, Eqs.~\eqref{eq:(3.2)}.
In~Appendix~\ref{sec:A}, we give the expression of the $-1$~variables in terms
of the original variables.

We note that Eq.~\eqref{eq:(3.1)}, written in terms of the $-1$~variables, is
invariant under the infinitesimal gauge transformation~\eqref{eq:(2.47)} and
thus can be an ansatz for the gauge-invariant part of the 1PI action. To see
this, for the infinitesimal gauge transformation~\eqref{eq:(2.47)}, we define
the flow of the gauge transformation function~$\chi(x)$ by
\begin{equation}
   \partial_t\chi_t(x)=\alpha_0\partial^2\chi_t(x),\qquad
   \chi_0(x)=\chi(x).
\label{eq:(3.4)}
\end{equation}
Then, it can be seen that $\mathcal{A}_{t\mu}(x)+\partial_\mu\chi_t(x)$,
$\Psi_t(x)+ie\chi_t(x)\Psi_t(x)$,
and~$\Bar{\Psi}_t(x)-ie\chi_t(x)\Bar{\Psi}_t(x)$ also solve
Eqs.~\eqref{eq:(3.2)}. This shows that the gauge
transformation~\eqref{eq:(2.47)} induces the gauge transformation on the
$-1$ variables as, $\delta\mathcal{A}_{-1\mu}(x)=\partial_\mu\chi_{-1}(x)$,
$\delta\Psi_{-1}(x)=ie\chi_{-1}(x)\Psi_{-1}(x)$,
and~$\delta\Bar{\Psi}_{-1}(x)=-ie\chi_{-1}(x)\Bar{\Psi}_{-1}(x)$. Therefore,
Eq.~\eqref{eq:(3.1)} is invariant under the gauge
transformation~\eqref{eq:(2.47)} after setting~$e\to e_\tau$.

In our ansatz~\eqref{eq:(3.1)}, we use the $-1$ variables instead of the
original variables. The reason is that in this way the ansatz~\eqref{eq:(3.1)}
contains the Gaussian fixed point as~$e_\tau\to0$ and~$\mathcal{K}_\tau\to1$.
In~Ref.~\cite{Sonoda:2022fmk}, GFERG in QED is solved in the one-loop order
perturbation theory and it was found that the 1PI action around the Gaussian
fixed point is succinctly described by the $-1$~variables. See~Eq.~(56)
of~Ref.~\cite{Sonoda:2022fmk}. We noted this fact. A naive simple
gauge-invariant combination of the original variables, on the other hand, does
not contain the Gaussian fixed point even as~$e_\tau\to0$. We think that the
existence of the Gaussian fixed point is a minimal requirement for any sensible
ansatz for the Wilson action. This is the reason for the usage of the
$-1$~variables in~Eq.~\eqref{eq:(3.1)}; this physically natural requirement,
however, makes the required burden rather heavy, because we have to solve the
flow equations, Eqs.~\eqref{eq:(3.2)} (see~Appendix~\ref{sec:A}).

For the chiral symmetry, we note that the chiral transformation,
$\delta\Psi(x)=i\alpha\gamma_5\Psi(x)$
and~$\delta\Bar{\Psi}(x)=\Bar{\Psi}(x)i\alpha\gamma_5$, induces the
same transformation on the $-1$ variables
$\Psi_{-1}(x)$ and~$\Bar{\Psi}_{-1}(x)$, because the flow
equations, Eqs.~\eqref{eq:(3.2)}, do not contain any Dirac matrix. Therefore,
when~$m_\tau=0$
, the ansatz~\eqref{eq:(3.1)} makes the first line of the
chiral WT identity~\eqref{eq:(2.49)} to vanish. Hence, $m_\tau=0$ case is
``classically'' chiral invariant. It is not fully chiral invariant, however,
and we do not study this issue in this paper.

From now on, we work in momentum space. In momentum space, the 1PI action,
Eq.~\eqref{eq:(2.46)} with the ansatz~\eqref{eq:(3.1)}, reads
\begin{align}
   \mathit{\Gamma}_\tau[\mathcal{A},\Bar{\Psi},\Psi]
   &=-\frac{1}{2}\int_k\,
   e^{k^2}\mathcal{A}_\mu(-k)e^{k^2}\mathcal{A}_\nu(k)
   \left[
   (k^2\delta_{\mu\nu}-k_\mu k_\nu)\mathcal{K}_\tau(k^2)
   +\frac{1}{\xi_\tau}k_\mu k_\nu
   \right]
\notag\\
   &\qquad{}
   -\int_p\,
   \Bar{\Psi}_{-1}(-p)(\Slash{p}+im_\tau)\Psi_{-1}(p)
   +e_\tau\int_{p,k}\,
   \Bar{\Psi}_{-1}(-p-k)e^{k^2}\Slash{\mathcal{A}}(k)\Psi_{-1}(p).
\label{eq:(3.5)}
\end{align}
Recall that the $-1$ variables $\Psi_{-1}$ and~$\Bar{\Psi}_{-1}$ depend on the
gauge field~$\mathcal{A}_\mu$ through the flow equations,
Eqs.~\eqref{eq:(3.2)}. Here, we impose the normalization of the gauge potential
by
\begin{equation}
   \mathcal{K}_\tau(k^2=0)=1.
\label{eq:(3.6)}
\end{equation}
The locality, i.e., the analyticity of the function~$\mathcal{K}_\tau(k^2)$,
on the other hand implies, e.g.
\begin{equation}
   \left.k^2\mathcal{K}_\tau'(k^2)\right|_{k^2=0}=0.
\label{eq:(3.7)}
\end{equation}

\subsection{GFERG for the ansatz: preparation}
\label{sec:3.2}
In momentum space, Eq.~\eqref{eq:(2.43)} reads
\begin{align}
   &\frac{\partial}{\partial\tau}
   \mathit{\Gamma}_\tau[\mathcal{A},\Bar{\Psi},\Psi]
\notag\\
   &=\int_p\,
   \biggl(
   \left[\frac{\delta}{\delta A_\mu(p)}
   +\frac{\delta\mathit{\Gamma}_\tau}{\delta\mathcal{A}_\mu(p)}\right]
   \left[
   \left(2p^2-\epsilon-\gamma_A\right)\mathcal{A}_\mu(p)
   +\frac{\delta\mathit{\Gamma}_\tau}{\delta\mathcal{A}_\mu(-p)}\right]
\notag\\
   &\qquad\qquad\qquad{}
   +\left(\frac{D+2}{2}+p\cdot\frac{\partial}{\partial p}\right)
   \left[\mathcal{A}_\mu(p)
   -\frac{\delta\mathit{\Gamma}_\tau}{\delta\mathcal{A}_\mu(-p)}\right]
   \cdot\frac{\delta\mathit{\Gamma}_\tau}{\delta\mathcal{A}_\mu(p)}
\notag\\
   &\qquad\qquad\qquad{}
   +\left[
   \frac{\delta}{\delta\Bar{\psi}(p)}
   +\frac{\delta}{\delta\Bar{\Psi}(p)}\mathit{\Gamma}_\tau
   \right](-2p^2+\gamma_\psi)
   \Bar{\Psi}(p)
\notag\\
   &\qquad\qquad\qquad{}
   +\left[
   \frac{\delta}{\delta\psi(p)}
   +\frac{\delta}{\delta\Psi(p)}\mathit{\Gamma}_\tau
   \right](-2p^2+\gamma_\psi)
   \Psi(p)
\notag\\
   &\qquad\qquad\qquad{}
   -i\left[
   \frac{\delta}{\delta\psi(p)}
   +\frac{\delta}{\delta\Psi(p)}\mathit{\Gamma}_\tau
   \right]
   \frac{\delta}{\delta\Bar{\Psi}(-p)}\mathit{\Gamma}_\tau
\notag\\
   &\qquad\qquad\qquad{}
   +\left(\frac{D+1}{2}+p\cdot\frac{\partial}{\partial p}\right)
   \left[
   \Psi(p)-i\frac{\delta}{\delta\Bar{\Psi}(-p)}\mathit{\Gamma}_\tau
   \right]
   \cdot\frac{\delta}{\delta\Psi(p)}\mathit{\Gamma}_\tau
\notag\\
   &\qquad\qquad\qquad{}
   +\left(\frac{D+1}{2}+p\cdot\frac{\partial}{\partial p}\right)
   \left[
   \Bar{\Psi}(p)+i\frac{\delta}{\delta\Psi(-p)}\mathit{\Gamma}_\tau
   \right]
   \cdot\frac{\delta}{\delta\Bar{\Psi}(p)}
   \mathit{\Gamma}_\tau
   \biggr)
\notag\\
   &\qquad{}
   +\int_{p,k}\,
   \left[
   \frac{\delta}{\delta\Bar{\psi}(p+k)}
   +\frac{\delta}{\delta\Bar{\Psi}(p+k)}\mathit{\Gamma}_\tau
   \right]
   (-4)p_\mu e_\tau\left[
   \mathcal{A}_\mu(k)
   +\frac{\delta}{\delta A_\mu(-k)}\right]
   \Bar{\Psi}(p)
\notag\\
   &\qquad{}
   +\int_{p,k}\,
   \left[
   \frac{\delta}{\delta\psi(p+k)}
   +\frac{\delta}{\delta\Psi(p+k)}\mathit{\Gamma}_\tau
   \right]
   4p_\mu e_\tau\left[
   \mathcal{A}_\mu(k)
   +\frac{\delta}{\delta A_\mu(-k)}\right]
   \Psi(p)
\notag\\
   &\qquad{}
   +\int_{p,k,l}\,
   \left[
   \frac{\delta}{\delta\Bar{\psi}(p+k+l)}
   +\frac{\delta}{\delta\Bar{\Psi}(p+k+l)}\mathit{\Gamma}_\tau
   \right]
   (-2)e_\tau^2\left[
   \mathcal{A}_\mu(k)
   +\frac{\delta}{\delta A_\mu(-k)}\right]
\notag\\
   &\qquad\qquad\qquad\qquad\qquad\qquad{}
   \times\left[
   \mathcal{A}_\mu(l)
   +\frac{\delta}{\delta A_\mu(-l)}\right]
   \Bar{\Psi}(p)
\notag\\
   &\qquad{}
   +\int_{p,k,l}\,
   \left[
   \frac{\delta}{\delta\psi(p+k+l)}
   +\frac{\delta}{\delta\Psi(p+k+l)}\mathit{\Gamma}_\tau
   \right]
   (-2)e_\tau^2\left[
   \mathcal{A}_\mu(k)
   +\frac{\delta}{\delta A_\mu(-k)}\right]
\notag\\
   &\qquad\qquad\qquad\qquad\qquad\qquad{}
   \times\left[
   \mathcal{A}_\mu(l)
   +\frac{\delta}{\delta A_\mu(-l)}\right]
   \Psi(p).
\label{eq:(3.8)}
\end{align}
This equation contains derivatives, such
as~$\delta\mathcal{A}_\nu(q)/\delta A_\mu(p)$. As noted
in~Eq.~\eqref{eq:(2.44)}, this requires the inversion of the matrix
in~Eq.~\eqref{eq:(2.45)}.

To write the inverse of the matrix in~Eq.~\eqref{eq:(2.45)}, we separate our
1PI action~\eqref{eq:(3.5)} into the free and interaction parts as
\begin{align}
   \mathit{\Gamma}_\tau&=\mathit{\Gamma}_\tau^0+\mathit{\Gamma}_\tau',
\notag\\
   \mathit{\Gamma}_\tau^0
   &:=\left.\mathit{\Gamma}_\tau\right|_{O(\mathcal{A}^2)}
   +\left.\mathit{\Gamma}_\tau\right|_{O(\Bar{\Psi}\Psi)},
\notag\\
   \mathit{\Gamma}_\tau'
   &:=\left.\mathit{\Gamma}_\tau\right|_{O(\Bar{\Psi}\mathcal{A}\Psi)}
   +\left.\mathit{\Gamma}_\tau\right|_{O(\Bar{\Psi}\mathcal{A}^2\Psi)}
   +\left.\mathit{\Gamma}_\tau\right|_{O(\Bar{\Psi}\mathcal{A}^3\Psi)}
   +\dotsb.
\label{eq:(3.9)}
\end{align}
Note that the latter is an infinite series due to our usage of the
$-1$~variables in the ansatz~\eqref{eq:(3.5)}. The matrix appearing
in~Eq.~\eqref{eq:(2.45)} in momentum space then takes the form
\begin{equation}
   M=M_0+M',
\label{eq:(3.10)}
\end{equation}
where $M_0$ is the free part,
\begin{align}
   M_0(q,r)
   =\begin{pmatrix}
   (M_0)_{\nu\rho}(q,r)&0&0\\
   0&[1-ie^{2q^2}(\Slash{q}-im)]_B{}^C\delta(q-r)&0\\
   0&0&[1+ie^{2q^2}({}^t\Slash{q}+im)]^B{}_C\delta(q-r)\\
   \end{pmatrix}
\label{eq:(3.11)}
\end{align}
(here, ${}^t$ denotes the matrix transposition) with
\begin{align}
   &(M_0)_{\nu\rho}(q,r)
\notag\\
   &=
   \left\{\left[1+e^{2q^2}q^2\mathcal{K}_\tau(q^2)\right]
   \left(\delta_{\nu\rho}-\frac{1}{q^2}q_\nu q_\rho\right)
   +\left(1+\frac{1}{\xi}e^{2q^2}q^2\right)
   \frac{1}{q^2}q_\nu q_\rho
   \right\}
   \delta(q-r),
\label{eq:(3.12)}
\end{align}
whereas $M'$ is the interaction part given by
\begin{equation}
   M'(q,r)
   :=
   \begin{pmatrix}
   m_{11}&m_{12}&m_{13}\\
   m_{21}&m_{22}&m_{23}\\
   m_{31}&m_{32}&m_{33}\\
   \end{pmatrix},
\label{eq:(3.13)}
\end{equation}
where
\begin{align}
   m_{11}(q,r)_{\nu\rho}
   &=-\frac{\delta^2\mathit{\Gamma}_\tau'}
   {\delta\mathcal{A}_\nu(q)\delta\mathcal{A}_\rho(-r)},
\notag\\
   m_{12}(q,r)_\nu{}^C
   &=-i\frac{\delta}{\delta\mathcal{A}_\nu(q)}
   \mathit{\Gamma}'_\tau
   \frac{\overleftarrow{\delta}}{\delta\Psi_C(-r)},
\notag\\
   m_{13}(q,r)_{\nu C}
   &=-i\frac{\delta}{\delta\mathcal{A}_\nu(q)}
   \frac{\delta}{\delta\Bar{\Psi}^C(-r)}
   \mathit{\Gamma}'_\tau,
\notag\\
   m_{21}(q,r)_{B\rho}
   &=-\frac{\delta}{\delta\Bar{\Psi}^B(q)}
   \frac{\delta\mathit{\Gamma}'_\tau}{\delta\mathcal{A}_\rho(-r)},
\notag\\
   m_{22}(q,r)_B{}^C
   &=-i\frac{\delta}{\delta\Bar{\Psi}^B(q)}
   \mathit{\Gamma}'_\tau
   \frac{\overleftarrow{\delta}}{\delta\Psi_C(-r)},
\notag\\
   m_{23}(q,r)_{BC}
   &=-i\frac{\delta}{\delta\Bar{\Psi}^B(q)}
   \frac{\delta}{\delta\Bar{\Psi}^C(-r)}
   \mathit{\Gamma}'_\tau,
\notag\\
   m_{31}(q,r)^B{}_\rho
   &=-\frac{\delta}{\delta\Psi_B(q)}
   \frac{\delta\mathit{\Gamma}'_\tau}{\delta\mathcal{A}_\rho(-r)},
\notag\\
   m_{32}(q,r)^{BC}
   &=i\mathit{\Gamma}'_\tau
   \frac{\overleftarrow{\delta}}{\delta\Psi_B(q)}
   \frac{\overleftarrow{\delta}}{\delta\Psi_C(-r)},
\notag\\
   m_{33}(q,r)^B{}_C
   &=-i\frac{\delta}{\delta\Psi_B(q)}
   \frac{\delta}{\delta\Bar{\Psi}^C(-r)}
   \mathit{\Gamma}'_\tau.
\label{eq:(3.14)}
\end{align}

The inverse of~$M$~\eqref{eq:(3.10)} is given by
\begin{equation}
   M^{-1}=M_0^{-1}
   -M_0^{-1}M'M_0^{-1}
   +M_0^{-1}M'M_0^{-1}M'M_0^{-1}
   +\dotsb,
\label{eq:(3.15)}
\end{equation}
where $M_0^{-1}$ is given explicitly by
\begin{align}
   M_0^{-1}(p,q)
   &=e^{-p^2}
   \begin{pmatrix}
   h_{\mu\nu}(p)&0&0\\
   0&\frac{1}{i}h_F(-p)_A{}^B&0\\
   0&0&\frac{1}{i}{}^th_F(p)^A{}_B\\
   \end{pmatrix}\delta(p-q)e^{-q^2}
   :=
   \begin{pmatrix}
   a&0&0\\
   0&b&0\\
   0&0&c\\
   \end{pmatrix},
\label{eq:(3.16)}
\end{align}
where
\begin{align}
   h_{\mu\nu}(p)
   &:=\left(\delta_{\mu\nu}-\frac{1}{p^2}p_\mu p_\nu\right)
   \frac{1}{e^{-2p^2}+p^2\mathcal{K}_\tau(p^2)}
   +\frac{1}{p^2}p_\mu p_\nu
   \frac{\xi_\tau}{\xi_\tau e^{-2p^2}+p^2},
\notag\\
   h_F(p)
   &:=\frac{i}{e^{-2p^2}+i(\Slash{p}+im_\tau)}.
\label{eq:(3.17)}
\end{align}

On the other hand, with the structure of~$\mathit{\Gamma}_\tau'$
in~Eq.~\eqref{eq:(3.9)}, each elements~$m_{ij}$ possess the following
dependences on fields,
\begin{align}
   m_{11}&=\Bar{\Psi}\Psi,\Bar{\Psi}\mathcal{A}\Psi,\Bar{\Psi}\mathcal{A}^2\Psi,
   \dotsc,
\notag\\
   m_{12},m_{31}&=\Bar{\Psi},\Bar{\Psi}\mathcal{A},\Bar{\Psi}\mathcal{A}^2,
   \dotsc,
\notag\\
   m_{13},m_{21}&=\Psi,\mathcal{A}\Psi,\mathcal{A}^2\Psi,
   \dotsc,
\notag\\
   m_{22},m_{33}&=\mathcal{A},\mathcal{A}^2,
   \dotsc,
\notag\\
   m_{23},m_{32}&=0.
\label{eq:(3.18)}
\end{align}

To find the flow of the ansatz~\eqref{eq:(3.5)} from GFERG~\eqref{eq:(3.8)},
we have to pick up terms of the structures, $\mathcal{A}^2$
and~$\Bar{\Psi}\Psi$ in the right-hand side of~Eq.~\eqref{eq:(3.8)}.
Fortunately, we do not need to keep terms of the
order~$\Bar{\Psi}\mathcal{A}\Psi$ to find the RG flow of the gauge
coupling~$e_\tau$, because the flow of~$e_\tau$ is determined by the anomalous
dimension~$\gamma_A$ as~Eq.~\eqref{eq:(2.48)} and $\gamma_A$ can be read off
from the flow of~$\xi_\tau$ which can be found from $O(\mathcal{A}^2)$
(nontransverse) terms in the right-hand side of~Eq.~\eqref{eq:(3.8)}; this
simplification is a consequence of the manifest gauge- or BRST-invariance.

As already mentioned, the GFERG equation~\eqref{eq:(3.8)} contains higher-order
functional derivatives. Higher-order functional derivatives are computed by
employing the chain rule such as, schematically,
\begin{equation}
   \frac{\delta}{\delta\psi}
   =\frac{\delta}{\delta\psi}\mathcal{A}\cdot\frac{\delta}{\delta\mathcal{A}}
   +\frac{\delta}{\delta\psi}\Bar{\Psi}\cdot\frac{\delta}{\delta\Bar{\Psi}}
   +\frac{\delta}{\delta\psi}\Psi\cdot\frac{\delta}{\delta\Psi},
\label{eq:(3.19)}
\end{equation}
where each coefficients such as~$\delta\mathcal{A}/\delta\psi$ are
given by~$M^{-1}$. Thus, higher-order functional derivatives require
the expansion~\eqref{eq:(3.15)} to sufficiently higher orders.

First, in~Eq.~\eqref{eq:(3.8)}, we find the the following derivatives appear
only in the very outside of the GFERG equation. Understanding schematic
representations, it turns out that we have to keep the following terms:
\begin{align}
   \frac{\delta}{\delta\Bar{\psi}}\mathcal{A}
   &=(M^{-1})_{21}
   =
   -bm_{21}a,
\notag\\
   \frac{\delta}{\delta\Bar{\psi}}\Bar{\Psi}
   &=(M^{-1})_{22}
   =b
   -bm_{22}b
   +bm_{21}am_{12}b
   +bm_{22}bm_{22}b,
\notag\\
   \frac{\delta}{\delta\Bar{\psi}}\Psi
   &=(M^{-1})_{23}=0,
\notag\\
   \frac{\delta}{\delta\psi}\mathcal{A}
   &=(M^{-1})_{31}
   =
   -cm_{31}a,
\notag\\
   \frac{\delta}{\delta\psi}\Bar{\Psi}
   &=(M^{-1})_{32}=0,
\notag\\
   \frac{\delta}{\delta\psi}\Psi
   &=(M^{-1})_{33}
   =c
   -cm_{33}c
   +cm_{31}am_{13}c
   +cm_{33}cm_{33}c.
\label{eq:(3.20)}
\end{align}

The derivative~$\delta\mathcal{A}/\delta A$ appears under the derivatives
$\delta/\delta\psi$ and~$\delta/\delta\Bar{\psi}$. We find that the
following terms should be kept:
\begin{equation}
   \frac{\delta}{\delta A}\mathcal{A}
   =(M^{-1})_{11}
   =a
   -am_{11}a
   +am_{12}bm_{21}a
   +am_{13}cm_{31}a.
\label{eq:(3.21)}
\end{equation}

Finally, in~Eq.~\eqref{eq:(3.8)}, $\delta\Bar{\Psi}/\delta A$
and~$\delta\Psi/\delta A$ appear under the second derivative such as
$\delta^2/(\delta\psi\delta A)$ and~$\delta^2/(\delta\Bar{\psi}\delta A)$. Some
consideration shows that the following terms become relevant:
\begin{align}
   \frac{\delta}{\delta A}\Bar{\Psi}
   &=(M^{-1})_{12}
\notag\\
   &=-am_{12}b
   +am_{11}am_{12}b
   +am_{12}bm_{22}b
\notag\\
   &\qquad{}
   -am_{11}am_{12}bm_{22}b
   -am_{12}bm_{21}am_{12}b
   -am_{12}bm_{22}bm_{22}b
   -am_{13}cm_{31}am_{12}b
\notag\\
   &\qquad{}
   +am_{11}am_{12}bm_{22}bm_{22}b
   +am_{12}bm_{21}am_{12}bm_{22}b
   +am_{12}bm_{22}bm_{21}am_{12}b
\notag\\
   &\qquad{}
   +am_{12}bm_{22}bm_{22}bm_{22}b
   +am_{13}cm_{31}am_{12}bm_{22}b
   +am_{13}cm_{33}cm_{31}am_{12}b,
\label{eq:(3.22)}
\end{align}
and
\begin{align}
   \frac{\delta}{\delta A}\Psi
   &=(M^{-1})_{13}
\notag\\
   &=-am_{13}c
   +am_{11}am_{13}c
   +am_{13}cm_{33}c
\notag\\
   &\qquad{}
   -am_{11}am_{13}cm_{33}c
   -am_{12}bm_{21}am_{13}c
   -am_{13}cm_{33}cm_{33}c
   -am_{13}cm_{31}am_{13}c
\notag\\
   &\qquad{}
   +am_{11}am_{13}cm_{33}cm_{33}c
   +am_{12}bm_{21}am_{13}cm_{33}c
   +am_{12}bm_{22}bm_{21}am_{13}c
\notag\\
   &\qquad{}
   +am_{13}cm_{33}cm_{33}cm_{33}c
   +am_{13}cm_{31}am_{13}cm_{33}c
   +am_{13}cm_{33}cm_{31}am_{13}c.
\label{eq:(3.23)}
\end{align}

We represent the 1PI action~\eqref{eq:(3.5)} by the following vertex functions,
$V_{\mu_1\mu_2\dotsb}(p,q;k_1,k_2,\dotsc)$, as
\begin{align}
   \left.\mathit{\Gamma}_\tau\right|_{O(\Bar{\Psi}\Psi)}
   &=-\int_p\,\Check{\Bar{\Psi}}(-p)(\Slash{p}+im_\tau)\Check{\Psi}(p),
\notag\\
   \mathit{\Gamma}_\tau'
   &=e_\tau\int_{p,k}\,
   \Check{\Bar{\Psi}}(-p-k)
   V_\mu(-p-k,p;k)\Check{\mathcal{A}}_\mu(k)
   \Check{\Psi}(p)
\notag\\
   &\qquad{}
   +e_\tau^2\int_{p,k_1,k_2}\,
   \Check{\Bar{\Psi}}(-p-k_1-k_2)
   V_{\mu\nu}(-p-k_1-k_2,p;k_1,k_2)
   \Check{\mathcal{A}}_\mu(k_1)\Check{\mathcal{A}}_\nu(k_2)
   \Check{\Psi}(p),
\notag\\
   &\qquad{}
   +e_\tau^3\int_{p,k_1,k_2,k_3}\,
   \Check{\Bar{\Psi}}(-p-k_1-k_2-k_3)
   V_{\mu\nu\rho}(-p-k_1-k_2-k_3,p;k_1,k_2,k_3)
\notag\\
   &\qquad\qquad\qquad{}
   \times
   \Check{\mathcal{A}}_\mu(k_1)\Check{\mathcal{A}}_\nu(k_2)
   \Check{\mathcal{A}}_\rho(k_3)
   \Check{\Psi}(p),
\notag\\
   &\qquad{}
   +e_\tau^4\int_{p,k_1,k_2,k_3,k_4}\,
   \Check{\Bar{\Psi}}(-p-k_1-k_2-k_3-k_4)
\notag\\
   &\qquad\qquad\qquad{}
   \times
   V_{\mu\nu\rho\sigma}(-p-k_1-k_2-k_3-k_4,p;k_1,k_2,k_3,k_4)
\notag\\
   &\qquad\qquad\qquad\qquad{}
   \times
   \Check{\mathcal{A}}_\mu(k_1)\Check{\mathcal{A}}_\nu(k_2)
   \Check{\mathcal{A}}_\rho(k_3)\Check{\mathcal{A}}_\sigma(k_4)
   \Check{\Psi}(p)
\notag\\
   &\qquad{}
   +\dotsb,
\label{eq:(3.24)}
\end{align}
where the checked variables are defined by
\begin{equation}
   \Check{\mathcal{A}}_\mu(p)
   :=e^{p^2}\mathcal{A}_\mu(p),\qquad
   \Check{\Psi}(p)
   :=e^{p^2}\Psi(p),\qquad
   \Check{\Bar{\Psi}}(p)
   :=e^{p^2}\Bar{\Psi}(p).
\label{eq:(3.25)}
\end{equation}
The explicit form of the vertex functions corresponding to~Eq.~\eqref{eq:(3.5)}
is studied in~Appendix~\ref{sec:A}. Because of the invariance of the
ansatz~\eqref{eq:(3.1)} under the local gauge transformation~\eqref{eq:(2.47)},
the vertex functions are not independent of each other and are related by the
following WT relations (see also~Refs.~\cite{Miyakawa:2021wus,Sonoda:2022fmk})
\begin{align}
   &V_\mu(-p-k,p;k)k_\mu e^{k^2}
   =-e^{p^2-(p+k)^2}(\Slash{p}+im)
   +(\Slash{p}+\Slash{k}+im)e^{(p+k)^2-p^2},
\notag\\
   &2V_{\mu\nu}(-p-k-\ell,p;k,\ell)k_\mu e^{k^2}
\notag\\
   &=e^{(p+\ell)^2-(p+k+\ell)^2}V_\nu(-p-\ell,p;\ell)
   -V_\nu(-p-k-\ell,p+k;\ell)e^{(p+k)^2-p^2},
\notag\\
   &3V_{\mu\nu\rho}(-p-k-\ell-m,p;k,\ell,m)k_\mu e^{k^2}
\notag\\
   &=e^{(p+\ell+m)^2-(p+k+\ell+m)^2}V_{\nu\rho}(-p-\ell-m,p;\ell,m)
\notag\\
   &\qquad\qquad{}
   -V_{\nu\rho}(-p-k-\ell-m,p+k;\ell,m)e^{(p+k)^2-p^2},
\notag\\
   &4V_{\mu\nu\rho\sigma}(-p-k-\ell-m-n,p;k,\ell,m,n)k_\mu e^{k^2}
\notag\\
   &=e^{(p+\ell+m+n)^2-(p+k+\ell+m+n)^2}V_{\nu\rho\sigma}(-p-\ell-m-n,p;\ell,m,n)
\notag\\
   &\qquad\qquad{}
   -V_{\nu\rho\sigma}(-p-k-\ell-m-n,p+k;\ell,m,n)e^{(p+k)^2-p^2}.
\label{eq:(3.26)}
\end{align}

In terms of vertex functions in~Eq.~\eqref{eq:(3.24)}, the matrix elements
in~Eq.~\eqref{eq:(3.14)} to the required orders are given as
\begin{align}
   &e^{-q^2}m_{11}(q,r)_{\nu\rho}e^{-r^2}
\notag\\
   &=-e^{-q^2}
   \frac{\delta^2\mathit{\Gamma}_\tau'}
   {\delta\mathcal{A}_\nu(q)\delta\mathcal{A}_\rho(-r)}
   e^{-r^2}
\notag\\
   &=-2e_\tau^2\int_p\,
   \Check{\Bar{\Psi}}(-p-q+r)
   V_{\nu\rho}(-p-q+r,p;q,-r)
   \Check{\Psi}(p)
\notag\\
   &\qquad{}
   -3\cdot2e_\tau^3\int_{p,k_3}\,
   \Check{\Bar{\Psi}}(-p-q+r-k_3)
   V_{\nu\rho\alpha}(-p-q+r-k_3,p;q,-r,k_3)
   \Check{\mathcal{A}}_\alpha(k_3)
   \Check{\Psi}(p)
\notag\\
   &\qquad{}
   -4\cdot3e_\tau^4\int_{p,k_3,k_4}\,
   \Check{\Bar{\Psi}}(-p-q+r-k_3-k_4)
   V_{\nu\rho\alpha\beta}(-p-q+r-k_3-k_4,p;q,-r,k_3,k_4)
\notag\\
   &\qquad\qquad\qquad\qquad{}
   \times
   \Check{\mathcal{A}}_\alpha(k_3)
   \Check{\mathcal{A}}_\beta(k_4)
   \Check{\Psi}(p).
\label{eq:(3.27)}
\end{align}
\begin{align}
   &e^{-q^2}m_{12}(q,r)_\nu{}^Ce^{-r^2}
\notag\\
   &=-ie^{-q^2}
   \frac{\delta}{\delta\mathcal{A}_\nu(q)}
   \mathit{\Gamma}'_\tau
   \frac{\overleftarrow{\delta}}{\delta\Psi_C(-r)}
   e^{-r^2}
\notag\\
   &=-ie_\tau
   \left[
   \Check{\Bar{\Psi}}(r-q)
   V_\nu(r-q,-r;q)
   \right]^C
\notag\\
   &\qquad{}
   -i2e_\tau^2\int_{k_2}\,
   \left[
   \Check{\Bar{\Psi}}(r-q-k_2)
   V_{\nu\alpha}(r-q-k_2,-r;q,k_2)
   \right]^C
   \Check{\mathcal{A}}_\alpha(k_2)
\notag\\
   &\qquad{}
   -i3e_\tau^3\int_{k_2,k_3}\,
   \left[\Check{\Bar{\Psi}}(r-q-k_2-k_3)
   V_{\nu\alpha\beta}(r-q-k_2-k_3,-r;q,k_2,k_3)\right]^C
   \Check{\mathcal{A}}_\alpha(k_2)
   \Check{\mathcal{A}}_\beta(k_3)
\notag\\
   &\qquad{}
   -i4e_\tau^4\int_{k_2,k_3,k_4}\,
   \left[\Check{\Bar{\Psi}}(r-q-k_2-k_3-k_4)
   V_{\nu\alpha\beta\gamma}(r-q-k_2-k_3-k_4,-r;q,k_2,k_3,k_4)\right]^C
\notag\\
   &\qquad\qquad\qquad\qquad\qquad{}
   \times
   \Check{\mathcal{A}}_\alpha(k_2)
   \Check{\mathcal{A}}_\beta(k_3)
   \Check{\mathcal{A}}_\gamma(k_4).
\label{eq:(3.28)}
\end{align}
\begin{align}
   &e^{-q^2}m_{13}(q,r)_{\nu C}e^{-r^2}
\notag\\
   &=-ie^{-q^2}
   \frac{\delta}{\delta\mathcal{A}_\nu(q)}
   \frac{\delta}{\delta\Bar{\Psi}^C(-r)}
   \mathit{\Gamma}'_\tau
   e^{-r^2}
\notag\\
   &=-ie_\tau
   \left[
   {}^t\Check{\Psi}(r-q)
   {}^tV_\nu(-r,r-q;q)
   \right]_C
\notag\\
   &\qquad{}
   -i2e_\tau^2\int_{k_2}\,
   \left[
   {}^t\Check{\Psi}(r-q-k_2)
   {}^tV_{\nu\alpha}(-r,r-q-k_2;q,k_2)
   \Check{\mathcal{A}}_\alpha(k_2)
   \right]_C
\notag\\
   &\qquad{}
   -i3e_\tau^3\int_{k_2,k_3}\,
   \left[
   {}^t\Check{\Psi}(r-q-k_2-k_3)
   {}^tV_{\nu\alpha\beta}(-r,r-q-k_2-k_3;q,k_2,k_3)
   \Check{\mathcal{A}}_\alpha(k_2)
   \Check{\mathcal{A}}_\beta(k_3)
   \right]_C
\notag\\
   &\qquad{}
   -i4e_\tau^4\int_{k_2,k_3,k_4}\,
   \bigl[
   {}^t\Check{\Psi}(r-q-k_2-k_3-k_4)
   {}^tV_{\nu\alpha\beta\gamma}(-r,r-q-k_2-k_3-k_4;q,k_2,k_3,k_4)
\notag\\
   &\qquad\qquad\qquad\qquad\qquad{}
   \times
   \Check{\mathcal{A}}_\alpha(k_2)
   \Check{\mathcal{A}}_\beta(k_3)
   \Check{\mathcal{A}}_\gamma(k_4)
   \bigr]_C.
\label{eq:(3.29)}
\end{align}
\begin{align}
   &e^{-q^2}m_{21}(q,r)_{B\rho}e^{-r^2}
\notag\\
   &=-e^{-q^2}
   \frac{\delta}{\delta\Bar{\Psi}^B(q)}
   \frac{\delta\mathit{\Gamma}'_\tau}{\delta\mathcal{A}_\rho(-r)}
   e^{-r^2}
\notag\\
   &=-e_\tau
   \left[
   V_\rho(q,-q+r;-r)
   \Check{\Psi}(-q+r)
   \right]_B
\notag\\
   &\qquad{}
   -2e_\tau^2\int_{k_2}\,
   \left[
   V_{\rho\alpha}(q,-q+r-k_2;-r,k_2)
   \Check{\mathcal{A}}_\alpha(k_2)
   \Check{\Psi}(-q+r-k_2)
   \right]_B
\notag\\
   &\qquad{}
   -3e_\tau^3\int_{k_2,k_3}\,
   \left[
   V_{\rho\alpha\beta}(q,-q+r-k_2-k_3;-r,k_2,k_3)
   \Check{\mathcal{A}}_\alpha(k_2)
   \Check{\mathcal{A}}_\beta(k_3)
   \Check{\Psi}(-q+r-k_2-k_3)\right]_B
\notag\\
   &\qquad{}
   -4e_\tau^4\int_{k_2,k_3,k_4}\,
   \bigl[
   V_{\rho\alpha\beta\gamma}(q,-q+r-k_2-k_3-k_4;-r,k_2,k_3,k_4)
   \Check{\mathcal{A}}_\alpha(k_2)
   \Check{\mathcal{A}}_\beta(k_3)
   \Check{\mathcal{A}}_\gamma(k_4)
\notag\\
   &\qquad\qquad\qquad\qquad\qquad{}
   \times
   \Check{\Psi}(-q+r-k_2-k_3-k_4)\bigr]_B.
\label{eq:(3.30)}
\end{align}
\begin{align}
   &e^{-q^2}m_{22}(q,r)_B{}^Ce^{-r^2}
\notag\\
   &=-ie^{-q^2}
   \frac{\delta}{\delta\Bar{\Psi}^B(q)}
   \mathit{\Gamma}'_\tau
   \frac{\overleftarrow{\delta}}{\delta\Psi_C(-r)}
   e^{-r^2}
\notag\\
   &=-ie_\tau
   V_\alpha(q,-r;-q+r)_B{}^C\Check{\mathcal{A}}_\alpha(-q+r)
\notag\\
   &\qquad{}
   -ie_\tau^2\int_{k_2}\,
   V_{\alpha\beta}(q,-r;-q+r-k_2,k_2)_B{}^C
   \Check{\mathcal{A}}_\alpha(-q+r-k_2)
   \Check{\mathcal{A}}_\beta(k_2)
\notag\\
   &\qquad{}
   -ie_\tau^3\int_{k_2,k_3}\,
   V_{\alpha\beta\gamma}(q,-r;-q+r-k_2-k_3,k_2,k_3)_B{}^C
   \Check{\mathcal{A}}_\alpha(-q+r-k_2-k_3)
   \Check{\mathcal{A}}_\beta(k_2)
   \Check{\mathcal{A}}_\gamma(k_3).
\label{eq:(3.31)}
\end{align}
\begin{equation}
   e^{-q^2}m_{23}(q,r)_{BC}e^{-r^2}
   =-ie^{-q^2}
   \frac{\delta}{\delta\Bar{\Psi}^B(q)}
   \frac{\delta}{\delta\Bar{\Psi}^C(-r)}
   \mathit{\Gamma}'_\tau
   e^{-r^2}
   =0.
\label{eq:(3.32)}
\end{equation}
\begin{align}
   &e^{-q^2}m_{31}(q,r)^B{}_\rho e^{-r^2}
\notag\\
   &=-e^{-q^2}
   \frac{\delta}{\delta\Psi_B(q)}
   \frac{\delta\mathit{\Gamma}'_\tau}{\delta\mathcal{A}_\rho(-r)}
   e^{-r^2}
\notag\\
   &=e_\tau
   \left[
   {}^tV_\rho(-q+r,q;-r)
   {}^t\Check{\Bar{\Psi}}(-q+r)
   \right]^B
\notag\\
   &\qquad{}
   +2e_\tau^2\int_{k_2}\,
   \left[
   {}^tV_{\rho\alpha}(-q+r-k_2,q;-r,k_2)
   \Check{\mathcal{A}}_\alpha(k_2)
   {}^t\Check{\Bar{\Psi}}(-q+r-k_2)
   \right]^B
\notag\\
   &\qquad{}
   +3e_\tau^3\int_{k_2,k_3}\,
   \left[
   {}^tV_{\rho\alpha\beta}(-q+r-k_2-k_3,q;-r,k_2,k_3)
   \Check{\mathcal{A}}_\alpha(k_2)
   \Check{\mathcal{A}}_\beta(k_3)
   {}^t\Check{\Bar{\Psi}}(-q+r-k_2-k_3)
   \right]^B
\notag\\
   &\qquad{}
   +4e_\tau^4\int_{k_2,k_3,k_4}\,
   \Bigl[
   {}^tV_{\rho\alpha\beta\gamma}(-q+r-k_2-k_3-k_4,q;-r,k_2,k_3,k_4)
   \Check{\mathcal{A}}_\alpha(k_2)
   \Check{\mathcal{A}}_\beta(k_3)
   \Check{\mathcal{A}}_\gamma(k_4)
\notag\\
   &\qquad\qquad\qquad\qquad\qquad{}
   \times
   {}^t\Check{\Bar{\Psi}}(-q+r-k_2-k_3-k_4)
   \Bigr]^B.
\label{eq:(3.33)}
\end{align}
\begin{equation}
   e^{-q^2}m_{32}(q,r)^{BC}e^{-r^2}
   =ie^{-q^2}
   \mathit{\Gamma}'_\tau
   \frac{\overleftarrow{\delta}}{\delta\Psi_B(q)}
   \frac{\overleftarrow{\delta}}{\delta\Psi_C(-r)}
   e^{-r^2}
   =0.
\label{eq:(3.34)}
\end{equation}
\begin{align}
   &e^{-q^2}m_{33}(q,r)^B{}_Ce^{-r^2}
\notag\\
   &=-ie^{-q^2}
   \frac{\delta}{\delta\Psi_B(q)}
   \frac{\delta}{\delta\Bar{\Psi}^C(-r)}
   \mathit{\Gamma}'_\tau
   e^{-r^2}
\notag\\
   &=-ie_\tau
   \,{}^tV_\alpha(-r,q;r-q)^B{}_C\Check{\mathcal{A}}_\alpha(r-q)
\notag\\
   &\qquad{}
   -ie_\tau^2\int_{k_2}\,
   {}^tV_{\alpha\beta}(-r,q;r-q-k_2,k_2)^B{}_C
   \Check{\mathcal{A}}_\alpha(r-q-k_2)
   \Check{\mathcal{A}}_\beta(k_2)
\notag\\
   &\qquad{}
   -ie_\tau^3\int_{k_2,k_3}\,
   {}^tV_{\alpha\beta\gamma}(-r,q;r-q-k_2-k_3,k_2,k_3)^B{}_C
   \Check{\mathcal{A}}_\alpha(r-q-k_2-k_3)
   \Check{\mathcal{A}}_\beta(k_2)
   \Check{\mathcal{A}}_\gamma(k_3).
\label{eq:(3.35)}
\end{align}

Finally, we note that the chiral WT identity~\eqref{eq:(2.49)} requires
\begin{align}
   &2m_\tau\gamma_5
\notag\\
   &\qquad
   +e_\tau^2\int_\ell\,
   h_{\rho\sigma}(\ell)
   \Bigl[
   e^{-2(p+\ell)^2}V_\rho(-p,p+\ell;-\ell)h_F(p+\ell)
   \gamma_5h_F(p+\ell)V_\sigma(-p-\ell,p;\ell)
\notag\\
   &\qquad\qquad{}
   +e^{-2(p-\ell)^2}V_\sigma(-p,p-\ell;\ell)h_F(p-\ell)
   \gamma_5h_F(p-\ell)V_\rho(-p+\ell,p;-\ell)
   \Bigr]=0.
\label{eq:(3.36)}
\end{align}
Since $m_\tau$ cannot depend on the momentum~$p$, it is clear that this chiral
WT identity cannot be fulfilled in general; this shows that our
ansatz~\eqref{eq:(3.5)} cannot be chiral invariant when~$e_\tau\neq0$. For the
exact chiral symmetry, a more refined construction is required.

\section{ERG flow}
\label{sec:4}
\subsection{GFERG equation}
\label{sec:4.1}
We now substitute our 1PI action~\eqref{eq:(3.5)} into the GFERG
equation, Eq.~\eqref{eq:(3.8)}. The left-hand side becomes
\begin{align}
   &\frac{\partial}{\partial\tau}
   \biggl\{
   -\frac{1}{2}\int_k\,\Check{\mathcal{A}}_\mu(-k)\Check{\mathcal{A}}_\nu(k)
   \left[
   (k^2\delta_{\mu\nu}-k_\mu k_\nu)\mathcal{K}_\tau(k^2)
   +\frac{1}{\xi_\tau}k_\mu k_\nu
   \right]
\notag\\
   &\qquad{}
   -\int_p\,\Check{\Bar{\Psi}}(-p)(\Slash{p}+im_\tau)\Check{\Psi}(p)
   +\text{higher orders in fields}\biggr\}.
\label{eq:(4.1)}
\end{align}
For the right-hand side of~Eq.~\eqref{eq:(3.8)}, we set
\begin{align}
   &-\frac{1}{2}\int_k\,\Check{\mathcal{A}}_\mu(-k)
   \Check{\mathcal{A}}_\nu(k)
   \left[
   \mathcal{R}_{\mu\nu}(k)
   -2(\epsilon+\gamma_A)\frac{1}{\xi_\tau}k_\mu k_\nu
   \right]
\notag\\
   &\qquad{}
   -\int_p\,
   \Check{\Bar{\Psi}}(-p)
   \left[
   -2\gamma_\psi(\Slash{p}+im_\tau)+im_\tau
   +\mathcal{R}(p)
   \right]
   \Check{\Psi}(p)
\notag\\
   &\qquad\qquad{}
   +\text{higher orders in fields}.
\label{eq:(4.2)}
\end{align}
Then, after a lengthy and patient calculation, we find the following
expressions: For the gauge kinetic term,
\begin{align}
   &\mathcal{R}_{\mu\nu}(k)
\notag\\
   &=-(k^2\delta_{\mu\nu}-k_\mu k_\nu)
   \left[
   (\epsilon+\gamma_A)\mathcal{K}_\tau(k^2)
   +k^2\mathcal{K}_\tau'(k^2)
   \right]
\notag\\
   &\qquad{}
   +ie_\tau^2\int_\ell\,
   (-4\ell^2+2\gamma_\psi-1)e^{-2\ell^2}
   \tr\left[h_F(\ell)
   (V_{\mu\nu}+V_\mu h_FV_\nu)h_F(\ell)\right]
\notag\\
   &\qquad{}
   +4ie_\tau^2\int_\ell\,e^{-(\ell+k)^2}e^{-\ell^2}e^{-k^2}
   \tr\left[h_F(\ell)V_\mu h_F(\ell+k)\right]
   (2\ell+k)_\nu
\notag\\
   &\qquad{}
   -4ie_\tau^2e^{-2k^2}\delta_{\mu\nu}\int_\ell\,e^{-2\ell^2}
   \tr\left[h_F(\ell)\right]
\notag\\
   &\qquad{}
   +4ie_\tau^4\int_{\ell,\ell'}\,e^{-(\ell+\ell')^2}e^{-\ell^2}e^{-(\ell')^2}
   (2\ell_\alpha+\ell_\alpha')h_{\alpha\beta}(\ell')
\notag\\
   &\qquad\qquad\qquad{}
   \times
   \tr\bigl\{h_F(\ell+\ell')
\notag\\
   &\qquad\qquad\qquad\qquad{}
   \times(3V_{\beta\mu\nu}
   +2V_{\beta\mu}h_FV_\nu+V_{\mu\nu}h_FV_\beta
   +V_\beta h_FV_{\mu\nu}+2V_\mu h_FV_{\beta\nu}
\notag\\
   &\qquad\qquad\qquad\qquad\qquad{}
   +V_\beta h_FV_\mu h_FV_\nu+V_\mu h_FV_\beta h_FV_\nu
   +V_\mu h_FV_\nu h_FV_\beta)
   h_F(\ell)\bigr\}
\notag\\
   &\qquad{}
   -8ie_\tau^4\int_{\ell,\ell'}\,e^{-(\ell+\ell'+k)^2}e^{-\ell^2}e^{-(\ell')^2}e^{-k^2}
   h_{\beta\nu}(\ell')
\notag\\
   &\qquad\qquad\qquad{}
   \times
   \tr\left[h_F(\ell)
   (2V_{\beta\mu}+V_\beta h_FV_\mu+V_\mu h_FV_\beta)
   h_F(\ell+\ell'+k)\right]
\notag\\
   &\qquad{}
   -4ie_\tau^4\int_\ell\,e^{-2\ell^2}h_{\rho\rho}(\ell)
   \int_{\ell'}\,e^{-2(\ell')^2}
   \tr\left[h_F(\ell')
   (V_{\mu\nu}+V_\mu h_FV_\nu)h_F(\ell')\right]
\notag\\
   &\qquad{}
   -4ie_\tau^6\int_{\ell,\ell',\ell''}\,e^{-(\ell+\ell'+\ell'')^2}
   e^{-\ell^2}e^{-(\ell')^2}e^{-(\ell'')^2}
   h_{\alpha\beta}(\ell')
   h_{\alpha\gamma}(\ell'')
\notag\\
   &\qquad\qquad\qquad{}
   \times
   \tr\bigl[h_F(\ell+\ell'+\ell'')
\notag\\
   &\qquad\qquad\qquad\qquad{}
   \times(
   12V_{\beta\gamma\mu\nu}
   +6V_{\beta\gamma\mu}h_FV_\nu+6V_{\beta\mu\nu}h_FV_\gamma
   +6V_\beta h_FV_{\gamma\mu\nu}+6V_\mu h_FV_{\beta\gamma\nu}
\notag\\
   &\qquad\qquad\qquad\qquad\qquad{}
   +2V_{\beta\gamma}h_FV_{\mu\nu}+8V_{\beta\mu}h_FV_{\gamma\nu}
   +2V_{\mu\nu}h_FV_{\beta\gamma}
\notag\\
   &\qquad\qquad\qquad\qquad\qquad{}
   +2V_{\beta\gamma}h_FV_\mu h_FV_\nu
   +4V_{\beta\mu}h_FV_\gamma h_FV_\nu
\notag\\
   &\qquad\qquad\qquad\qquad\qquad{}
   +4V_{\beta\mu}h_FV_\nu h_FV_\gamma
   +2V_{\mu\nu}h_FV_\beta h_FV_\gamma
\notag\\
   &\qquad\qquad\qquad\qquad\qquad{}
   +4V_\beta h_FV_{\gamma\mu}h_FV_\nu
   +2V_\beta h_FV_{\mu\nu}h_FV_\gamma
\notag\\
   &\qquad\qquad\qquad\qquad\qquad{}
   +2V_\mu h_FV_{\beta\gamma}h_FV_\nu
   +4V_\mu h_FV_{\beta\nu}h_FV_\gamma
\notag\\
   &\qquad\qquad\qquad\qquad\qquad{}
   +2V_\beta h_FV_\gamma h_FV_{\mu\nu}
   +4V_\beta h_FV_\mu h_FV_{\gamma\nu}
\notag\\
   &\qquad\qquad\qquad\qquad\qquad{}
   +4V_\mu h_FV_\beta h_FV_{\gamma\nu}
   +2V_\mu h_FV_\nu h_FV_{\beta\gamma}
\notag\\
   &\qquad\qquad\qquad\qquad\qquad{}
   +2V_\beta h_FV_\gamma h_FV_\mu h_F V_\nu
   +2V_\beta h_FV_\mu h_FV_\gamma h_F V_\nu
\notag\\
   &\qquad\qquad\qquad\qquad\qquad{}
   +2V_\beta h_FV_\mu h_FV_\nu h_F V_\gamma
   +2V_\mu h_FV_\beta h_FV_\nu h_F V_\gamma
\notag\\
   &\qquad\qquad\qquad\qquad\qquad{}
   +2V_\mu h_FV_\nu h_FV_\beta h_F V_\gamma
   +2V_\mu h_FV_\beta h_FV_\gamma h_F V_\nu
   )h_F(\ell)\bigr]
\notag\\
   &\qquad{}
   +(\mu\leftrightarrow\nu,k\leftrightarrow-k).
\label{eq:(4.3)}
\end{align}
In this expression, the first line consists of terms counting the dimension of
the photon and the scale change of the momentum variable. The order~$e_\tau^2$
terms are fermion one-loop contributions; the first $O(e_\tau^2)$ term is
expected also in the conventional ERG. The remaining two $O(e_\tau^2)$ terms are
peculiar to GFERG and they arise from higher functional derivative terms
in~Eq.~\eqref{eq:(3.8)} which are absent in the conventional ERG; these are
required for the gauge invariance. See also descriptions
below~Eq.~\eqref{eq:(4.11)}. For the fermion kinetic term, we have
\begin{align}
   &\mathcal{R}(p)
\notag\\
   &=-e_\tau^2\int_\ell\,
   (2\ell^2+1-\epsilon-\gamma_A)e^{-2\ell^2}h_{\mu\rho}(\ell)h_{\sigma\mu}(\ell)
   \mathcal{A}_{\rho\sigma}
\notag\\
   &\qquad{}
   +ie_\tau^2\int_\ell\,
   \left[4(p+\ell)^2+1-2\gamma_\psi\right]
   e^{-2(p+\ell)^2}h_{\mu\nu}(\ell)V_\mu h_F(p+\ell)^2V_\nu
\notag\\
   &\qquad{}
   -4ie_\tau^2\int_\ell\,
   e^{-(p+\ell)^2}e^{-p^2}e^{-\ell^2}
   h_{\mu\nu}(\ell)p_\mu h_F(p+\ell)V_\nu
\notag\\
   &\qquad{}
   -4e_\tau^2\int_\ell\,
   e^{p^2}e^{-(p+\ell)^2}e^{-\ell^2}
   h_{\mu\nu}(\ell)V_\mu h_F(p+\ell)(\Slash{p}+im)(p+\ell)_\nu
\notag\\
   &\qquad{}
   -4ie_\tau^2\int_\ell\,
   e^{-(p+\ell)^2}e^{-p^2}e^{-\ell^2}
   h_{\mu\nu}(\ell)V_\mu h_F(p+\ell)p_\nu
\notag\\
   &\qquad{}
   -4e_\tau^2\int_\ell\,
   e^{p^2}e^{-(p+\ell)^2}e^{-\ell^2}
   h_{\mu\nu}(\ell)(p+\ell)_\mu(\Slash{p}+im)h_F(p+\ell)V_\nu
\notag\\
   &\qquad{}
   +4e_\tau^2\int_\ell\,
   e^{-2\ell^2}
   h_{\mu\mu}(\ell)(\Slash{p}+im)
\notag\\
   &\qquad{}
   -4ie_\tau^4
   \int_{\ell,\ell'}\,(2p+\ell+2\ell')_\mu
   e^{-(p+\ell+\ell')^2}e^{-(p+\ell')^2}e^{-\ell^2}
   h_{\mu\nu}(-\ell)h_{\rho\sigma}(+\ell')
\notag\\
   &\qquad\qquad\qquad{}
   \times
   \mathcal{A}_{\nu\rho}
   h_F(p+\ell+\ell')h_F(p+\ell')V_\sigma
\notag\\
   &\qquad{}
   -4ie_\tau^4
   \int_{\ell,\ell'}\,(2p+\ell+2\ell')_\mu
   e^{-(p+\ell+\ell')^2}e^{-(p+\ell')^2}e^{-\ell^2}
   h_{\mu\nu}(+\ell)h_{\rho\sigma}(-\ell')
\notag\\
   &\qquad\qquad\qquad{}
   \times V_\sigma h_F(p+\ell')h_F(p+\ell+\ell')
   \mathcal{A}_{\nu\rho}
\notag\\
   &\qquad{}
   +8ie_\tau^4
   \int_{\ell,\ell'}\,\ell_\mu
   e^{-(\ell+\ell')^2}e^{-\ell^2}e^{-(\ell')^2}
   h_{\mu\nu}(+\ell')h_{\rho\sigma}(+\ell')
\notag\\
   &\qquad\qquad\qquad{}
   \times
   \tr[h_F(\ell+\ell')V_\sigma h_F(\ell)]
   \mathcal{A}_{\nu\rho}
\notag\\
   &\qquad{}
   +8ie_\tau^4
   \int_{\ell,\ell'}\,
   e^{-(p+\ell)^2}e^{-(p+\ell')^2}e^{-\ell^2}e^{-(\ell')^2}
   h_{\mu\nu}(-\ell)h_{\rho\mu}(-\ell')
\notag\\
   &\qquad\qquad\qquad{}
   \times
   V_\nu h_F(p+\ell)h_F(p+\ell')V_\rho
\notag\\
   &\qquad{}
   +4ie_\tau^4
   \int_{\ell,\ell'}\,
   e^{-2\ell^2}h_{\mu\mu}(+\ell)
   e^{-2(p+\ell')^2}
   h_{\rho\sigma}(+\ell')
   V_\rho h_F(p+\ell')^2V_\sigma
\notag\\
   &\qquad{}
   -4ie_\tau^4
   \int_{\ell,\ell'}\,
   e^{-2\ell^2}\tr[h_F(+\ell)]
   e^{-2(\ell')^2}
   h_{\mu\nu}(-\ell')h_{\rho\mu}(-\ell')\mathcal{A}_{\nu\rho}
\notag\\
   &\qquad{}
   +2ie_\tau^4
   \int_{\ell,\ell'}\,
   e^{-(p+\ell+\ell')^2}e^{-p^2}e^{-\ell^2}e^{-(\ell')^2}
   h_{\mu\nu}(+\ell)h_{\rho\mu}(-\ell')
\notag\\
   &\qquad\qquad\qquad{}
   \times
   h_F(p+\ell+\ell')\mathcal{A}_{\nu\rho}
\notag\\
   &\qquad{}
   +2ie_\tau^4
   \int_{\ell,\ell'}\,
   e^{-(p+\ell+\ell')^2}e^{-p^2}e^{-\ell^2}e^{-(\ell')^2}
   h_{\mu\nu}(-\ell)h_{\rho\mu}(+\ell')
\notag\\
   &\qquad\qquad\qquad{}
   \times
   \mathcal{A}_{\nu\rho}
   h_F(p+\ell+\ell')
\notag\\
   &\qquad{}
   +2e_\tau^4
   \int_{\ell,\ell'}\,
   e^{p^2}e^{-(p+\ell+\ell')^2}e^{-\ell^2}e^{-(\ell')^2}
   h_{\mu\nu}(-\ell)h_{\rho\mu}(+\ell')
\notag\\
   &\qquad\qquad\qquad{}
   \times
   \mathcal{A}_{\nu\rho}
   h_F(p+\ell+\ell')(\Slash{p}+im)
\notag\\
   &\qquad{}
   +2e_\tau^4
   \int_{\ell,\ell'}\,
   e^{p^2}e^{-(p+\ell+\ell')^2}e^{-\ell^2}e^{-(\ell')^2}
   h_{\mu\nu}(+\ell)h_{\rho\mu}(-\ell')
\notag\\
   &\qquad\qquad\qquad{}
   \times
   (\Slash{p}+im)
   h_F(p+\ell+\ell')\mathcal{A}_{\nu\rho}
\notag\\
   &\qquad{}
   +4ie_\tau^6\int_{\ell,\ell',\ell''}\,
   e^{-\ell^2}e^{-(\ell')^2}e^{-(p+\ell'')^2}e^{-(p+\ell+\ell'+\ell'')^2}
   h_{\mu\nu}(+\ell)h_{\mu\rho}(+\ell')h_{\sigma\tau}(-\ell'')
\notag\\
   &\qquad\qquad\qquad{}
   \times V_\tau h_F(p+\ell'')h_F(p+\ell+\ell'+\ell'')
   \mathcal{B}_{\rho\sigma\nu}
\notag\\
   &\qquad{}
   +4ie_\tau^6\int_{\ell,\ell',\ell''}\,
   e^{-\ell^2}e^{-(\ell')^2}e^{-(p+\ell+\ell'+\ell'')^2}e^{-(p+\ell'')^2}
   h_{\mu\nu}(-\ell)h_{\mu\rho}(-\ell')h_{\sigma\tau}(-\ell'')
\notag\\
   &\qquad\qquad\qquad{}
   \times
   \mathcal{B}_{\nu\rho\tau}
   h_F(p+\ell+\ell'+\ell'')
   h_F(p+\ell'')V_\sigma
\notag\\
   &\qquad{}
   -4ie_\tau^6\int_{\ell,\ell',\ell''}\,
   e^{-\ell^2}e^{-(\ell')^2}e^{-(\ell+\ell'+\ell'')^2}e^{-(\ell'')^2}
   h_{\mu\nu}(+\ell)h_{\mu\rho}(+\ell')h_{\sigma\tau}(+\ell+\ell')
\notag\\
   &\qquad\qquad\qquad{}
   \times
   \tr[h_F(\ell+\ell'+\ell'')V_\tau h_F(\ell'')]
   \mathcal{B}_{\rho\sigma\nu}
\notag\\
   &\qquad{}
   +8ie_\tau^6\int_{\ell,\ell',\ell''}\,
   e^{-\ell^2}e^{-(\ell')^2}e^{-(p+\ell+\ell'')^2}e^{-(p+\ell'+\ell'')^2}
   h_{\mu\nu}(-\ell)h_{\mu\rho}(+\ell')h_{\sigma\tau}(-\ell'')
\notag\\
   &\qquad\qquad\qquad{}
   \times
   \mathcal{A}_{\tau\nu}h_F(p+\ell+\ell'')h_F(p+\ell'+\ell'')
   \mathcal{A}_{\rho\sigma}
\notag\\
   &\qquad{}
   -8ie_\tau^6\int_{\ell,\ell',\ell''}\,
   e^{-\ell^2}e^{-(\ell')^2}e^{-(\ell+\ell'+\ell'')^2}e^{-(\ell'')^2}
   h_{\mu\nu}(+\ell)h_{\mu\rho}(+\ell')h_{\sigma\tau}(+\ell')
\notag\\
   &\qquad\qquad\qquad{}
   \times
   \tr[h_F(\ell+\ell'+\ell'')\mathcal{A}_{\tau\nu}h_F(\ell'')]
   \mathcal{A}_{\rho\sigma}.
\label{eq:(4.4)}
\end{align}
In the above expression,
\begin{equation}
   \mathcal{A}_{\rho\sigma}
   :=2V_{\rho\sigma}+V_\rho h_FV_\sigma+V_\sigma h_FV_\rho,
\label{eq:(4.5)}
\end{equation}
and
\begin{align}
   \mathcal{B}_{\rho\sigma\nu}
   &:=
   6V_{\rho\sigma\nu}
\notag\\
   &\qquad{}
   +2V_{\rho\sigma}h_FV_\nu
   +2V_{\sigma\nu}h_FV_\rho
   +2V_{\nu\rho}h_FV_\sigma
   +2V_\rho h_FV_{\sigma\nu}
   +2V_\sigma h_FV_{\nu\rho}
   +2V_\nu h_FV_{\rho\sigma}
\notag\\
   &\qquad{}
   +V_\rho h_FV_\sigma h_FV_\nu
   +V_\sigma h_FV_\nu h_FV_\rho
   +V_\nu h_FV_\rho h_FV_\sigma
\notag\\
   &\qquad{}
   +V_\rho h_FV_\nu h_FV_\sigma
   +V_\nu h_FV_\sigma h_FV_\rho
   +V_\sigma h_FV_\rho h_FV_\nu.
\label{eq:(4.6)}
\end{align}
The charge conjugation trick in~Appendix~\ref{sec:B} is helpful in the
computation of~$\mathcal{R}(k)$. In these expressions, we have adopted the
following abbreviation rule for momentum variable assignments. First, the
momentum variables of a vertex function are identified with the momenta of the
fields attached to the corresponding indices. That is,
\begin{equation}
   \Check{\mathcal{A}}_\mu(k)\Check{\Bar{\Psi}}(p)V_{\mu\nu\dotsb}\Check{\Psi}(q)
   \to
   \Check{\mathcal{A}}_\mu(k)\Check{\Bar{\Psi}}(p)
   V_{\mu\nu\dotsb}(p,q;k,\dotsc)\Check{\Psi}(q).
\label{eq:(4.7)}
\end{equation}
Similarly, when propagators are attached to a vertex function, the momentum
variables are taken to be the corresponding momenta. For example,
\begin{equation}
   h_{\rho\mu}(k)h_F(-p)V_{\mu\nu\dotsb}h_F(q)
   \to h_{\rho\mu}(k)h_F(-p)V_{\mu\nu\dotsb}(p,q;k,\dotsc)h_F(q).
\label{eq:(4.8)}
\end{equation}
It is also understood that
\begin{equation}
   h_{\mu\rho}(k)h_F(-p)V_{\mu\nu\dotsb}h_F(q)
   \to h_{\mu\rho}(k)h_F(-p)V_{\mu\nu\dotsb}(p,q;-k,\dotsc)h_F(q).
\label{eq:(4.9)}
\end{equation}
Finally, arguments of functions not fixed by the above rules are determined by
momentum conservation at the vertex. For instance, in the third line
of~Eq.~\eqref{eq:(4.3)}, we understand
\begin{align}
   &\Check{\mathcal{A}}_\mu(-k)\Check{\mathcal{A}}_\nu(k)
   h_F(-\ell)V_\mu h_FV_\nu h_F(-\ell)
\notag\\
   &\to\Check{\mathcal{A}}_\mu(-k)\Check{\mathcal{A}}_\nu(k)
   h_F(-\ell)V_\mu(\ell,-\ell+k;-k)h_F(-\ell+k)
   V_\nu(\ell-k,-\ell;k)h_F(-\ell).
\label{eq:(4.10)}
\end{align}
By these rules, the assignment of the momentum variables can be uniquely
reproduced.

The structures in~Eqs.~\eqref{eq:(4.3)} and~\eqref{eq:(4.4)} strongly indicate
that there exists a diagrammatic method which leads to the above results; we do
not investigate this issue in this paper.

We note that the above expressions, Eqs.~\eqref{eq:(4.3)} and~\eqref{eq:(4.4)},
themselves may be valid even for a more general ansatz other
than~\eqref{eq:(3.5)} as far as the ansatz possesses the structure
in~Eq.~\eqref{eq:(3.24)}; we may even include a change in the free parts of
the 1PI action by modifying the form of propagators, $h_{\mu\nu}(p)$
and~$h_F(p)$.

Now, the GFERG equation, Eq.~\eqref{eq:(3.8)}, equates Eqs.~\eqref{eq:(4.1)}
and~\eqref{eq:(4.2)}. The equality of the $O(\Check{\mathcal{A}}^2)$ terms
implies
$(k^2\delta_{\mu\nu}-k_\mu k_\nu)\frac{\partial}{\partial\tau}\mathcal{K}_\tau(k^2)
+\frac{d}{d\tau}\frac{1}{\xi_\tau}k_\mu k_\nu=\mathcal{R}_{\mu\nu}(k)
-2(\epsilon+\gamma_A)\frac{1}{\xi_\tau}k_\mu k_\nu$. Then the latter
of~Eqs.~\eqref{eq:(2.48)} shows that the tensor~$\mathcal{R}_{\mu\nu}(k)$ is
transverse:
\begin{equation}
   k_\mu\mathcal{R}_{\mu\nu}(k)=0.
\label{eq:(4.11)}
\end{equation}
This is a consequence of the underlying manifest gauge- or BRST-invariance of
GFERG.\footnote{%
We can explicitly confirm that the expression~\eqref{eq:(4.3)} is transverse
[see~Eq.~\eqref{eq:(4.11)}] using the WT identities~\eqref{eq:(3.26)}. See
Appendix~\ref{sec:C}.} In the gauge field kinetic term, besides the
gauge-fixing term whose renormalization is completely controlled by the
anomalous dimension~$\gamma_A$, only the transverse part receives nontrivial
renormalization; no gauge-noninvariant term such as the photon mass term is
induced by GFERG.

Let us elaborate on differences between the conventional ERG and GFERG. A
possible conventional ERG may be defined by adopting simple diffusion
equations such as $\partial_t\psi'(t,x)=\partial_\mu\partial_\mu\psi'(t,x)$
in~Eq.~\eqref{eq:(2.1)}, instead of gauge-covariant diffusion equations
in~Eqs.~\eqref{eq:(2.4)}. Then, repeating the above procedures, corresponding to
those simple diffusion equations, one ends up with an ERG equation that is
given by setting~$e_\tau=0$ in~Eq.~\eqref{eq:(3.8)}. As noted
below~Eq.~\eqref{eq:(2.45)}, the resulting ERG equation contains only the
first-order functional derivatives. The computation of those first-order
functional derivatives eventually reduces to the computation
of~$\delta/\delta A_\mu(p)\cdot\mathcal{A}_\nu(q)$,
$\delta/\delta\psi(p)\Psi(q)$, and~$\delta/\delta\Bar{\psi}(p)\Bar{\Psi}(q)$.
These combinations are given by the inverse of~Eq.~\eqref{eq:(2.45)}, the
second-order functional derivatives of the 1PI action. Since the inverse of the
second-order functional derivative of the 1PI action is the full propagator,
the right-hand side of the ERG equation in this conventional formulation takes
the ``one-loop structure,'' i.e., vertex functions are connected by the full
propagator. Also, corresponding to the use of simple diffusion equations,
the $-1$ variables in the ansatz~\eqref{eq:(3.5)} are simply given by
$\Psi_{-1}(p)=e^{p^2}\Psi(p)$ and~$\Bar{\Psi}_{-1}(p)=e^{p^2}\Bar{\Psi}(p)$
[only the first term of~Eq.~\eqref{eq:(A2)}] without containing any gauge
potential nor gauge coupling. Consequently, vertex functions are simply given
by~$V_\mu=\gamma_\mu$ and others vanish, $V_{\mu\nu\dotsb}=0$. Because of
these differences of the conventional ERG from GFERG, the combinations
$\mathcal{R}_{\mu\nu}(k)$ and~$\mathcal{R}(p)$ in~Eqs.~\eqref{eq:(4.3)}
and~\eqref{eq:(4.4)} become quite simple. $\mathcal{R}_{\mu\nu}(k)$ is given by
only the first two lines on the right-hand side of~Eq.~\eqref{eq:(4.3)} whereas
$\mathcal{R}(p)$ is given by only the first two lines on the right-hand side
of~Eq.~\eqref{eq:(4.4)}; also $V_{\mu\nu}=0$. Of course, there is no
guarantee that $\mathcal{R}_{\mu\nu}(k)$ is transverse in this case. Note that
the expansion of the full propagator in powers of field variables contains
terms that are higher order in the ``tree-level propagators''
in~Eq.~\eqref{eq:(3.17)} as indicated
in~Eqs.~\eqref{eq:(3.20)}--\eqref{eq:(3.23)}; the factors $a$, $b$ and~$c$
correspond to the tree-level propagators. Thus, the flow equation in GFERG
is quite complicated compared to that of the conventional ERG. This complexity
arises solely from the requirement to maintain the manifest gauge invariance.
We believe that, in nonperturbative analyses in ERG, the advantages of
manifest gauge invariance more than compensate for this complexity.

At the cost of the manifest gauge invariance, the combinations
$\mathcal{R}_{\mu\nu}(k)$ and~$\mathcal{R}(p)$ in~Eqs.~\eqref{eq:(4.3)}
and~\eqref{eq:(4.4)} are quite complicated and the solution to GFERG requires
three-loop momentum integrations: We would like to revisit this problem in the
near future. In the present paper, as a readily tractable approach, we solve
the GFERG equation in the leading and partially next-to-leading orders in the
large-$N_f$ approximation, where $N_f$ is the number of the fermion flavor.

\subsection{Leading order in the large-$N_f$ approximation}
\label{sec:4.2}
The large-$N_f$ approximation does not affect the gauge invariance of GFERG.
To define this approximation, we introduce the ``flavor'' index by
\begin{equation}
   \Psi\to\Psi^i\qquad
   \Bar{\Psi}\to\Bar{\Psi}_i\qquad
   i=1,\dotsc,N_f,
\label{eq:(4.12)}
\end{equation}
and set
\begin{equation}
   e_\tau^2:=\frac{\Tilde{e}_\tau^2}{\tr(1)N_f},
\label{eq:(4.13)}
\end{equation}
where $\tr(1)$ denotes the trace over the Dirac index\footnote{%
In this way, the leading large-$N_f$ approximation becomes independent of
a normalization convention of the Dirac trace in $D$~dimensions.} and
$\Tilde{e}_\tau^2$ is regarded as order~$1$ as~$N_f\to\infty$. In this setup,
the Dirac trace~$\tr$ in~Eq.~\eqref{eq:(4.3)} is replaced by the sum over the
flavor indices as well as that over the Dirac indices; the Dirac trace gains
the factor~$N_f$, Therefore, in the leading order in this approximation,
from~Eqs.~\eqref{eq:(4.1)}--\eqref{eq:(4.4)}, the GFERG equation reduces to
\begin{align}
   &\frac{\partial}{\partial\tau}
   \biggl\{
   -\frac{1}{2}\int_k\,\Check{\mathcal{A}}_\mu(-k)\Check{\mathcal{A}}_\nu(k)
   \left[
   (k^2\delta_{\mu\nu}-k_\mu k_\nu)\mathcal{K}_\tau(k^2)
   +\frac{1}{\xi_\tau}k_\mu k_\nu
   \right]
\notag\\
   &\qquad{}
   -\int_p\,\Check{\Bar{\Psi}}(-p)(\Slash{p}+im_\tau)\Check{\Psi}(p)\biggr\}
\notag\\
   &=-\frac{1}{2}\int_k\,\Check{\mathcal{A}}_\mu(-k)\Check{\mathcal{A}}_\nu(k)
\notag\\
   &\qquad\qquad{}
   \times\biggl\{(k^2\delta_{\mu\nu}-k_\mu k_\nu)
   (-2)\left[
   (\epsilon+\gamma_A)\mathcal{K}_\tau(k^2)
   +k^2\mathcal{K}_\tau'(k^2)
   \right]
   +\Tilde{e}_\tau^2\mathcal{C}_{\mu\nu}(k)
\notag\\
   &\qquad\qquad\qquad{}
   -2(\epsilon+\gamma_A)\frac{1}{\xi_\tau}k_\mu k_\nu
   \biggr\}
\notag\\
   &\qquad{}
   -\int_p\,\Check{\Bar{\Psi}}(-p)
   \left[
   -2\gamma_\psi(\Slash{p}+im_\tau)+im_\tau
   \right]
   \Check{\Psi}(p),
\label{eq:(4.14)}
\end{align}
where
\begin{align}
   \tr(1)\mathcal{C}_{\mu\nu}(k)
   &:=-i\int_\ell\,
   (4\ell^2+1-2\gamma_\psi)e^{-2\ell^2}
   \tr\left[h_F(\ell)
   (V_{\mu\nu}+V_\mu h_FV_\nu)h_F(\ell)\right]
\notag\\
   &\qquad{}
   +4i\int_\ell\,e^{-(\ell+k)^2}e^{-\ell^2}e^{-k^2}
   \tr\left[h_F(\ell)V_\mu h_F(\ell+k)\right]
   (2\ell+k)_\nu
\notag\\
   &\qquad{}
   -4ie^{-2k^2}\delta_{\mu\nu}\int_\ell\,e^{-2\ell^2}
   \tr\left[h_F(\ell)\right]
\notag\\
   &\qquad{}
   +(\mu\leftrightarrow\nu,k\to-k).
\label{eq:(4.15)}
\end{align}

Comparison of both sides of Eq.~\eqref{eq:(4.14)} first yields
\begin{equation}
   \gamma_\psi=0,\qquad\frac{d}{d\tau}m_\tau=m_\tau.
\label{eq:(4.16)}
\end{equation}
That is, the fermion anomalous dimension is trivial and $m_\tau$ is always
marginal. In the present order of approximation, the chiral WT
identity~\eqref{eq:(3.36)} is satisfied by~$m_\tau=0$ because we have
$e_\tau^2\to0$ in the large-$N_f$ limit. Thus, $m_\tau=0$ is chiral symmetric.

Equation~\eqref{eq:(4.11)} implies that $\mathcal{C}_{\mu\nu}(k)$
in~Eq.~\eqref{eq:(4.15)} is transverse, $k_\mu\mathcal{C}_{\mu\nu}(k)=0$.
Therefore, setting
\begin{equation}
   \mathcal{C}_{\mu\nu}(k)=\mathcal{C}(k^2)(k^2\delta_{\mu\nu}-k_\mu k_\nu),\qquad
   \mathcal{C}(k^2)=\frac{1}{D-1}\frac{1}{k^2}\mathcal{C}_{\mu\mu}(k),
\label{eq:(4.17)}
\end{equation}
from~Eq.~\eqref{eq:(4.14)}, we have
\begin{equation}
   \frac{\partial}{\partial\tau}\mathcal{K}_\tau(k^2)
   =-2\left[
   (\epsilon+\gamma_A)\mathcal{K}_\tau(k^2)
   +k^2\mathcal{K}_\tau'(k^2)
   \right]
   +\Tilde{e}_\tau^2\mathcal{C}(k^2).
\label{eq:(4.18)}
\end{equation}
From this equation with~$k^2=0$ and Eqs.~\eqref{eq:(3.6)} and~\eqref{eq:(3.7)},
the anomalous dimension~$\gamma_A$ is given by
\begin{equation}
   \gamma_A=-\epsilon
   +\frac{1}{2}\mathcal{C}(k^2=0)\Tilde{e}_\tau^2.
\label{eq:(4.19)}
\end{equation}
Equation~\eqref{eq:(2.48)} thus yields
\begin{equation}
   \frac{d}{d\tau}\Tilde{e}_\tau
   =\epsilon\Tilde{e}_\tau
   -\frac{1}{2}\mathcal{C}(0)\Tilde{e}_\tau^3,\qquad
   \frac{d}{d\tau}\xi_\tau
   =\mathcal{C}(0)\Tilde{e}_\tau^2\xi_\tau.
\label{eq:(4.20)}
\end{equation}
From the above relations, in the present order of approximation, the fixed
point exists at
\begin{equation}
   \Tilde{e}_*=0,\qquad
   \xi_*=\text{arbitrary},\qquad
   \mathcal{K}_*(k^2)=1,\qquad
   m_*=0,
\label{eq:(4.21)}
\end{equation}
with~$\gamma_{A*}=-\epsilon$ and~$\gamma_{\psi*}=0$ (this is the Gaussian fixed
point). For $\epsilon>0$, i.e., $D<4$, the fixed point exists also at
\begin{equation}
   \Tilde{e}_*^2=\frac{2\epsilon}{\mathcal{C}(0)},\qquad
   \xi_*=0,\qquad
   k^2\mathcal{K}_*'(k^2)+\epsilon\mathcal{K}_*(k^2)
   =\epsilon\frac{\mathcal{C}(k^2)}{\mathcal{C}(0)},\qquad
   m_*=0,
\label{eq:(4.22)}
\end{equation}
and $\gamma_{A*}=0$ and~$\gamma_{\psi*}=0$.

We encountered the function~$\mathcal{C}_{\mu\nu}(k)$~\eqref{eq:(4.15)} in the
one-loop analysis of GFERG in QED~\cite{Miyakawa:2021wus}. According to the
analysis of~Ref.~\cite{Miyakawa:2021wus}, this function enjoys the following
representation:
\begin{align}
   \tr(1)\mathcal{C}_{\mu\nu}(k)
   &=\int_\ell\,\left(
   \ell\cdot\frac{\partial}{\partial\ell}
   +k\cdot\frac{\partial}{\partial k}
   +m_\tau\frac{\partial}{\partial m_\tau}
   +2
   \right)
\notag\\
   &\qquad{}
   \times\frac{1}{2}\tr\Bigl[
   h_F(\ell)V_\mu h_F(\ell+k)V_\nu+h_F(\ell)V_\nu h_F(\ell-k)V_\mu
\notag\\
   &\qquad\qquad\qquad{}
   -16h_F(\ell)X_{\mu\nu}(\ell,-k,k)
   -8i\delta_{\mu\nu}F(-2k^2)e^{-2\ell^2}h_F(\ell)
   \Bigr],
\label{eq:(4.23)}
\end{align}
where functions~$X_{\mu\nu}(\ell,-k,k)$ and~$F(x)$ are reproduced
in~Appendix~\ref{sec:A}. For $D=4$ ($\epsilon=0$), a simplification occurs and,
from~Eq.~\eqref{eq:(4.23)}, we can deduce~\cite{Miyakawa:2021wus}
\begin{equation}
   \mathcal{C}(k^2=0)
   =\frac{1}{24\pi^2}=0.00422172\qquad D=4,
\label{eq:(4.24)}
\end{equation}
and thus from~Eq.~\eqref{eq:(4.19)},
\begin{equation}
   \gamma_A=\frac{1}{48\pi^2}\Tilde{e}_\tau^2\qquad D=4,
\label{eq:(4.25)}
\end{equation}
which coincides with the photon wave-function anomalous dimension in the
conventional one-loop approximation.

\begin{figure}[htbp]
\centering
\includegraphics[width=13cm]{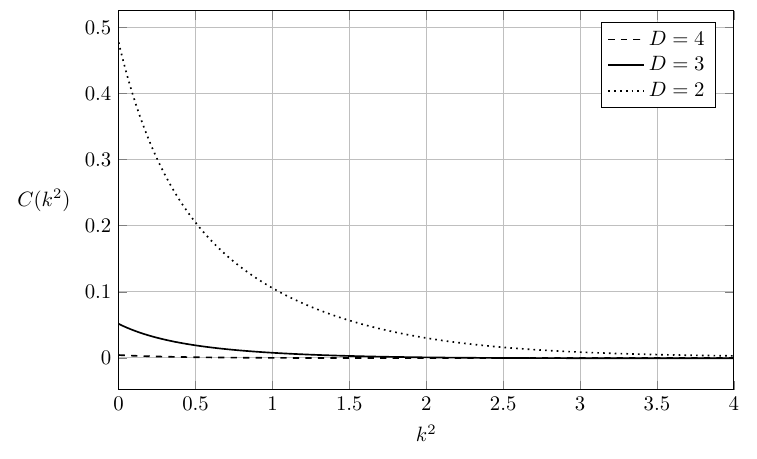}
\caption{The function~$\mathcal{C}(k^2)$~\eqref{eq:(4.17)} for~$m_\tau=0$ as the
function of~$k^2$.}
\label{fig:1}
\end{figure}

For~$D\neq4$ and/or for~$k^2\neq0$, it appears that the
function~$\mathcal{C}_{\mu\nu}(k^2)$~\eqref{eq:(4.15)} does not allow a simple
analytic treatment. We thus resort to a numerical integration. Setting
$\gamma_\psi=0$ and computing the Dirac trace in~Eq.~\eqref{eq:(4.15)} by using
expressions in~Appendix~\ref{sec:A}, we find
\begin{align}
   &\left.C_{\mu\mu}(k^2)\right|_{m_\tau=0}
\notag\\
   &=
   -\int_\ell\,(4\ell^2+1)\frac{e^{-4\ell^2}}{(\ell^2+e^{-4\ell^2})^2}
\notag\\
   &\qquad{}
   \times
   \biggl\{
   8D\ell^2F(-2k^2)
   +16\ell^2F(2\ell\cdot k)
   +16\ell^2\ell\cdot(\ell+k)F(2\ell\cdot k)^2
\notag\\
   &\qquad\qquad{}
   -16\frac{\ell^2\ell\cdot(\ell+k)}{\ell\cdot k+k^2}
   \left[F(-2k^2)-F(2\ell\cdot k)\right]
   \biggr\}
\notag\\
   &\qquad{}
   -\int_\ell\,(4\ell^2+1)
   \frac{e^{-2\ell^2}}{(\ell^2+e^{-4\ell^2})^2[(\ell+k)^2+e^{-4(\ell+k)^2}]}
\notag\\
   &\qquad\qquad{}
   \times
   \biggl(
   e^{-2(\ell+k)^2}\left(-\ell^2+e^{-4\ell^2}\right)
\notag\\
   &\qquad\qquad\qquad{}
   \times\bigl\{
   2D
   +8\ell\cdot(\ell+k)
   \left[F(2\ell\cdot k)+F(-2\ell\cdot k-2k^2)\right]
\notag\\
   &\qquad\qquad\qquad\qquad{}
   +8\ell^2(\ell+k)^2
   \left[F(2\ell\cdot k)^2+F(-2\ell\cdot k-2k^2)^2\right]
\notag\\
   &\qquad\qquad\qquad\qquad{}
   +16[\ell\cdot(\ell+k)]^2
   F(2\ell\cdot k)F(-2\ell\cdot k-2k^2)
   \bigr\}
\notag\\
   &\qquad\qquad\qquad{}
   +e^{-2\ell^2}
   \bigl\{
   -4(2-D)\ell\cdot(\ell+k)
\notag\\
   &\qquad\qquad\qquad\qquad\qquad{}
   -16\ell^2(\ell+k)^2
   \left[F(2\ell\cdot k)+F(-2\ell\cdot k-2k^2)\right]
\notag\\
   &\qquad\qquad\qquad\qquad\qquad{}
   -16\ell^2\ell\cdot(\ell+k)(\ell+k)^2
   \left[F(2\ell\cdot k)+F(-2\ell\cdot k-2k^2)\right]^2
   \bigr\}\biggr)
\notag\\
   &\qquad{}
   -e^{-k^2}\int_\ell\,
   \frac{e^{-(\ell+k)^2}e^{-\ell^2}}
   {(\ell^2+e^{-4\ell^2})[(\ell+k)^2+e^{-4(\ell+k)^2}]}
\notag\\
   &\qquad\qquad{}
   \times
   \Bigl(
   e^{-2\ell^2}
   \left\{
   16\ell\cdot(\ell+k)
   +32\ell^2(\ell+k)^2F(2\ell\cdot k)
   +32[\ell\cdot(\ell+k)]^2F(-2\ell\cdot k-2k^2)
   \right\}
\notag\\
   &\qquad\qquad\qquad{}
   +e^{-2(\ell+k)^2}
   \left\{
   16\ell^2
   +32\ell^2\ell\cdot(\ell+k)
   \left[F(2\ell\cdot k)+F(-2\ell\cdot k-2k^2)\right]
   \right\}
   \Bigr)
\notag\\
   &\qquad{}
   +8De^{-2k^2}\int_\ell\,
   \frac{e^{-4\ell^2}}{\ell^2+e^{-4\ell^2}},
\label{eq:(4.26)}
\end{align}
where we have set $m_\tau=0$ for simplicity. The
function~$\mathcal{C}(k^2)$~\eqref{eq:(4.17)} for $D=4$, $3$, and~$2$ obtained
by a numerical integration of~Eq.~\eqref{eq:(4.26)} is depicted
in~Fig.~\ref{fig:1}.
We also have
\begin{equation}
   \left.\mathcal{C}(k^2=0)\right|_{m_\tau=0}=\begin{cases}
   0.00422172&D=4,\\
   0.0514226&D=3,\\
   0.477465&D=2.\\
   \end{cases}
\label{eq:(4.27)}
\end{equation}

Also, from~Eq.~\eqref{eq:(4.26)}, we can find the asymptotic form as\footnote{%
Each term of the integrand in~Eq.~\eqref{eq:(4.26)} contains an exponential
factor of the form~$e^{-\alpha\ell^2-\beta(\ell+k)^2-\gamma\ell\cdot k-\delta k^2}$. The
completion of square shows that if
$\varepsilon:=\beta-(2\beta+\gamma)^2/[4(\alpha+\beta)]+\delta>0$, 
the integral is proportional to~$e^{-\varepsilon k^2}$ with~$\varepsilon>0$, which
decays exponentially as~$k^2\to\infty$. Most of terms in~Eq.~\eqref{eq:(4.26)}
are such exponentially small terms; whereas terms with~$\varepsilon<0$ are
canceled out. We find that only few terms contribute to the power-like decay
(corresponding to~$\varepsilon=0$) and some examination then yields the
leading-order behavior in~Eq.~\eqref{eq:(4.28)}.}
\begin{equation}
   \left.\mathcal{C}(k^2)\right|_{m_\tau=0}
   \stackrel{k^2\to\infty}{\to}
   \begin{cases}
   -\frac{8}{D}(D-2)
   \int_\ell\,\frac{e^{-4\ell^2}}{(\ell^2+e^{-4\ell^2})^2}
   (4\ell^2+1)\ell^2\,\frac{1}{(k^2)^2}&D\neq2,\\
   8\int_\ell\,\frac{e^{-4\ell^2}}{(\ell^2+e^{-4\ell^2})^2}
   (4\ell^2+1)(\ell^2)^2\,\frac{1}{(k^2)^3}&D=2,\\
   \end{cases}
\label{eq:(4.28)}
\end{equation}
and
\begin{equation}
   \left.\mathcal{C}(k^2)\right|_{m_\tau=0}
   \stackrel{k^2\to\infty}{\to}
   \begin{cases}
   -0.00445715\,\frac{1}{(k^2)^2}&D=4,\\
   -0.0159836\,\frac{1}{(k^2)^2}&D=3,\\
   +0.11202\,\frac{1}{(k^2)^3}&D=2.\\
   \end{cases}
\label{eq:(4.29)}
\end{equation}

\begin{figure}[htbp]
\centering
\includegraphics[width=13cm]{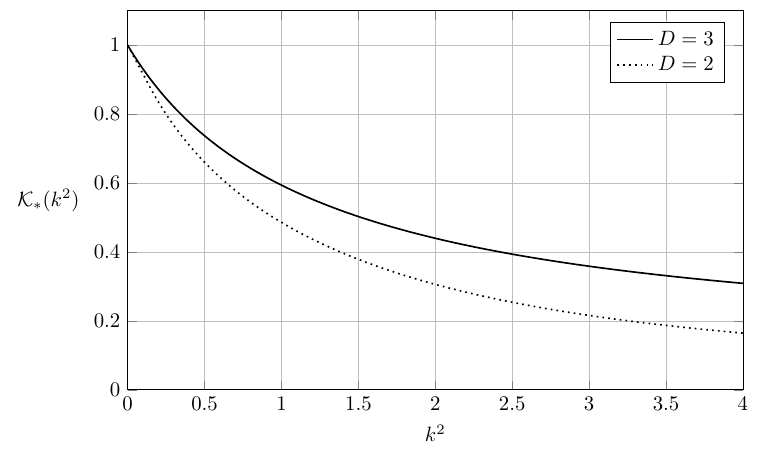}
\caption{The function~$\mathcal{K}_\tau(k^2)$ in the 1PI Wilson
action~\eqref{eq:(3.5)} at the IR fixed point in the leading large-$N_f$
approximation; the horizontal axis is~$k^2$.}
\label{fig:2}
\end{figure}

Once the function~$\mathcal{C}(k^2)$ is obtained, we can obtain the
function~$\mathcal{K}_\tau(k^2)$ in~Eq.~\eqref{eq:(3.5)} by solving the flow
equation, Eq.~\eqref{eq:(4.18)}. In particular, the fixed point
($\partial_\tau\mathcal{K}_\tau(k^2)=0$) form is given by~$\mathcal{K}_*(k^2)=1$
at the Gaussian fixed point~\eqref{eq:(4.21)} and, when $\epsilon>0$ ($D<4$),
\begin{equation}
   \mathcal{K}_*(k^2)
   =\epsilon\int_0^{k^2}dx\,
   x^{\epsilon-1}\frac{\mathcal{C}(x)}{\mathcal{C}(0)}
   \frac{1}{(k^2)^\epsilon},\qquad
   \mathcal{K}_*(0)=1,
\label{eq:(4.30)}
\end{equation}
at the IR fixed point~\eqref{eq:(4.22)}. This fixed-point
function~$\mathcal{K}_*(k^2)$ obtained by a numerical integration is depicted
in~Fig.~\ref{fig:2}. Note that $m_\tau=0$ at the fixed point~\eqref{eq:(4.22)}
and we may use the expression of~$\mathcal{C}_{\mu\nu}(k^2)$
in~Eq.~\eqref{eq:(4.26)}. 
Because of the asymptotic behaviors in~Eq.~\eqref{eq:(4.28)}, for~$D>0$, the
integral in~Eq.~\eqref{eq:(4.30)} converges as~$k^2\to\infty$ and thus the
asymptotic behavior of~$\mathcal{K}_*(k^2)$ as~$k^2\to\infty$ is given by
\begin{equation}
   \mathcal{K}_*(k^2)\stackrel{k^2\to\infty}{\to}
   \epsilon\int_0^\infty dx\,
   x^{\epsilon-1}\frac{\mathcal{C}(x)}{\mathcal{C}(0)}\frac{1}{(k^2)^\epsilon}
   =\begin{cases}
   0.608\,\frac{1}{(k^2)^{1/2}}&D=3,\\
   0.667\,\frac{1}{k^2}&D=2.\\
   \end{cases}
\label{eq:(4.31)}
\end{equation}
Since the gauge-invariant part of the gauge
potential two-point function at the fixed point is given by
\begin{equation}
   \left.\left\langle
   \mathcal{A}_\mu(k)\mathcal{A}_\nu(p)\right\rangle
   \right|_{\text{transverse}}
   =\frac{1}{k^2}\left(\delta_{\mu\nu}-\frac{k_\mu k_\nu}{k^2}\right)
   \frac{1}{\mathcal{K}_*(k^2)}e^{-k^2-p^2}\delta(k+p),
\label{eq:(4.32)}
\end{equation}
the fact that $\mathcal{K}_*(k^2)>0$ in~Fig.~\ref{fig:2} appears consistent with
unitarity of the 1PI action at the IR fixed point.\footnote{%
The factor~$e^{-k^2-p^2}$ in this expression is expected from the correspondence
to the gradient flow formalism~\cite{Sonoda:2025tyu}. See, e.g.\ Eqs.~(3.6)
and~(3.7) of~Ref.~\cite{Luscher:2011bx}. Note that the flow
time~$t$~\eqref{eq:(2.2)} with~$\Lambda_0\to\infty$ becomes~$1$ in the present
dimensionless formulation.} From this and~Eq.~\eqref{eq:(4.31)}, we have
\begin{equation}
   \left.\left\langle
   \mathcal{A}_\mu(k)\mathcal{A}_\nu(p)\right\rangle
   \right|_{\text{transverse}}
   \stackrel{k^2\to\infty}{\to}
   \frac{1}{(k^2)^{1-\epsilon}}
   \left(\delta_{\mu\nu}-\frac{k_\mu k_\nu}{k^2}\right)
   e^{-k^2-p^2}\delta(k+p),
\label{eq:(4.33)}
\end{equation}
up to a proportionality constant. This shows that the scaling dimension of
(the gauge-invariant part of) the gauge potential at the IR fixed point
is~$1-\epsilon$.\footnote{%
We may go back to the dimensionful formulation before Section~\ref{sec:2.5} by
setting $k^2\to k^2/\Lambda^2$. The Wilson action in the $\Lambda\to0$ limit
amounts to the functional integrations over all momentum modes. The 1PI action
in the limit~$\Lambda\to0$ thus amounts to the full 1PI effective action. This
$\Lambda\to0$ limit can be read off from the $k^2\to\infty$ limit in the
present dimensionless formulation. In this way, Eq.~\eqref{eq:(4.33)} gives
the exact two-point function at the IR fixed point. We would like to thank
Hidenori Sonoda for pointing this fact out to us.}

Considering linear fluctuations of~$\mathcal{K}_\tau(k^2)$ around the fixed
point~$\mathcal{K}_*(k^2)$, it is easy to find the critical
exponents~$\lambda_n$ in
\begin{equation}
   \frac{\partial}{\partial\tau}\mathcal{O}_{n,\tau}(k^2)
   =\lambda_n\mathcal{O}_{n,\tau}(k^2)
\label{eq:(4.34)}
\end{equation}
are given by
\begin{equation}
   \lambda_n=-2\begin{cases}
   n&\text{at the Gaussian fixed point~\eqref{eq:(4.21)}},\\
   n+\epsilon&\text{at the IR fixed point~\eqref{eq:(4.22)}},\\
   \end{cases}\qquad
   \mathcal{O}_{n,\tau}(k^2)=e^{\lambda_n\tau}(k^2)^n,
\label{eq:(4.35)}
\end{equation}
where the locality implies $n=0$, $1$, $2$, \dots\ All these operators (except
the identity) are therefore irrelevant.

Most of the above results are not surprising because they can be immediately
obtained by the one-loop level perturbation theory; the large-$N_f$
approximation justifies such a weak-coupling approximation. What is new in the
present study is the explicit form of the fixed point
function~$\mathcal{K}_*(k^2)$. We may further obtain the ERG flow
of~$\mathcal{K}_\tau(k^2)$ by numerically solving~Eq.~\eqref{eq:(4.18)}. The
determination of these objects would require a nonperturbative ERG framework
and we studied them while perfectly preserving the gauge WT identity. We
illustrated this point in the leading large-$N_f$ approximation.

\subsection{Next-to-leading order in the large-$N_f$ approximation}
\label{sec:4.3}
In the next-to-leading order of the large-$N_f$~approximation, the fermion
anomalous dimension and the mass renormalization become nontrivial. They arise
from $O(e_\tau^2)$~terms in~Eq.~\eqref{eq:(4.4)}. We thus set
\begin{equation}
   \left.\mathcal{R}(p)\right|_{O(\Tilde{e}_\tau^2)}
   :=\frac{\Tilde{e}_\tau^2}{\tr(1)N_f}\mathcal{D}(p).
\label{eq:(4.36)}
\end{equation}
Remarkably, as noticed in~Ref.~\cite{Miyakawa:2021wus}, $\mathcal{D}(p)$
with~$\gamma_\psi=0$ can be represented as
\begin{align}
   \mathcal{D}(p)
   &=-\frac{1}{2}\int_\ell\,
   \left(\ell\cdot\frac{\partial}{\partial\ell}
   +p\cdot\frac{\partial}{\partial p}
   +m_\tau\frac{\partial}{\partial m_\tau}+3\right)
   \left[
   h_{\mu\nu}(\ell)\mathcal{A}_{\mu\nu}(-p,p;-\ell,\ell)\right]
\notag\\
   &\qquad{}
   +(\epsilon+\gamma_A)\int_\ell\,
   e^{-2\ell^2}h_{\mu\rho}(\ell)h_{\rho\nu}(\ell)
   \mathcal{A}_{\mu\nu}(-p,p;-\ell,\ell)
\notag\\
   &\qquad{}
   -\int_\ell\,
   \frac{(\ell^2)^2\mathcal{K}_\tau'(\ell^2)}
   {[e^{-2\ell^2}+\ell^2\mathcal{K}_\tau(\ell^2)]^2}
   \left(\delta_{\mu\nu}-\frac{\ell_\mu\ell_\nu}{\ell^2}\right)
   \mathcal{A}_{\mu\nu}(-p,p;-\ell,\ell),
\label{eq:(4.37)}
\end{align}
where $\mathcal{A}_{\mu\nu}$ is given by~Eq.~\eqref{eq:(4.5)}. In fact,
in~Ref.~\cite{Miyakawa:2021wus}, only the case of $\mathcal{K}_\tau(k^2)=1$
(see~Eq.~\eqref{eq:(3.17)}) and $\epsilon=\gamma_A=0$ is considered;
in~Eq.~\eqref{eq:(4.37)}, we have relaxed these assumptions. When
$\mathcal{K}_\tau(k^2)\neq1$ in~Eq.~\eqref{eq:(3.17)}, we have to use the
identity
\begin{align}
   &(2\ell^2+1)e^{-2\ell^2}h_{\mu\rho}(\ell)h_{\rho\nu}(\ell)
\notag\\
   &=\frac{1}{2}\left(\ell\cdot\frac{\partial}{\partial\ell}+2\right)
   h_{\mu\nu}(\ell)
   +\frac{(\ell^2)^2\mathcal{K}_\tau'(\ell^2)}
   {[e^{-2\ell^2}+\ell^2\mathcal{K}_\tau(\ell^2)]^2}
   \left(\delta_{\mu\nu}-\frac{\ell_\mu\ell_\nu}{\ell^2}\right).
\label{eq:(4.38)}
\end{align}
The last term is absent in the corresponding identity
in~Ref.~\cite{Miyakawa:2021wus}.

To read off the renormalization of the fermion kinetic term, we extract terms
linear in~$p$ and~$m_\tau$ from~$\mathcal{A}_{\mu\nu}(-p,p;-\ell,\ell)$
in~Eq.~\eqref{eq:(4.37)}, $\mathcal{A}_{\mu\nu}|_{O(p,m_\tau)}$. Then, since
\begin{equation}
   \left(p\cdot\frac{\partial}{\partial p}
   +m_\tau\frac{\partial}{\partial m_\tau}\right)
   \left.\mathcal{A}_{\mu\nu}\right|_{O(p,m_\tau)}
   =\left.\mathcal{A}_{\mu\nu}\right|_{O(p,m_\tau)},
\label{eq:(4.39)}
\end{equation}
integration by parts yields
\begin{align}
   \left.\mathcal{D}(p)\right|_{O(p,m_\tau)}
   &=-\frac{1}{2}\frac{\Omega_D}{(2\pi)^D}
   \lim_{|\ell|\to\infty}
   \left[(\ell^2)^{D/2}h_{\mu\nu}(\ell)
   \left.\mathcal{A}_{\mu\nu}(-p,p;-\ell,\ell)\right|_{O(p,m_\tau)}
   \right]
\notag\\
   &\qquad{}
   +\gamma_A\int_\ell\,
   h_{\mu\nu}(\ell)
   \left.\mathcal{A}_{\mu\nu}\right|_{O(p,m_\tau)}
\notag\\
   &\qquad{}
   +\frac{1}{2}\int_\ell\,
   \frac{\ell^2[\partial_\tau\mathcal{K}_\tau(\ell^2)
   -\Tilde{e}_\tau^2\mathcal{C}(\ell^2)]}
   {[e^{-2\ell^2}+\ell^2\mathcal{K}_\tau(\ell^2)]^2}
   \left(\delta_{\mu\nu}-\frac{\ell_\mu\ell_\nu}{\ell^2}\right)
   \left.\mathcal{A}_{\mu\nu}\right|_{O(p,m_\tau)}
\notag\\
   &\qquad{}
   -(\epsilon+\gamma_A)\xi_\tau\int_\ell\,
   \frac{\ell^2}{(\xi_\tau e^{-2\ell^2}+\ell^2)^2}
   \frac{\ell_\mu\ell_\nu}{\ell^2}
   \left.\mathcal{A}_{\mu\nu}\right|_{O(p,m_\tau)},
\label{eq:(4.40)}
\end{align}
where
\begin{equation}
  \frac{\Omega_D}{(2\pi)^D}:=\frac{2}{(4\pi)^{D/2}\Gamma(D/2)}.
\label{eq:(4.41)}
\end{equation}
In deriving Eq.~\eqref{eq:(4.40)}, we have also noted
\begin{align}
   &e^{-2\ell^2}h_{\mu\rho}(\ell)h_{\rho\nu}(\ell)
\notag\\
   &=h_{\mu\nu}(\ell)
   -\frac{\ell^2\mathcal{K}_\tau(\ell^2)}
   {[e^{-2\ell^2}+\ell^2\mathcal{K}_\tau(\ell^2)]^2}
   \left(\delta_{\mu\nu}-\frac{\ell_\mu\ell_\nu}{\ell^2}\right)
   -\frac{\xi_\tau\ell^2}{(\xi_\tau e^{-2\ell^2}+\ell^2)^2}
   \frac{\ell_\mu\ell_\nu}{\ell^2},
\label{eq:(4.42)}
\end{align}
and employed the flow equation for~$\mathcal{K}_\tau(\ell^2)$,
Eq.~\eqref{eq:(4.18)}.

A straightforward calculation using~Eq.~\eqref{eq:(4.5)} and expressions
in~Appendix~\ref{sec:A} yields
\begin{align}
   &\left(\delta_{\mu\nu}-\frac{\ell_\mu\ell_\nu}{\ell^2}\right)
   \left.\mathcal{A}_{\mu\nu}(-p,p;-\ell,\ell)\right|_{O(p,m_\tau)}
\notag\\
   &=2(1-D)\frac{1-e^{-2\ell^2}}{\ell^2}(\Slash{p}+im_\tau)
   +2(4-D)\left(1-\frac{1}{D}\right)\frac{1}{\ell^2+e^{-4\ell^2}}
   \Slash{p}
\notag\\
   &\qquad{}
   -24\left(1-\frac{1}{D}\right)\frac{e^{-4\ell^2}}{\ell^2+e^{-4\ell^2}}
   \Slash{p}
   -4\left(1-\frac{1}{D}\right)
   \frac{(1-4e^{-4\ell^2})e^{-4\ell^2}}{(\ell^2+e^{-4\ell^2})^2}
   \Slash{p}
\notag\\
   &\qquad{}
   +2(1-D)\frac{\ell^2-e^{-4\ell^2}}{(\ell^2+e^{-4\ell^2})^2}im_\tau,
\label{eq:(4.43)}
\end{align}
and
\begin{align}
   &\frac{\ell_\mu\ell_\nu}{\ell^2}
   \left.\mathcal{A}_{\mu\nu}(-p,p;-\ell,\ell)\right|_{O(p,m_\tau)}
\notag\\
   &=\left[
   2\frac{e^{-2\ell^2}}{\ell^2}
   -4\frac{e^{-2\ell^2}}{\ell^2+e^{-4\ell^2}}
   -2\frac{e^{-4\ell^2}}{\ell^2(\ell^2+e^{-4\ell^2})}
   \right](\Slash{p}+im_\tau)
\notag\\
   &\qquad{}
   +\left[
   -\frac{8}{D}\frac{e^{-4\ell^2}}{\ell^2+e^{-4\ell^2}}
   +\frac{4}{D}\frac{e^{-4\ell^2}}{(\ell^2+e^{-4\ell^2})^2}
   +\frac{16}{D}\frac{\ell^2e^{-4\ell^2}}{(\ell^2+e^{-4\ell^2})^2}
   \right]\Slash{p}
\notag\\
   &\qquad{}
   +4\frac{e^{-4\ell^2}}{(\ell^2+e^{-4\ell^2})^2}im_\tau,
\label{eq:(4.44)}
\end{align}
where we kept only terms even under~$\ell\to-\ell$ and set
$\ell_\alpha\ell_\beta\to\ell^2\delta_{\alpha\beta}/D$ as these operations are valid
under the $|\ell|\to\infty$ limit and the integrations
in~Eq.~\eqref{eq:(4.40)}.

We now classify the cases depending on~$D$.

\subsubsection{$D=4$}
\label{sec:4.3.1}
For~$D=4$ ($\epsilon=0$), in the leading-order $N_f$ approximation
in~Section~\ref{sec:4.2}, the unique fixed point was the Gaussian fixed point
in~Eq.~\eqref{eq:(4.21)}. At this fixed point, since
$\gamma_{A*}=-\epsilon=0$ and~$\Tilde{e}_*=0$, in~Eq.~\eqref{eq:(4.40)}, only
the first surface term survives. That is, recalling Eq.~\eqref{eq:(3.17)},
$\mathcal{D}(p)|_{O(p,m_\tau)}$ is immediately given by the $|\ell|\to\infty$
limit of~Eqs.~\eqref{eq:(4.43)} and~\eqref{eq:(4.44)}. In this way, we
have~\cite{Miyakawa:2021wus}
\begin{equation}
   \mathcal{D}(p)=\frac{3}{8\pi^2}[(\Slash{p}+im_\tau)+im_\tau],
\label{eq:(4.45)}
\end{equation}
and
\begin{equation}
   \gamma_\psi=\frac{3}{16\pi^2}\frac{\Tilde{e}_\tau^2}{\tr(1)N_f},\qquad
   \frac{d}{d\tau}\ln m_\tau
   =1+\frac{3}{8\pi^2}\frac{\Tilde{e}_\tau^2}{\tr(1)N_f},
   \qquad D=4.
\label{eq:(4.46)}
\end{equation}
The latter is the one-loop mass renormalization in QED and the former is the
one-loop anomalous dimension for the fermion field in the gradient flow
formalism~\cite{Luscher:2013cpa}. This $\gamma_\psi$ differs from the
conventional fermion wave-function anomalous dimension; refer
to~Refs.~\cite{Miyakawa:2021wus,Sonoda:2025tyu} on this point.

\subsubsection{$D<4$}
\label{sec:4.3.2}
Let us first consider the Gaussian fixed point~\eqref{eq:(4.21)}.
From~Eqs.~\eqref{eq:(3.17)} with~$\mathcal{K}_*(p^2)=1$, \eqref{eq:(4.43)}
and~\eqref{eq:(4.44)}, we see that the
combination~$h_{\mu\nu}(\ell)\mathcal{A}_{\mu\nu}|_{O(p,m_\tau)}$ decreases
as~$1/(\ell^2)^2$ and the first surface term in~Eq.~\eqref{eq:(4.40)} vanishes
for~$D<4$. Instead, for~$D<4$, the integration in the second term
of~Eq.~\eqref{eq:(4.40)} converges and contributes. Note that
$\Tilde{e}_*=0$ and~$\epsilon+\gamma_{A*}=0$ at the Gaussian fixed point. By a
numerical integration of~Eq.~\eqref{eq:(4.40)}, we have
\begin{equation}
   \left.\mathcal{D}(p)\right|_{O(p,m_\tau)}
   =\begin{cases}
   0.124622(\Slash{p}+im_\tau)+0.16619im_\tau&D=3,\\
   0.283878(\Slash{p}+im_\tau)+0.488842im_\tau&D=2,\\
   \end{cases}
\label{eq:(4.47)}
\end{equation}
and thus
\begin{align}
   \gamma_\psi&=0.0623108\frac{\Tilde{e}_\tau^2}{\tr(1)N_f},&
   \frac{d}{d\tau}\ln m_\tau
   &=1+0.16619\frac{\Tilde{e}_\tau^2}{\tr(1)N_f},&
   D=3,
\notag\\
   \gamma_\psi&=0.141939\frac{\Tilde{e}_\tau^2}{\tr(1)N_f},&
   \frac{d}{d\tau}\ln m_\tau
   &=1+0.488842\frac{\Tilde{e}_\tau^2}{\tr(1)N_f},&
   D=2.
\label{eq:(4.48)}
\end{align}
In fact, we see that $\gamma_\psi$ and the mass running generally depend also on
the gauge parameter~$\xi_\tau$ and the above results are for~$\xi_\tau=0$. Since
$d\xi_\tau/d\tau=\mathcal{C}(0)\Tilde{e}_\tau^2\xi_\tau$
as~Eq.~\eqref{eq:(4.20)}, $\xi_\tau$ is a marginally relevant parameter.
Therefore, the running behavior around the Gaussian fixed point can be
determined by setting~$\xi_\tau=0$.

Now, at the the IR fixed point~\eqref{eq:(4.22)}, the
combination~\eqref{eq:(4.40)} contains the nontrivial
function~$\mathcal{K}_*(\ell^2)$ whose form is depicted in~Fig.~\ref{fig:2}.
From the asymptotic behavior in~Eq.~\eqref{eq:(4.31)},
and Eqs.~\eqref{eq:(4.43)} and~\eqref{eq:(4.44)}, we find that the surface term
possesses a nontrivial contribution and
\begin{align}
   &\left.\mathcal{D}(p)\right|_{O(p,m_\tau)}
\notag\\
   &=\frac{\Omega_D}{(2\pi)^D}
   \frac{\mathcal{C}(0)}{\epsilon\int_0^\infty dx\,x^{\epsilon-1}\mathcal{C}(x)}
   \left[
   \frac{2(D-1)(D-2)}{D}(\Slash{p}+im)+\frac{4(D-1)}{D}im
   \right]
\notag\\
   &\qquad{}
   -\frac{\epsilon}{\mathcal{C}(0)}
   \int_\ell\,
   \frac{\ell^2\mathcal{C}(\ell^2)}
   {[e^{-2\ell^2}+\ell^2\mathcal{K}_\tau(\ell^2)]^2}
   \left(\delta_{\mu\nu}-\frac{\ell_\mu\ell_\nu}{\ell^2}\right)
   \left.\mathcal{A}_{\mu\nu}\right|_{O(p,m_\tau)},
\label{eq:(4.49)}
\end{align}
where we have used~Eq.~\eqref{eq:(4.22)} and~$\gamma_{A*}=0$. Then a numerical
integration of~Eq.~\eqref{eq:(4.49)} gives
\begin{align}
   \left.\mathcal{D}(p)\right|_{O(p,m_\tau)}
   =\begin{cases}
   0.127(\Slash{p}+im_\tau)+0.232im_\tau
   &D=3,\\
   0.05635(\Slash{p}+im_\tau)+0.70im_\tau
   &D=2.\\
   \end{cases}
\label{eq:(4.50)}
\end{align}
From this, using $\Tilde{e}_*^2=2\epsilon/\mathcal{C}(0)$
and~Eq.~\eqref{eq:(4.27)}, at the IR fixed point~\eqref{eq:(4.22)},
\begin{align}
   \gamma_{\psi*}
   &=1.24\,\frac{1}{\tr(1)N_f},&
   \frac{d}{d\tau}\ln m_\tau
   &=1+4.52\,\frac{1}{\tr(1)N_f},&
   D=3,
\notag\\
   \gamma_{\psi*}
   &=0.1180\,\frac{1}{\tr(1)N_f},&
   \frac{d}{d\tau}\ln m_\tau
   &=1+2.9\,\frac{1}{\tr(1)N_f},&
   D=2.
\label{eq:(4.51)}
\end{align}

So far, we have considered the multiplicative renormalization of the mass
parameter~$m_\tau$. However, relating to the absence of the chiral symmetry with
the present ansatz, which we have noted at~Eq.~\eqref{eq:(3.36)}, the
parameter~$m_\tau$ receives also the additive renormalization. In what follows,
we study this effect.

In the same way as deriving~Eq.~\eqref{eq:(4.40)}, we have
\begin{align}
   \left.\mathcal{D}(0)\right|_{m_\tau=0}
   &=-\frac{1}{2}\frac{\Omega_D}{(2\pi)^D}
   \lim_{|\ell|\to\infty}
   \left[(\ell^2)^{D/2}h_{\mu\nu}(\ell)
   \left.\mathcal{A}_{\mu\nu}(0,0;-\ell,\ell)\right|_{m_\tau=0}
   \right]
\notag\\
   &\qquad{}
   +\left(\frac{1}{2}+\gamma_A\right)\int_\ell\,
   h_{\mu\nu}(\ell)
   \left.\mathcal{A}_{\mu\nu}(0,0;-\ell,\ell)\right|_{m_\tau=0}
\notag\\
   &\qquad{}
   +\frac{1}{2}\int_\ell\,
   \frac{\ell^2[\partial_\tau\mathcal{K}_\tau(\ell^2)
   -\Tilde{e}_\tau^2\mathcal{C}(\ell^2)]}
   {[e^{-2\ell^2}+\ell^2\mathcal{K}_\tau(\ell^2)]^2}
   \left(\delta_{\mu\nu}-\frac{\ell_\mu\ell_\nu}{\ell^2}\right)
   \left.\mathcal{A}_{\mu\nu}(0,0;-\ell,\ell)\right|_{m_\tau=0}
\notag\\
   &\qquad{}
   -(\epsilon+\gamma_A)\xi_\tau\int_\ell\,
   \frac{\ell^2}{(\xi_\tau e^{-2\ell^2}+\ell^2)^2}
   \frac{\ell_\mu\ell_\nu}{\ell^2}
   \left.\mathcal{A}_{\mu\nu}(0,0;-\ell,\ell)\right|_{m_\tau=0}.
\label{eq:(4.52)}
\end{align}
Noting
\begin{align}
   \left(\delta_{\mu\nu}-\frac{\ell_\mu\ell_\nu}{\ell^2}\right)
   \left.\mathcal{A}_{\mu\nu}(0,0;-\ell,\ell)\right|_{m_\tau=0}
   &=2i(D-1)\frac{e^{-2\ell^2}}{\ell^2+e^{-4\ell^2}},
\notag\\
   \frac{\ell_\mu\ell_\nu}{\ell^2}
   \left.\mathcal{A}_{\mu\nu}(0,0;-\ell,\ell)\right|_{m_\tau=0}
   &=2i\frac{e^{-2\ell^2}}{\ell^2+e^{-4\ell^2}},
\label{eq:(4.53)}
\end{align}
we see that the first surface term in~Eq.~\eqref{eq:(4.52)} vanishes and
\begin{align}
   \left.\mathcal{D}(0)\right|_{m_\tau=0}
   &=2i\left(\frac{1}{2}+\gamma_A\right)
   \int_\ell\,
   \left[
   \frac{D-1}{e^{-2\ell^2}+\ell^2\mathcal{K}_\tau(\ell^2)}
   +\frac{\xi_\tau}{\xi_\tau e^{-2\ell^2}+\ell^2}
   \right]
   \frac{e^{-2\ell^2}}{e^{-4\ell^2}+\ell^2}
\notag\\
   &\qquad{}
   +i(D-1)\int_\ell\,
   \frac{\ell^2[\partial_\tau\mathcal{K}_\tau(\ell^2)
   -\Tilde{e}_\tau^2\mathcal{C}(\ell^2)]}
   {[e^{-2\ell^2}+\ell^2\mathcal{K}_\tau(\ell^2)]^2}
   \frac{e^{-2\ell^2}}{e^{-4\ell^2}+\ell^2}
\notag\\
   &\qquad{}
   -2i(\epsilon+\gamma_A)\xi_\tau\int_\ell\,
   \frac{\ell^2}{(\xi_\tau e^{-2\ell^2}+\ell^2)^2}
   \frac{e^{-2\ell^2}}{e^{-4\ell^2}+\ell^2}.
\label{eq:(4.54)}
\end{align}

At the Gaussian fixed point~\eqref{eq:(4.21)}, only the first line of this
expression survives and
\begin{equation}
   \left.\mathcal{D}(0)\right|_{m_\tau=0}
   =\begin{cases}
   0.00524184i&D=4,\\
   0&D=3,\\
   -0.0566923i&D=2,\\
   \end{cases}
\label{eq:(4.55)}
\end{equation}
where we have set~$\xi_\tau=0$.

At the IR fixed point~\eqref{eq:(4.22)}, on the other hand,
Eq.~\eqref{eq:(4.54)} becomes
\begin{align}
   \left.\mathcal{D}(0)\right|_{m_\tau=0}
   &=i(D-1)
   \int_\ell\,
   \frac{1}{e^{-2\ell^2}+\ell^2\mathcal{K}_*(\ell^2)}
   \frac{e^{-2\ell^2}}{e^{-4\ell^2}+\ell^2}
\notag\\
   &\qquad{}
   -i(D-1)\frac{2\epsilon}{\mathcal{C}(0)}
   \int_\ell\,
   \frac{\ell^2\mathcal{C}(\ell^2)}
   {[e^{-2\ell^2}+\ell^2\mathcal{K}_\tau(\ell^2)]^2}
   \frac{e^{-2\ell^2}}{e^{-4\ell^2}+\ell^2}.
\label{eq:(4.56)}
\end{align}
A numerical integration then yields
\begin{equation}
   \left.\mathcal{D}(0)\right|_{m_\tau=0}
   =\begin{cases}
   0.0202686i&D=3,\\
   0.0366615i&D=2.\\
   \end{cases}
\label{eq:(4.57)}
\end{equation}

\section{Conclusion}
\label{sec:5}
In this paper, we have considered the RG flow of a manifestly gauge-invariant
nonperturbative ansatz of the 1PI Wilson action in QED by employing GFERG. The
gauge invariance is exactly preserved in every steps in our treatment. Since
the resulting ERG equation, Eqs.~\eqref{eq:(4.1)}--\eqref{eq:(4.4)}, is
quite complicated, so far we do not solve the RG flow of the
ansatz~\eqref{eq:(3.5)} in full generality; we would like to revisit this issue
in the near future. In this paper, as a tractable approach, we explicitly
solve the GFERG equation in the leading and partially next-to-leading orders of
the large-$N_f$ approximation. We obtain gauge-invariant critical exponents and
the gauge-invariant 1PI Wilson action at an IR fixed point for~$D<4$. An
interesting and important question to be clarified is which quantities are
independent of the gauge-fixing parameter~$\xi_\tau$ under the gauge WT
identity.

As a subject for future work, we want to include four-Fermi interactions in
our 1PI Wilson action; it should be possible to study possible nontrivial fixed
points and a nontrivial realization of chiral symmetry in~$D=4$
QED~\cite{Aoki:1996fh,Gies:2004hy,Igarashi:2016gcf,Gies:2020xuh,Igarashi:2021zml,Echigo:2025dia} in a manifestly gauge-invariant manner. We also want to extend
our work to non-Abelian gauge theory; first we want to see whether there
exists a principal difficulty in such an extension. Also, a recent study on the
Ricci flow in quantum gravity~\cite{Harlander:2026aav} is quite suggestive for
a possible generalization of GFERG to general coordinate-invariant systems.

\section*{Acknowledgments}
We would like to thank Hidenori Sonoda for many years of valuable discussions.
The work of H.S. was partially supported by Japan Society for the Promotion of
Science (JSPS) Grant-in-Aid for Scientific Research, JP23K03418.
This work was partially supported by the Quantum and Spacetime Research
Institute, Kyushu University.

\section*{Note added}
Somewhat surprisingly, from~Eq.~\eqref{eq:(4.3)}, it can be seen that the
term in~$\mathcal{R}_{\mu\nu}(k)$ explicitly contains $\xi_\tau$ is simply given
by
\begin{equation}
   +4ie_\tau^4\int_\ell\,e^{-2\ell^2}
   \frac{\xi_\tau}{\xi_\tau e^{-2\ell^2}+\ell^2}
   \int_{\ell'}\,e^{-2(\ell')^2}
   \tr\left[h_F(\ell')
   (2V_{\mu\nu}+V_\mu h_FV_\nu+V_\nu h_FV_\mu)h_F(\ell')\right].
\end{equation}
Since this itself is transverse (see~Sect.~\ref{sec:C}),
$k_\mu\mathcal{R}_{\mu\nu}(k)=0$, we write this as
\begin{equation*}
   -4e_\tau^4
   (k^2\delta_{\mu\nu}-k_\mu k_\nu)\int_\ell\,e^{-2\ell^2}
   \frac{\xi_\tau}{\xi_\tau e^{-2\ell^2}+\ell^2}\,\mathcal{J}(k^2).
\end{equation*}
Then, for~$m_\tau=0$, we find
\begin{align*}
   &\mathcal{J}(k^2)
\notag\\
   &=\frac{1}{D-1}\frac{1}{k^2}
   (\text{sum of first 2 integrals in~Eq.~\eqref{eq:(4.26)},
   but without $\ell^2$ in the integrands}).
\end{align*}
that is,
\begin{align}
   &\mathcal{J}(k^2)
\notag\\
   &=\frac{1}{D-1}\frac{1}{k^2}
   \biggl[
   -\int_\ell\,\frac{e^{-4\ell^2}}{(\ell^2+e^{-4\ell^2})^2}
\notag\\
   &\qquad{}
   \times
   \biggl\{
   8D\ell^2F(-2k^2)
   +16\ell^2F(2\ell\cdot k)
   +16\ell^2\ell\cdot(\ell+k)F(2\ell\cdot k)^2
\notag\\
   &\qquad\qquad{}
   -16\frac{\ell^2\ell\cdot(\ell+k)}{\ell\cdot k+k^2}
   \left[F(-2k^2)-F(2\ell\cdot k)\right]
   \biggr\}
\notag\\
   &\qquad{}
   -\int_\ell\,
   \frac{e^{-2\ell^2}}{(\ell^2+e^{-4\ell^2})^2[(\ell+k)^2+e^{-4(\ell+k)^2}]}
\notag\\
   &\qquad\qquad{}
   \times
   \biggl(
   e^{-2(\ell+k)^2}\left(-\ell^2+e^{-4\ell^2}\right)
\notag\\
   &\qquad\qquad\qquad{}
   \times\bigl\{
   2D
   +8\ell\cdot(\ell+k)
   \left[F(2\ell\cdot k)+F(-2\ell\cdot k-2k^2)\right]
\notag\\
   &\qquad\qquad\qquad\qquad{}
   +8\ell^2(\ell+k)^2
   \left[F(2\ell\cdot k)^2+F(-2\ell\cdot k-2k^2)^2\right]
\notag\\
   &\qquad\qquad\qquad\qquad{}
   +16[\ell\cdot(\ell+k)]^2
   F(2\ell\cdot k)F(-2\ell\cdot k-2k^2)
   \bigr\}
\notag\\
   &\qquad\qquad\qquad{}
   +e^{-2\ell^2}
   \bigl\{
   -4(2-D)\ell\cdot(\ell+k)
\notag\\
   &\qquad\qquad\qquad\qquad\qquad{}
   -16\ell^2(\ell+k)^2
   \left[F(2\ell\cdot k)+F(-2\ell\cdot k-2k^2)\right]
\notag\\
   &\qquad\qquad\qquad\qquad\qquad{}
   -16\ell^2\ell\cdot(\ell+k)(\ell+k)^2
   \left[F(2\ell\cdot k)+F(-2\ell\cdot k-2k^2)\right]^2
   \bigr\}\biggr)
   \biggr]
\end{align}

A numerical integration shows that $\mathcal{J}(k^2=0)\neq0$. This implies
that $\gamma_A$ depends on~$\xi_\tau$ by
\begin{equation*}
   -4e_\tau^4\int_\ell\,e^{-2\ell^2}
   \frac{\xi_\tau}{\xi_\tau e^{-2\ell^2}+\ell^2}\,\mathcal{J}(k^2=0).
\end{equation*}

\appendix

\section{Vertex functions associated with the $-1$~variables}
\label{sec:A}
To find vertex functions in~Eq.~\eqref{eq:(3.24)} corresponding to our
ansatz~\eqref{eq:(3.5)}, we have to solve the flow equation~\eqref{eq:(3.2)} to
express $\Psi_{-1}$ and~$\Bar{\Psi}_{-1}$ in terms of $\Psi$, $\Bar{\Psi}$,
and~$\mathcal{A}_\mu$. For this, we note that the solution
to~Eq.~\eqref{eq:(3.2)} (with~$\alpha_0=1$) can be expressed by the integral
representation as~\cite{Luscher:2013cpa}
\begin{align}
   \left\{\genfrac{}{}{0pt}{}{\Psi_t(p)}{\Bar{\Psi}_t(p)}\right\}
   &=e^{-tp^2}
   \left\{\genfrac{}{}{0pt}{}{\Psi(p)}{\Bar{\Psi}(p)}\right\}
\notag\\
   &\qquad{}
   +\int_0^tds\,e^{-(t-s)p^2}
   \biggl[
   \pm2e\int_k\,e^{-sk^2}\mathcal{A}_\mu(k)(p-k)_\mu
   \left\{\genfrac{}{}{0pt}{}{\Psi_s(p-k)}{\Bar{\Psi}_s(p-k)}\right\}
\notag\\
   &\qquad\qquad\qquad\qquad\qquad\qquad{}
   -e^2\int_{k,l}\,e^{-sk^2}\mathcal{A}_\mu(k)e^{-sl^2}\mathcal{A}_\mu(l)
   \left\{\genfrac{}{}{0pt}{}{\Psi_s(p-k-l)}{\Bar{\Psi}_s(p-k-l)}\right\}
   \biggr],
\label{eq:(A1)}
\end{align}
where we have noted $\mathcal{A}_{t\mu}(k)=e^{-tk^2}\mathcal{A}_\mu(k)$. By
iteratively solving this equation, after integrations over flow times, we have
\begin{align}
   &\Psi_{-1}(p)
\notag\\
   &=e^{p^2}\Psi(p)
   -2e\int_k\,F(p^2-k^2-(p-k)^2)e^{k^2}\mathcal{A}_\mu(k)
   (p-k)_\mu e^{(p-k^2)}\Psi(p-k)
\notag\\
   &\qquad{}
   +e^2\int_{k_1,k_2}\,
   F(p^2-k_1^2-k_2^2-(p-k_1-k_2)^2)
   e^{k_1^2}\mathcal{A}_\mu(k_1)e^{k_2^2}\mathcal{A}_\mu(k_2)
\notag\\
   &\qquad\qquad\qquad\qquad{}
   \times
   e^{(p-k_1-k_2)^2}\Psi(p-k_1-k_2)
\notag\\
   &\qquad{}
   +4e^2\int_{k_1,k_2}\,
   \frac{1}{p^2-k_1^2-(p-k_1)^2}
\notag\\
   &\qquad\qquad\qquad\qquad{}
   \times
   \Bigl[
   F(p^2-k_1^2-k_2^2-(p-k_1-k_2)^2)
\notag\\
   &\qquad\qquad\qquad\qquad\qquad{}
   -F((p-k_1)^2-k_2^2-(p-k_1-k_2)^2)
   \Bigr]
\notag\\
   &\qquad\qquad\qquad\qquad\qquad{}
   \times
   e^{k_1^2}\mathcal{A}_\mu(k_1)(p-k_1)_\mu
   e^{k_2^2}\mathcal{A}_\nu(k_2)(p-k_1-k_2)_\nu
\notag\\
   &\qquad\qquad\qquad\qquad\qquad\qquad{}
   \times
   e^{(p-k_1-k_2)^2}\Psi(p-k_1-k_2)
\notag\\
   &\qquad{}
   -2e^3\int_{k_1,k_2,k_3}\,
   \frac{1}{p^2-k_1^2-(p-k_1)^2}
\notag\\
   &\qquad\qquad\qquad\qquad{}
   \times
   \bigl[
   F(p^2-k_1^2-k_2^2-k_3^2-(p-k_1-k_2-k_3)^2)
\notag\\
   &\qquad\qquad\qquad\qquad\qquad{}
   -F((p-k_1)^2-k_2^2-k_3^2-(p-k_1-k_2-k_3)^2)
   \bigr]
\notag\\
   &\qquad\qquad{}
   \times
   e^{k_1^2}\mathcal{A}_\mu(k_1)(p-k_1)_\mu
   e^{k_2^2}\mathcal{A}_\nu(k_2)
   e^{k_3^2}\mathcal{A}_\nu(k_3)
\notag\\
   &\qquad\qquad\qquad\qquad\qquad\qquad\qquad\qquad{}
   \times e^{(p-k_1-k_2-k_3)^2}\Psi(p-k_1-k_2-k_3)
\notag\\
   &\qquad{}
   -2e^3\int_{k_1,k_2,k_3}\,
   \frac{1}{p^2-k_1^2-k_2^2-(p-k_1-k_2)^2}
\notag\\
   &\qquad\qquad\qquad\qquad{}
   \times
   \bigl[
   F(p^2-k_1^2-k_2^2-k_3^2-(p-k_1-k_2-k_3)^2)
\notag\\
   &\qquad\qquad\qquad\qquad\qquad{}
   -F((p-k_1-k_2)^2-k_3^2-(p-k_1-k_2-k_3)^2)
   \bigr]
\notag\\
   &\qquad\qquad{}
   \times
   e^{k_1^2}\mathcal{A}_\mu(k_1)
   e^{k_2^2}\mathcal{A}_\mu(k_2)
   e^{k_3^2}\mathcal{A}_\nu(k_3)(p-k_1-k_2-k_3)_\nu
\notag\\
   &\qquad\qquad\qquad\qquad\qquad\qquad\qquad\qquad{}
   \times e^{(p-k_1-k_2-k_3)^2}\Psi(p-k_1-k_2-k_3)
\notag\\
   &\qquad{}
   -8e^3\int_{k_1,k_2,k_3}\,
   \frac{1}{p^2-k_1^2-(p-k_1)^2}
\notag\\
   &\qquad\qquad{}
   \times\biggl\{
   \frac{1}{p^2-k_1^2-k_2^2-(p-k_1-k_2)^2}
\notag\\
   &\qquad\qquad\qquad\qquad{}
   \times
   \bigl[
   F(p^2-k_1^2-k_2^2-k_3^2-(p-k_1-k_2-k_3)^2)
\notag\\
   &\qquad\qquad\qquad\qquad\qquad{}
   -F((p-k_1-k_2)^2-k_3^2-(p-k_1-k_2-k_3)^2)
   \bigr]
\notag\\
   &\qquad\qquad\qquad{}
   -\frac{1}{(p-k_1)^2-k_2^2-(p-k_1-k_2)^2}
\notag\\
   &\qquad\qquad\qquad\qquad{}
   \times
   \bigl[
   F((p-k_1)^2-k_2^2-k_3^2-(p-k_1-k_2-k_3)^2)
\notag\\
   &\qquad\qquad\qquad\qquad\qquad{}
   -F((p-k_1-k_2)^2-k_3^2-(p-k_1-k_2-k_3)^2)
   \bigr]\biggr\}
\notag\\
   &\qquad\qquad\qquad{}
   \times
   e^{k_1^2}\mathcal{A}_\mu(k_1)(p-k_1)_\mu
   e^{k_2^2}\mathcal{A}_\nu(k_2)(p-k_1-k_2)_\nu
   e^{k_3^2}\mathcal{A}_\rho(k_3)(p-k_1-k_2-k_3)_\rho
\notag\\
   &\qquad\qquad\qquad\qquad\qquad\qquad\qquad\qquad{}
   \times e^{(p-k_1-k_2-k_3)^2}\Psi(p-k_1-k_2-k_3)
\notag\\
   &\qquad{}
   +O(\mathcal{A}^4),
\label{eq:(A2)}
\end{align}
where the function~$F(x)$ is given by\footnote{In writing down
Eq.~\eqref{eq:(A2)}, we have used the identity,
$F(x)e^y=(1+y/x)F(x+y)-(y/x)F(y)$.}
\begin{equation}
   F(x):=\frac{e^x-1}{x}.
\label{eq:(A3)}
\end{equation}
$\Bar{\Psi}_{-1}(p)$ is similarly obtained by setting~$e\to-e$ and
$\Psi\to\Bar{\Psi}$ in~Eq.~\eqref{eq:(A2)}.

Substituting the above expressions into~Eq.~\eqref{eq:(3.5)}, we have vertex
functions in~Eq.~\eqref{eq:(3.24)},
\begin{align}
   V_\mu(-p-k,p;k)
   &:=\gamma_\mu+2(\Slash{p}+\Slash{k}+im)p_\mu F((p+k)^2-k^2-p^2)
\notag\\
   &\qquad{}
   +2(\Slash{p}+im)(p+k)_\mu F(p^2-k^2-(p+k)^2),
\label{eq:(A4)}
\end{align}
and
\begin{align}
   &V_{\mu\nu}(-p-k_1-k_2,p;k_1,k_2)
\notag\\
   &:=-\delta_{\mu\nu}
   \bigl[
   (\Slash{p}+\Slash{k}_1+\Slash{k}_2+im)
   F((p+k_1+k_2)^2-k_1^2-k_2^2-p^2)
\notag\\
   &\qquad\qquad{}
   +(\Slash{p}+im)
   F(p^2-k_1^2-k_2^2-(p+k_1+k_2)^2)
   \bigr]
\notag\\
   &\qquad{}-4X_{\mu\nu}(p,k_1,k_2),
\label{eq:(A5)}
\end{align}
where
\begin{align}
   &X_{\mu\nu}(p,k_1,k_2)=X_{\nu\mu}(p,k_2,k_1)
\notag\\
   &:=
   \frac{1}{4}\gamma_\mu p_\nu F((p+k_2)^2-k_2^2-p^2)
   +\frac{1}{4}(p+k_1+k_2)_\mu\gamma_\nu F((p+k_2)^2-k_1^2-(p+k_1+k_2)^2)
\notag\\
   &\qquad{}
   +\frac{1}{2}(\Slash{p}+\Slash{k}_2+im)(p+k_1+k_2)_\mu p_\nu
\notag\\
   &\qquad\qquad{}
   \times
   F((p+k_2)^2-k_1^2-(p+k_1+k_2)^2)
   F((p+k_2)^2-k_2^2-p^2)
\notag\\
   &\qquad{}
   +\frac{1}{2}\frac{(\Slash{p}+\Slash{k}_1+\Slash{k_2}+im)(p+k_2)_\mu p_\nu}
   {(p+k_1+k_2)^2-k_1^2-(p+k_2)^2}
\notag\\
   &\qquad\qquad\times
   \left[F((p+k_1+k_2)^2-k_1^2-k_2^2-p^2)
   -F((p+k_2)^2-k_2^2-p^2)\right]
\notag\\
   &\qquad{}
   +\frac{1}{2}\frac{(\Slash{p}+im)(p+k_1+k_2)_\mu(p+k_2)_\nu}
   {p^2-k_2^2-(p+k_2)^2}
\notag\\
   &\qquad\qquad\times
   \left[F(p^2-k_1^2-k_2^2-(p+k_1+k_2)^2)
   -F((p+k_2)^2-k_1^2-(p+k_1+k_2)^2)\right]
\notag\\
   &\qquad{}
   +(\mu\leftrightarrow\nu,k_1\leftrightarrow k_2).
\label{eq:(A6)}
\end{align}
The above vertex functions $V_\mu$ and~$V_{\mu\nu}$ are identical to the
interaction vertices~$\widetilde{V}_\mu$ and~$\overline{V}_{\mu\nu}$ given
in~Eqs.~(48) and~(50) of~Ref.~\cite{Sonoda:2022fmk}, respectively; these
functions were obtained~\cite{Miyakawa:2021wus} by directly solving the GFERG
equation in perturbation theory.

By further pushing the above calculation, we may obtain higher vertex
functions, such as $V_{\mu\nu\rho}(-p-k_1-k_2-k_3,p;k_1,k_2,k_3)$
and~$V_{\mu\nu\rho\sigma}(-p-k_1-k_2-k_3-k_4,p;k_1,k_2,k_3,k_4)$.

\section{Charge conjugation trick}
\label{sec:B}
The GFERG equation~\eqref{eq:(3.8)} contains pairs of terms which are related
by the exchanges, $\Psi\leftrightarrow\Bar{\Psi}$
and~$\psi\leftrightarrow\Bar{\psi}$. The following charge conjugation trick
enables us to obtain one in a pair from the other in the pair and somewhat
reduces the labor for the calculation.

We note that the 1PI action $\mathit{\Gamma}_\tau$~\eqref{eq:(3.5)} is invariant
under the following charge conjugation transformation:
\begin{equation}
   \Psi_A\to (C)_{AB}\Bar{\Psi}^B,\qquad
   \Bar{\Psi}^A\to-\Psi_B(C^{-1})^{BA},\qquad
   \mathcal{A}_\mu\to-\mathcal{A}_\mu,
\label{eq:B1}
\end{equation}
where
\begin{equation}
   C^{-1}\gamma_\mu C=-{}^t\gamma^\mu.
\end{equation}
In terms of vertex functions in~Eq.~\eqref{eq:(3.24)}, this invariance implies
\begin{align}
   C^{-1}h_F(p)C&={}^th_F(-p),
\notag\\
   C^{-1}V_{\mu_1\dotsb\mu_n}(p,q;\dotsb)C
   &=(-1)^n\,{}^tV_{\mu_1\dotsb\mu_n}(q,p;\dotsb).
\label{eq:B2}
\end{align}
Through the Legendre transformation~\eqref{eq:(2.25)}, the charge conjugation
also induces
\begin{equation}
   \psi_A\to (C)_{AB}\Bar{\psi}^B,\qquad
   \Bar{\psi}^A\to-\psi_B(C^{-1})^{BA},\qquad
   A_\mu\to-A_\mu,
\label{eq:B3}
\end{equation}
and, on functional derivatives,
\begin{align}
   &\frac{\delta}{\delta\psi_A}
   \to\frac{\delta}{\delta\Bar{\psi}^B}(C^{-1})^{BA},\qquad
   \frac{\delta}{\delta\Bar{\psi}^A}
   \to-(C)_{AB}\frac{\delta}{\delta\psi_B},\qquad
   \frac{\delta}{\delta A_\mu}
   \to-\frac{\delta}{\delta A_\mu},
\notag\\
   &\frac{\delta}{\delta\Psi_A}
   \to\frac{\delta}{\delta\Bar{\Psi}^B}(C^{-1})^{BA},\qquad
   \frac{\delta}{\delta\Bar{\Psi}^A}
   \to-(C)_{AB}\frac{\delta}{\delta\Psi_B},\qquad
   \frac{\delta}{\delta\mathcal{A}_\mu}
   \to-\frac{\delta}{\delta\mathcal{A}_\mu}.
\label{eq:B4}
\end{align}
These substitution rules between terms in a pair related by
$\Psi\leftrightarrow\Bar{\Psi}$ and~$\psi\leftrightarrow\Bar{\psi}$ provide a
quick way to find one in a pair from the other in the pair.

\section{Transversality of~$\mathcal{R}_{\mu\nu}(k)$~\eqref{eq:(4.3)}}
\label{sec:C}
In this appendix, we give a neat way to see the
transversality~\eqref{eq:(4.11)} of~$\mathcal{R}_{\mu\nu}(k)$~\eqref{eq:(4.3)}
under the gauge WT identities of the vertex functions~\eqref{eq:(3.26)}.

For this, it is convenient to first introduce
\begin{align}
   &\Check{V}_{\mu_1\dotsb\mu_n}(-p-k_1-\dotsb-k_n,p;k_1,\dotsc,k_n)
\notag\\
   &:=e^{(p+k_1+\dotsb+k_n)^2}
   V_{\mu_1\dotsb\mu_n}(-p-k_1-\dotsb-k_n,p;k_1,\dotsc,k_n)
   e^{p^2}
\label{eq:(C1)}
\end{align}
and
\begin{equation}
   \Check{h}_F(p):=e^{-p^2}h_F(p)e^{-p^2}.
\label{eq:(C2)}
\end{equation}
Then, the WT identities~\eqref{eq:(3.26)} read
\begin{equation}
   \Check{h}_F(p+k)\Check{V}_\mu(-p-k,p;k)\Check{h}_F(p)\cdot k_\mu e^{k^2}
   =\bm{1}_k\Check{h}_F(p)-\Check{h}_F(p+k)\bm{1}_k,
\label{eq:(C3)}
\end{equation}
and
\begin{align}
   &n\Check{V}_{\mu_1\mu_2\dotsb\mu_n}(-p-k_1-\dotsb-k_n,p;k_1,\dotsc,k_n)
   \cdot(k_1)_{\mu_1}e^{k_1^2}
\notag\\
   &=\bm{1}_{k_1}
   \Check{V}_{\mu_2\dotsb\mu_n}(-p-k_2-\dotsb-k_n,p;k_2,\dotsc,k_n)
\notag\\
   &\qquad{}
   -\Check{V}_{\mu_2\dotsb\mu_n}(-p-k_1-\dotsb-k_n,p+k_1;k_2,\dotsc,k_n)
   \bm{1}_{k_1}
\label{eq:(C4)}
\end{align}
for $n\geq2$. In these expressions, the symbol~$\bm{1}_k$ indicates that the
momentum of the fermion line jumps at the inserted point by~$k$ when going from
right to left and otherwise it is simply a unit matrix. For instance,
Eq.~\eqref{eq:(C4)} with~$n=2$, sandwitched by $\Check{h}_F(p+k_1+k_2)$
and~$\Check{h}_F(p)$, yields
\begin{align}
   &\Check{h}_F(p+k_1+k_2)2\Check{V}_{\mu\nu}(-p-k_1-k_2,p;k_1,k_2)\Check{h}_F(p)
   \cdot(k_1)_{\mu_1}e^{k_1^2}
\notag\\
   &=\Check{h}_F(p+k_1+k_2)\bm{1}_{k_1}\Check{V}_\nu(-p-k_2,p;k_2)\Check{h}_F(p)
\notag\\
   &\qquad{}
   -\Check{h}_F(p+k_1+k_2)\Check{V}_\nu(-p-k_2,p+k_1;k_2)\bm{1}_{k_1}\Check{h}_F(p).
\label{eq:(C5)}
\end{align}
The symbol~$\bm{1}_k$ thus keeps track how momenta are assigned for arguments.

To illustrate the idea, let us consider the following example,
\begin{align}
   r_{\mu\nu}(k)
   &:=\int_\ell\,f(\ell)e^{-2\ell^2}
   \tr\{
   h_F(\ell)
   [
   2V_{\mu\nu}(-\ell,\ell;-k,k)
\notag\\
   &\qquad\qquad\qquad\qquad{}
   +V_\mu(-\ell,\ell+k;-k)h_F(\ell+k)V_\nu(-\ell-k,\ell;k)
\notag\\
   &\qquad\qquad\qquad\qquad{}
   +V_\nu(-\ell,\ell-k;k)h_F(\ell-k)V_\mu(-\ell+k,\ell;-k)
   ]
   h_F(\ell)
   \}
\notag\\
   &=\int_\ell\,f(\ell)
   \tr\left[\Check{h}_F(\ell)
   (2\Check{V}_{\mu\nu}
   +\Check{V}_\mu\Check{h}_F\Check{V}_\nu
   +\Check{V}_\nu\Check{h}_F\Check{V}_\mu
   )\Check{h}_F(\ell)\right],
\label{eq:(C6)}
\end{align}
where $f(\ell)$ is a certain function of~$\ell$. Here, we follow the same
momentum assignment abbreviation rule as the main text. The
combination~\eqref{eq:(C6)} actually appears at several places
in~Eq.~\eqref{eq:(4.3)}. Then, Eqs.~\eqref{eq:(C3)} and~\eqref{eq:(C4)}
with~$n=2$ immediately lead to
\begin{align}
   &r_{\mu\nu}(k)(-k)_\mu e^{k^2}
\notag\\
   &=\int_\ell\,f(\ell)
   \tr\left[
   \Check{h}_F(\ell)
   \left(\bm{1}\Check{V}_\nu-\Check{V}_\nu\bm{1}\right)
   \Check{h}_F(\ell)
   +\left(\bm{1}\Check{h}_F-\Check{h}_F\bm{1}\right)
   \Check{V}_\nu\Check{h}_F(\ell)
   +\Check{h}_F(\ell)\Check{V}_\nu
   \left(\bm{1}\Check{h}_F-\Check{h}_F\bm{1}\right)
   \right]
\notag\\
   &=\int_\ell\,f(\ell)
   \tr\left[
   \bm{1}\Check{h}_F\Check{V}_\nu\Check{h}_F(\ell)
   -\Check{h}_F(\ell)\Check{V}_\nu\Check{h}_F\bm{1}
   \right],
\label{eq:(C7)}
\end{align}
where $\bm{1}$ is the abbreviation of~$\bm{1}_{-k}$. Restoring the momentum
variables, this is
\begin{align}
   &r_{\mu\nu}(k)(-k)_\mu e^{k^2}
\notag\\
   &=\int_\ell\,f(\ell)
   \tr\left[
   \Check{h}_F(\ell+k)\Check{V}_\nu(-\ell-k,\ell;k)\Check{h}_F(\ell)
   -\Check{h}_F(\ell)\Check{V}_\nu(-\ell,\ell-k;k)\Check{h}_F(\ell-k)
   \right]
\notag\\
   &=\int_\ell\,[f(\ell)-f(\ell+k)]
   \tr\left[
   \Check{h}_F(\ell+k)\Check{V}_\nu(-\ell-k,\ell;k)\Check{h}_F(\ell)
   \right],
\label{eq:(C8)}
\end{align}
 where, in the last equality, we have made the shift~$\ell\to\ell+k$ in the
second term. Therefore, if $f(\ell)$ does not depend on~$\ell$,
$r_{\mu\nu}(k)$~\eqref{eq:(C6)} is transverse. The point is that we may avoid to
keep track of the explicit momentum variable assignment in this method
(although we did so for the sake of elucidation). Similar calculations can be
repeated for all terms of~Eq.~\eqref{eq:(4.3)} and we finally find that it is
transverse.



%



\let\doi\relax










\end{document}